\documentclass[11pt,a4paper]{article}
\usepackage{jheppub}
\usepackage{subcaption}
\usepackage{slashed}

\usepackage[normalem]{ulem}
\usepackage{bbold}
\usepackage{tikz}
\usepackage{tikz-feynman}
\usepackage{contour}
\makeatletter
\gdef\@fpheader{}
\makeatother
\definecolor{lime}{HTML}{A6CE39}
\DeclareRobustCommand{\orcidicon}{\hspace{-4pt}
	\begin{tikzpicture}
	\draw[lime, fill=lime] (0,0) 
	circle [radius=0.16] 
	node[white] {\hspace{0.1mm}{\fontfamily{qag}\selectfont \tiny ID}};
	\draw[white, fill=white] (-0.07,0.1) 
	circle [radius=0.01];
	\end{tikzpicture}
	\hspace{-3.2mm}
}

\definecolor{darkgreen}{rgb}{0,0.5,0}

\foreach \x in {A, ..., Z}{\expandafter\xdef\csname 
	orcid\x\endcsname{\noexpand\href{https://orcid.org/\csname orcidauthor\x\endcsname}
		{\noexpand\orcidicon}}
}

\newcommand{\ket}[1]{\ensuremath{\left|#1\right\rangle}}

\def \MET{\rm E{\!\!\!/}_T}
\newcommand{\opqdh}{\mathcal{O}_{qdH\phi^2}}
\newcommand{\opquh}{\mathcal{O}_{quH\phi^2}}
\newcommand{\opgh}{\mathcal{O}_{G\phi}}
\newcommand{\opbh}{\mathcal{O}_{B\phi}}
\newcommand{\opwh}{\mathcal{O}_{W\phi}}
\newcommand{\wcqdh}{\mathcal{C}_{qdH\phi^2}}
\newcommand{\wcquh}{\mathcal{C}_{quH\phi^2}}
\newcommand{\wcgh}{\mathcal{C}_{G\phi}}
\newcommand{\wcqq}{\mathcal{C}_{qq}}
\newcommand{\wc}{\mathcal{C}}

\title{Monojet and direct detection constraints on real scalar dark matter: EFT and a simple UV completion} 

\author[a]{Arnab Roy\orcidA{},\,}
\author[b]{Michael A.~Schmidt\orcidB{},\,}
\author[a]{and German Valencia\orcidC{}}
\affiliation[a]{
	School of Physics and Astronomy, Monash University, Wellington Road, Clayton, Victoria 3800, Australia}
\affiliation[b]{
	Sydney Consortium for Particle Physics and Cosmology, School of Physics, The University of New South Wales, Sydney, NSW 2052, Australia}

\emailAdd{arnab.roy1@monash.edu}
\emailAdd{m.schmidt@unsw.edu.au}
\emailAdd{german.valencia@monash.edu}

\abstract{We consider constraints that can be placed on certain invisible scalar particles through monojet studies at the LHC and compare them with those from direct detection experiments when interpreted as dark matter. Whereas direct detection constraints are typically more restrictive, we identify regions of parameter space where monojet studies provide important complementary bounds. We carry out our analysis using both a $\phi$SMEFT for real scalar particle pairs coupled to standard-model fields through operators of up to dimension six, and a simple UV completion with vector-like quarks, with both the scalars and the vector-like quarks being odd under a $\mathbb{Z}_2$ symmetry, while the SM particles are even. The vector-like quarks can only decay into a jet and an invisible scalar, and we recast the current ATLAS monojet data to constrain their parameter space. Comparison of the two descriptions yields some insight into interpreting dark matter constraints obtained with EFTs.
}
\arxivnumber{}
\preprint{CPPC-25-07}

\begin{document}
	
	\maketitle

	\newpage

	\section{Introduction}
	
	Bounds on dark matter from monojet searches at the LHC have been extensively discussed in the literature, both in the context of simplified models~\cite{Papucci:2014iwa,Chala:2015ama,Jacques:2016dqz,Liew:2016oon,Arina:2020udz,Arina:2025zpi} and of effective field theories (EFTs)~\cite{Racco:2015dxa,DeSimone:2016fbz,Bauer:2016pug,Bruggisser:2016nzw,Pobbe:2017wrj,Belwal:2017nkw,Belyaev:2018pqr,Roy:2024ear}.  Some of these studies have discussed the region of validity of the dark matter (DM) EFTs~\cite{Racco:2015dxa,Bauer:2016pug,Pobbe:2017wrj,Belyaev:2018pqr}, and resolved it mostly by relying on analyzing events with partonic centre-of-mass energy less than a hypothetical cut-off scale for the validity of the EFT. 
	In contrast, monojet constraints on UV-complete models have received less attention.
	
	In this paper, we recast the ATLAS monojet search to constrain both a $\phi$SMEFT for real scalar particle pairs coupled to SM fields through operators of up to dimension six and a simple UV completion, 
	the scalar-DM extended vector-like quark (VLQS) model introduced in Ref.~\cite{He:2024iju} in which both the scalars and the vector-like quarks are odd under a $\mathbb{Z}_2$ symmetry, while the SM particles are even. 
	This specific model was motivated by the apparent excess observed by Belle II in the rare decay mode $B\to K\nu\bar\nu$ \cite{Belle-II:2023esi}, and the possibility of interpreting such an excess as deriving from unobserved scalar particles. 
	Our study provides monojet bounds specific to this model, and in particular, we estimate the reach of the LHC in searches for vector-like quarks that decay to jets plus missing energy.
	
	Although general prescriptions to ensure EFT validity have been proposed in the literature~\cite{Busoni:2013lha,Busoni:2014haa,Busoni:2014sya,Racco:2015dxa,Bell:2016obu,DeSimone:2016fbz,Bauer:2016pug,Pobbe:2017wrj} , their robustness has not been systematically tested within specific UV-complete DM models.  
	Our step-by-step comparison of the EFT and its UV complete model serves as a tool to explore the potential pitfalls of relying on EFTs to constrain theories where the new scale is just above the reach of LHC. In particular, we identify three potential issues: 
	\begin{itemize}
		\item The energy scale at which the EFT begins to differ significantly from its completion while still obeying unitarity in terms of the $\MET$. 
		Previous studies have assessed the EFT validity based on the partonic centre-of-mass energy, employing proxy variables such as the invariant mass of the DM pair~\cite{Belyaev:2018pqr} or the invariant mass of the DM pair and the jet~\cite{Pobbe:2017wrj}. They rely on parton-level information, such as the four-momenta of the individual DM particles, which is inaccessible in experimental analyses. In this study, we use the $\MET$ of the event to discuss EFT validity, which is directly measurable at the detector level, and compare it to the partonic centre-of-mass energy. While an EFT validity criterion based on $\MET$ is inherently sensitive to the specific UV completion, this study nevertheless provides a useful estimate of the $\MET$ range over which an EFT description remains trustworthy for new physics (NP) in the few-TeV region. 
		
		\item The possibility of on-shell production of a heavy mediator affects the assumed jet plus $\MET$ topology. VLQS of masses at the assumed EFT scale produce large differences between differential cross-sections in the EFT and UV complete descriptions at or above the EFT scale. In our example, this results in peculiar differences between the two descriptions that can be attributed to a large fluctuation in a single bin of the ATLAS data.
		
		\item The way the kinematic differences between the EFT and its UV completion lead to different results in automated calculations. Specifically, in this case, we pinpoint a new source of mismatch between the EFT and its UV completion in the default choice of the factorisation scale used by {\tt Madgraph5}. 
	\end{itemize}
	
	Our paper is organised as follows. In Section~\ref{sec:framework}, we present the EFT for the SM plus a real scalar with a $\mathbb{Z}_2$ symmetry up to dimension six, as well as the UV completion of \cite{He:2024iju} and the corresponding matching equations. In Section~\ref{sec:monojetanalysis}, we provide details of our recast of \cite{ATLAS:2021kxv} to obtain monojet constraints on both the EFT and VLQS and compare them, detailing the kinematic regions where they disagree. We also obtain the corresponding future sensitivity for the HL-LHC.
	In Section~\ref{sec:directdetection}, we study the constraints on these EFT operators from direct detection (DD) experiments and the requirement to explain the observed DM relic density, and point out the complementarity of the collider and DD approaches. Finally, in Section~\ref{sec:conclusions}, we present our conclusions. Technical details are included in two appendices.

	\section{Theoretical framework} \label{sec:framework}
	
	We consider a real scalar DM candidate, $\phi$, singlet under the SM gauge group, i.e., $\phi \sim (\mathbf{1}, \mathbf{1}, 0)$, which interacts with SM particles via beyond-the-Standard-Model (BSM) mediators. To stabilise this DM particle, we introduce a $\mathbb{Z}_2$ symmetry under which the SM fields are even, but $\phi$  and the heavy mediators are odd as well. We choose the mediators to be heavy compared to the real scalar $\phi$, with mass $\rm m_{med} \gtrsim 1 TeV$.
	In contrast, the DM mass can lie below the electroweak (EW) scale, potentially even below the GeV scale, yet remain undetected due to its suppressed production rate. We implement these basic assumptions both through an EFT framework and a simple UV completion in the following.
	
	\subsection{Effective field theory with operators up to dimension six}
	
	The possible interactions between DM and SM particles can be encapsulated into higher-dimensional interaction terms in the Lagrangian. To this end, we include an extra dark degree of freedom in the dimension-6 Standard Model effective field theory (SMEFT), known as DSMEFT~\cite{Criado:2021trs,Aebischer:2022wnl}. These DM fields, being gauge singlets, interact with the SM fields in a way that preserves the gauge symmetry of the SM, $\rm SU(3)_C \otimes SU(2)_L \otimes U(1)_Y$. The field we have chosen, a real gauge-singlet scalar field that is odd under a $\mathbb{Z}_2$ symmetry, restricts DSMEFT to a subset, which we call $\rm \phi SMEFT$. The $\rm \phi SMEFT$ Lagrangian up to dimension six has the form:
	\begin{align}
	\label{eq:DSMEFT-Lagrangian}
	\mathcal{L}_{\mathrm{\phi SMEFT}} &= \mathcal{L}_{SM}+ \sum_i \mathcal{C}_i \mathcal{O}_i^{(4)}+\frac{1}{\Lambda^2} \sum_i \mathcal{C}_i \mathcal{O}_i^{(6)},
	\end{align}
	where the parameter $\Lambda$ denotes the scale of new physics, above which additional states, such as the heavy mediators in this study, may emerge and interact with both SM and DM particles, $\mathcal{C}_i$  are the dimensionless Wilson coefficients (WC) specifying the strength of each interaction, and $\mathcal{O}_i^{(d)}$ are operators of dimension four or six.  The $\mathbb{Z}_2$ symmetry restricts the EFT operators, $\mathcal{O}_i^{(d)}$, to contain an even number of DM fields, and they are~\cite{Criado:2021trs,Aebischer:2022wnl}:
	\begin{equation}
	\label{eq:phiSMEFT_operators}
	\begin{aligned}
	\opqdh^{pr(6)} &= (\bar{q}_{Lp} d_{Rr} H)\, \phi^2,
	&
	\opquh^{pr(6)}  &= (\bar q_{Lp} u_{Rr} \tilde H) \phi^2,\\
	\mathcal{O}_{B\phi}^{(6)} &= (B_{\mu\nu}B^{\mu\nu})\, \phi^2, & \mathcal{O}_{\widetilde{B}\phi}^{\prime(6)} &= (B_{\mu\nu}\widetilde{B}^{\mu\nu})\, \phi^2,\\
	\mathcal{O}_{W\phi}^{(6)} &= (W_{\mu\nu}^{A}W^{A\mu\nu})\, \phi^2, & \mathcal{O}_{\widetilde{W}\phi}^{\prime(6)} &= (W^{A}_{\mu\nu}\widetilde{W}^{A\mu\nu})\, \phi^2,\\
	\mathcal{O}_{G\phi}^{(6)} &= (G^{A}_{\mu\nu}G^{A\mu\nu})\, \phi^2, & \mathcal{O}_{\widetilde{G}\phi}^{\prime(6)} &= (G^{A}_{\mu\nu}\widetilde{G}^{A\mu\nu})\, \phi^2,\\
	\mathcal{O}_{H\phi 1}^{(4)} &= (H^\dagger H)\, \phi^2, & 
	\mathcal{O}_{H\phi 2}^{(6)} &= (H^\dagger H)^2\, \phi^2,\\
	\mathcal{O}_{H\phi 3}^{(6)} &= (H^\dagger H)\, \phi^4, & 
	\mathcal{O}_{\Box\phi}^{(6)} &= (H^\dagger H) \Box\, \phi^2,\\
	\mathcal{O}_{e\phi}{}  &= (\overline l_i e_j H) \phi^2 .
	\end{aligned}
	\end{equation}
	In these operators $H$ is the SM Higgs doublet, $q_{L}$ and $u_{R}/d_{R}$ are the left-handed quark doublets and right-handed singlets with generation indices $p,r$; $B^{\mu\nu}, W^{\mu\nu}, G^{\mu\nu}$ are the field-strength tensors corresponding to the gauge groups $U(1), SU(2)$, and $SU(3)$ respectively, $\Box\equiv \partial^{\mu}\partial_{\mu}$ is the d'Alembert operator. Operators with a structure $(\phi^\dagger i\overleftrightarrow{\partial}^{\mu}\phi)$  do not appear because they vanish for a \textit{real} scalar field $\phi$. The last operator in the list, with leptons, does not contribute to monojet events. It can, however, be constrained by semi-visible Higgs decay \cite{Dawson:2025dmi}.
	
	Of this list, those operators involving only the scalar dark matter and the SM Higgs fields are already tightly constrained by existing data. Specifically, Higgs-portal interactions are limited by measurements of the invisible decay width of the SM Higgs boson, with current bounds requiring $\rm BR(H_{SM} \to invisible) \lesssim 10\%$~\cite{CMS:2022qva,ATLAS:2022yvh}. For scalar DM with masses below 10 GeV, this constraint leads to DM-nucleon scattering cross-sections that approach the neutrino floor~\cite{Strigari:2009bq,Billard:2013qya,Monroe:2007xp}, rendering the scenario uninteresting to direct detection (DD) experiments~\cite{Arcadi:2021mag}. Although heavier DM masses might still lie above the neutrino floor, they are already excluded by existing DD limits~\cite{Arcadi:2021mag}. As a result, the Higgs portal scenario for scalar DM offers limited phenomenological interest. An exception to this is $\mathcal{O}_{\Box\phi}^{(6)}$ with momentum-suppressed DM-nucleon interations, and a future collider sensitivity study can be found in Ref.~\cite{Ruhdorfer:2024dgz}. The operators with electroweak field strength tensors do not contribute to monojet events, and the operator with the dual gluon field strength tensor violates CP, so we do not include them in our study. We  therefore focus on the remaining three operators (with various flavour indices) from Eq.~\eqref{eq:phiSMEFT_operators} in this study. 
	
	\subsection{Simple UV-completion}
	
	A simple UV completion for the operators in Eq.~\eqref{eq:phiSMEFT_operators} is the one introduced in Ref.~\cite{He:2024iju} (see Ref.~\cite{Moretti:2017qby,Borah:2020nsz,Babu:2021hef,Ghosh:2022rta,Ghosh:2023xhs,Ghosh:2024nkj,Ghosh:2024boo,Das:2024xle,Bhattacharya:2025mlg,Ghosh:2025agw,Olgoso:2025jot} for some other VLQ-related DM analyses). It consists of two heavy vector-like quarks, $Q \sim (\mathbf{3}, \mathbf{2}, 1/6)$ and $D \sim (\mathbf{3}, \mathbf{1}, -1/3)$, with respective masses $m_Q$ and $m_D$ which are also odd under the $\mathbb{Z}_2$ symmetry. The singlet scalar DM field $\phi$ interacts with these vector-like quarks through Yukawa interactions.  
	The Lagrangian for the model can be written as $\mathcal{L}^{\tt NP}=\mathcal{L}_{\tt kinetic}^{\tt NP}-V_{\tt Yukawa}^{\tt NP}-V_{\tt potential}^{\tt NP}$~\cite{He:2024iju}, where
	\begin{subequations} 
		\label{eq:modelL}
		\begin{align}
		\mathcal{L}_{\tt kinetic}^{\tt NP} & = 
		\bar Q i\slashed{D} Q - m_Q \,\bar Q Q 
		+ \bar D i\slashed{D} D - m_D\, \bar D D 
		+ \frac12 \partial_\mu \phi \partial^\mu \phi - \frac12 m_\phi^2 \,\phi^2,
		\\
		V_{\tt Yukawa}^{\tt NP} & = 
		y_{q}^p \,\bar{q}_{Lp} Q_{R} \phi 
		+ y_{d}^p \,\bar{D}_L d_{Rp} \phi 
		- y_1 \,\bar{Q}_L D_R H 
		- y_2 \,\bar{Q}_R D_L H + {\rm h.c.}\;,    
		\\
		V_{\tt potential}^{\tt NP}  & = \frac14 \lambda_\phi\, \phi^4 
		+ \frac12 \kappa\, \phi^2 H^\dagger H\;.
		\end{align}
	\end{subequations}
	Here, the indices $p$ and $r$ label the SM quark generation indices, $Q_R \equiv P_R Q$ and $D_L \equiv P_L D$. We also assume that the lightest $\mathbb{Z}_2$-odd particle is $\phi$, so that $m_\phi < m_{Q/D}$. This scenario leads to decays of $Q$ and $D$ having missing energy. We denote this model by VLQS in the following.  With this $\mathbb{Z}_2$ symmetry assignment, the vector-like quarks in our model always decay with missing energy, and are therefore unconstrained by current ATLAS/CMS bounds on vector-like quarks (see \cite{Benbrik:2024fku} for a review). Our study provides the first constraint on these objects.
	
	\subsection{EFT-UV Matching}
	
	To discuss the correspondence between the two theories, the VLQS model of Eq.~\eqref{eq:modelL} can be matched to the WCs in Eq.~\eqref{eq:phiSMEFT_operators}. The operators $\opqdh$ and $\opquh$ are generated at tree level, and the operators $\opbh$, $\opwh$, and $\opgh$ at 1-loop level: 
	\begin{equation} 
	\label{eq:matching1} 
	\begin{aligned}
	\frac{[\wc_{qdH\phi^2}]_{pr}}{\Lambda^2} & = \frac{ y_q^p y_d^r y_1}{m_Q m_D }
	+ \frac{ y_q^p y_q^{x*} (Y_d)_{xr}}{2 m_Q^2}
	+ \frac{ (Y_d)_{px} y_d^{x*} y_d^r}{2 m_D^2}, 
	\\
	\frac{[\wc_{quH\phi^2}]_{pr}}{\Lambda^2} & =  \frac{ y_q^p y_q^{x*} (Y_u)_{xr}}{2 m_Q^2},\\
	\frac{\wc_{G\phi}}{\Lambda^2} &= {\frac{1}{16\pi^2}} 
	\frac{g_s^2}{6} \left(\frac{(y_d^p)^2}{m_D^2}+\frac{2(y_q^p)^2}{m_Q^2}\right), \\
	\frac{\wc_{B\phi}}{\Lambda^2} &= {\frac{1}{16\pi^2}}     \frac{g_Y^2}{18} \left(\frac{2(y_d^p)^2}{m_D^2}+\frac{(y_q^p)^2}{m_Q^2}\right),\\
	\frac{\wc_{W\phi}}{\Lambda^2} &={\frac{1}{16\pi^2}} 
	\frac{g_L^2}{2} \frac{(y_q^p)^2}{m_Q^2}.
	\end{aligned}
	\end{equation}
	
	From the perspective of monojet searches, the primary focus of this study, the WCs $\wc_{qdH\phi^2}$, $\wc_{quH\phi^2}$, and $\wc_{G\phi}$ are particularly relevant. The WCs $\wc_{B\phi}$ and $\wc_{W\phi}$ mainly contribute to mono-$Z$ or mono-$W$ signatures as already mentioned, and we include their matching for future reference. The matching in Eq.~\eqref{eq:matching1}, at tree-level for the first two couplings and one-loop level for the last three, was obtained with the aid of {\tt Matchete} \cite{Fuentes-Martin:2022jrf}.
	
	In this study, we focus on constraints on $\opqdh$ and $\opgh$ from monojet searches, while also presenting bounds on $\opquh$ as a byproduct of our analysis. Bounds on $\opquh$ can be used for comparison with different UV completions, which may be sensitive to it.\footnote{An example being the two Higgs doublet model plus a real scalar as in \cite{He:2025jfc} or an additional vector-like up quark $U\sim(\mathbf{3},\mathbf{1},2/3)$ with Lagrangian
		\begin{align*}
		\mathcal{L} = \bar U i \slashed{D} U - m_U \bar U U - y_u^p \bar U_L u_{Rp} \phi + y_1^u \bar Q_L U_R \tilde H + y_2^u \bar Q_R U_L \tilde H + \mathrm{h.c.}\;,
		\end{align*}
		which results in additional contributions to the matching
		\begin{align*}
		\delta\frac{[\wc_{quH\phi^2}]_{pr}}{\Lambda^2} & = 
		\frac{ y_q^p y_u^r y_1^u}{m_Q m_U }
		+ \frac{ (Y_u)_{px} y_u^{x*} y_u^r}{2 m_U^2}, 
		& 
		\delta \frac{\wc_{G\phi}}{\Lambda^2} &= {\frac{1}{16\pi^2}} 
		\frac{g_s^2}{6} \frac{(y_u^p)^2}{m_U^2}, 
		& 
		\delta \frac{\wc_{B\phi}}{\Lambda^2} &= {\frac{1}{16\pi^2}}     \frac{g_Y^2}{18} \frac{8(y_u^p)^2}{m_U^2} .
		\end{align*}} 
	We include the operator $\opgh$ but not ${O}_{W\phi}$ nor ${O}_{B\phi}$ in our study because even though all three have coefficients that are loop suppressed, the former is enhanced in gluon-initiated processes at LHC and has much larger hadronic matrix elements relevant for direct detection. For the UV complete case, we restrict our study to the model of \cite{He:2024iju}, without the additional vector-like up-quark
	because its low energy phenomenology has not been studied. For all the computations in this study, we adopt the diagonal up-quark basis, this is the usual basis in which all flavor mixing is shifted entirely into the down-type quark sector.

	\section{Monojet constraints from the LHC} \label{sec:monojetanalysis}
	
	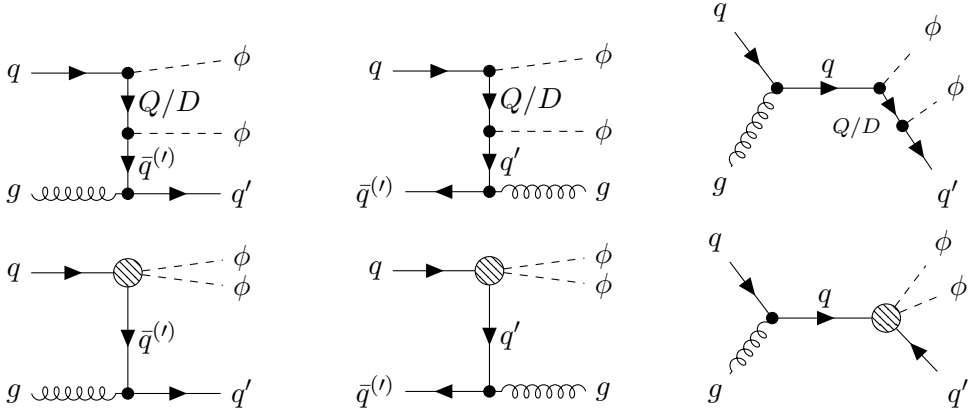
\begin{figure}
        \centering
        \begin{subfigure}[b]{0.3\textwidth}
		\centering
		\begin{tikzpicture}
			\begin{feynman}
				\vertex (a) at (-1.5, 0.8) {${q}$};
				\vertex (b) at (-1.5, -0.8) {$g$};
				\vertex [dot](m1) at (0, 0.8) {};
                \vertex[dot] (m0) at (0,0) {};
				\vertex [dot](m2) at (0, -0.8) {};
				\vertex (c1) at (1.5, 1.0) {$\phi$};
                \vertex (c2) at (1.5, 0.0) {$\phi$};
				\vertex (d) at (1.5, -0.8) {$q^{\prime}$};
				\diagram* {
					(a) -- [fermion] (m1) -- [scalar] (c1),
                    (m0) -- [scalar] (c2),
					(b) -- [gluon] (m2) -- [fermion] (d),
					(m1) -- [fermion, edge label=$Q/D$] (m0) -- [fermion, edge label=$\bar{q}^{(\prime)}$] (m2),
				};
			\end{feynman}
		\end{tikzpicture}
	\end{subfigure}
    \begin{subfigure}[b]{0.3\textwidth}
		\centering
		\begin{tikzpicture}
			\begin{feynman}
				\vertex (a) at (-1.5, 0.8) {${q}$};
				\vertex (b) at (-1.5, -0.8) {$\bar{q}^{(\prime)}$};
				\vertex [dot](m1) at (0, 0.8) {};
                \vertex[dot] (m0) at (0,0) {};
				\vertex [dot](m2) at (0, -0.8) {};
				\vertex (c1) at (1.5, 1.0) {$\phi$};
                \vertex (c2) at (1.5, 0.0) {$\phi$};
				\vertex (d) at (1.5, -0.8) {$g$};
				\diagram* {
					(a) -- [fermion] (m1) -- [scalar] (c1),
                    (m0) -- [scalar] (c2),
					(d) -- [gluon] (m2) -- [fermion] (b),
					(m1) -- [fermion, edge label=$Q/D$] (m0) -- [fermion, edge label=$q^{\prime}$] (m2),
				};
			\end{feynman}
		\end{tikzpicture}
	\end{subfigure}
    \begin{subfigure}[b]{0.3\textwidth}
		\centering
		\begin{tikzpicture}
		\begin{feynman}
        \vertex (a) at (-1.5, 1.0) {$q$};
		\vertex (b) at (-1.5, -1.3) {$g$};
        \vertex (c1) at ( 1.3, 0.8) {$\phi$};
        \vertex (c2) at ( 1.6, 0.0) {$\phi$};
		\vertex (d) at ( 1.5, -1.4) {$q^{\prime}$};
		\vertex[dot] (e) at (-0.75, 0){};
		\vertex[dot] (f) at ( 0.6, 0){};
        \vertex[dot] (g) at ( 0.9, -0.5){};
        
		\diagram* {
			(a) -- [fermion] (e) -- [gluon] (b),
			(f) -- [fermion, edge label'=$\scriptstyle{Q/D}$] (g) -- [fermion] (d),
            (f) -- [scalar] (c1),
            (g) -- [scalar] (c2),
			(e) -- [fermion, edge label=$q$] (f)
		};
		\end{feynman}
		\end{tikzpicture}
        \end{subfigure}
        \begin{subfigure}[b]{0.3\textwidth}
		\centering
		\begin{tikzpicture}
			\begin{feynman}
				\vertex (a) at (-1.5, 0.8) {${q}$};
				\vertex (b) at (-1.5, -0.8) {$g$};
				\vertex [blob, scale=0.5](m1) at (0, 0.8) {\contour{white}{}} {};
				\vertex [dot](m2) at (0, -0.8) {};
				\vertex (c1) at (1.5, 1.0) {$\phi$};
                \vertex (c2) at (1.5, 0.6) {$\phi$};
				\vertex (d) at (1.5, -0.8) {$q^{\prime}$};
				\diagram* {
					(a) -- [fermion] (m1) -- [scalar] (c1),
                    (m1) -- [scalar] (c2),
					(b) -- [gluon] (m2) -- [fermion] (d),
					(m1) -- [fermion, edge label=$\bar{q}^{(\prime)}$] (m2),
				};
			\end{feynman}
		\end{tikzpicture}
	\end{subfigure}
    \begin{subfigure}[b]{0.3\textwidth}
		\centering
		\begin{tikzpicture}
			\begin{feynman}
				\vertex (a) at (-1.5, 0.8) {${q}$};
				\vertex (b) at (-1.5, -0.8) {$\bar{q}^{(\prime)}$};
				\vertex [blob, scale=0.5](m1) at (0, 0.8) {\contour{white}{}} {};
				\vertex [dot](m2) at (0, -0.8) {};
				\vertex (c1) at (1.5, 1.0) {$\phi$};
                \vertex (c2) at (1.5, 0.6) {$\phi$};
				\vertex (d) at (1.5, -0.8) {$g$};
				\diagram* {
					(a) -- [fermion] (m1) -- [scalar] (c1),
                    (m1) -- [scalar] (c2),
					(d) -- [gluon] (m2) -- [fermion] (b),
					(m1) -- [fermion, edge label=$q^{\prime}$] (m2),
				};
			\end{feynman}
		\end{tikzpicture}
	\end{subfigure}
    \begin{subfigure}[b]{0.3\textwidth}
		\centering
		\begin{tikzpicture}
		\begin{feynman}
        \vertex (a) at (-1.5, 1.0) {$q$};
		\vertex (b) at (-1.5, -1.0) {$g$};
        \vertex (c1) at ( 1.5, 1.0) {$\phi$};
        \vertex (c2) at ( 1.7, 0.4) {$\phi$};
		\vertex (d) at ( 1.7, -1.0) {$q^{\prime}$};
		\vertex[dot] (e) at (-0.75, 0){};
		\vertex[blob, scale=0.5] (f) at ( 0.75, 0) {\contour{white}{}};
		\diagram* {
			(a) -- [fermion] (e) -- [gluon] (b),
			(d) -- [fermion] (f) -- [scalar] (c1),
            (f) -- [scalar] (c2),
			(e) -- [fermion, edge label=$q$] (f)
		};
		\end{feynman}
		\end{tikzpicture}
        \end{subfigure}
        \caption{\small A few representative Feynman diagrams of the process $p p \to \phi \phi j$ (monojet) for VLQS (upper pannel) and $\rm \phi SMEFT$ (bottom pannel).}
        \label{fig:monojet_feynman}
    \end{figure}
	
	The operators $\opquh$, $\opqdh$, and $\opgh$, as well as the UV model itself, contribute to the monojet signal at the LHC through the production of DM in association with a jet ($pp \to \phi\phi j$), where the DM appears as missing transverse energy $\rm \MET$ in the detector. Consequently, the absence of any NP signal in monojet searches at the LHC imposes constraints on these EFT operators and on the UV model parameters. A few representative Feynman diagrams of the monojet process are presented in Fig.~\ref{fig:monojet_feynman} both for VLQS (top panel) and $\rm \phi SMEFT$ (bottom panel). 
	
	We will use the latest ATLAS measurement~\cite{ATLAS:2021kxv} at an integrated luminosity of $\rm \mathcal{L} = 140~fb^{-1}$. Our analysis is based on the recast presented in Ref.~\cite{Roy:2024avj}, which uses UFO model files generated by \texttt{FeynRules}~\cite{Alloul:2013bka} into \nolinkurl{Madgraph5} \cite{Alwall:2014hca} for matrix element generation with the \nolinkurl{PDF4LHC15_nlo_mc} PDF set~\cite{Butterworth:2015oua}, followed by parton showering and hadronization via \nolinkurl{Pythia8}~\cite{Sjostrand:2006za,Sjostrand:2007gs}, and detector simulation using \nolinkurl{Delphes3}~\cite{deFavereau:2013fsa} with the ATLAS detector card. The simulation includes pile-up, object reconstruction, resolutions, and efficiencies. The recast has been validated by comparing SM background yields across different $\rm \MET$ bins with those reported by ATLAS~\cite{ATLAS:2021kxv}. For further details, see Section III.B of Ref.~\cite{Roy:2024avj}. The ATLAS results~\cite{ATLAS:2021kxv} are presented  in two forms: inclusive $\MET$ bins ($\text{bin}(i): \MET $$> X_i$ GeV) and exclusive $\MET$ bins ($\text{bin}(i): X_{i+1} > \MET $$> X_i$ GeV). Since the EFT breaks down at some scale in the missing energy range probed, the inclusive bins are not suitable for that analysis. For the UV complete model, we performed the analysis using both sets of bins, obtaining similar results. Since we are interested in comparing the two scenarios, we will only present the analysis that uses the exclusive bins.
	
	Within this setup, we compute the predicted BSM event yield in each $\rm \MET$ bin as a polynomial in the WCs or in the new Yukawa couplings (for the benchmark case $y_q^p=y_d^p$), respectively. This takes the form,
	\begin{align}
	\mathcal{N}^{NP}(\vec{\wc}) &= \sum_{j,k}^{n} \wc_j \wc_k \gamma_{jk},\nonumber \\
	\mathcal{N}^{NP}(\vec{y}) & = \sum_{i=dd,ds,ss} \mathcal{N}_{i}^{NP}(\vec{y}) \\ 
	&\mathrm{with}\;\;\mathcal{N}_{mn}^{NP}(\vec{y}) =  \frac{(y_{q/d}^m)^2(y_{q/d}^n)^2}{(y_{q/d}^d)^2 + (y_{q/d}^s)^2}\beta_{mn} + (y_{q/d}^m)^2 (y_{q/d}^n)^2 \kappa_{mn},
	\label{eq:fit_polynomial}
	\end{align}
	where $\mathcal{N}^{NP}$ denotes the deviation from the SM expectation. The first of these two forms follows from the matrix element being linearly proportional to one of the $\wc_i$. The second form follows from there being contributions from the on-shell production of a VLQ (top-right diagram in Figure~\ref{fig:monojet_feynman}) in addition to quadratic terms in the matrix element.
	The two terms have minimal interference, as we have verified numerically, explaining the absence of a cubic term in Eq.~\eqref{eq:fit_polynomial}.   In the VLQS fit we impose the constraints $\beta_{dd,ss}>0,~\kappa_{dd,ss}>0$ as these terms cannot originate from interference between different diagrams. In addition, for both fits, we impose the constraint that the number of events vanishes when all the coefficients are 0 as there is no SM contribution.
	
	To extract the coefficients ($\gamma_{jk}$, $\beta_{mn}$, $\kappa_{mn}$), we fit the simulated event yields by varying the model parameters. See Ref.~\cite{Guchait:2022ktz} for details on the fitting methodology. The fitted coefficients for the plots in Figure~\ref{fig:region95CL1} are presented in Appendix~\ref{sec:fitted_coefficients}. We then construct a $\chi^2$ for each  $\rm \MET$ bin using these fitted polynomials and compare the prediction to the ATLAS data as,
	\begin{align}
	\chi^2 = \sum_{\rm \MET\,bins}\frac{\left(\Delta\mathcal{N}^{exp} - \mathcal{N}^{NP}(\vec{\wc})\right)^2}{\left[\delta(\Delta\mathcal{N})\right]^2}.
	\end{align}
	In the above expression, $\Delta\mathcal{N}^{exp}$ denotes the observed excess (deficit) over the SM background in the ATLAS data, and $\delta(\Delta\mathcal{N})$ accounts for the experimental uncertainty on the background estimate.

	\subsection{Allowed parameter regions}
	
	For the case of the simple UV completion, we restrict ourselves to the VLQS benchmark of \cite{He:2024iju}, $y_1=1,  \kappa=\lambda=0, M_{Q/D}=3~\rm TeV$, and also assume $y_q^p=y_d^p$ and additionally set $y_2$ to zero for simplicity, as they do not appear in the matching equations.  
	This benchmark\footnote{Recall that the quark Yukawa couplings are negligible as we are not including top in our study.} leads to reduced matching conditions and only two independent couplings at tree level,
	\begin{align}
	&\frac{[\wcqdh]_{11}}{\Lambda^2}  = \frac{ (y_q^d)^2}{m_Q m_D },~    \frac{[\wcqdh]_{22}}{\Lambda^2}  = \frac{ (y_q^s)^2}{m_Q m_D },\\\nonumber 
	&[\wcqdh]_{12}=[\wcqdh]_{21} =\left([\wcqdh]_{11}[\wcqdh]_{22}\right)^{1/2}.
	\end{align}
	When we  compare directly the two sets of constraints, we will only concern ourselves with the parameter space $([\wcqdh]_{11}-[\wcqdh]_{22})\leftrightarrow (y_{q/d}^d-y_{q/d}^s)$.
	
	The two VLQ masses are chosen near the reach of the LHC and slightly different to avoid numerical issues in the simulation. 
	
	Since the initial states $d\bar{s},s\bar{d},d\bar{d},s\bar{s}$ all occur through the couplings of s- and d-quarks to vector-like quarks, we need to include four operators of the type ${\cal{O}}_{qdH\phi^2}$, with coefficients $[\wcqdh]_{ij}/\Lambda^2$ , $i,j\in\{1,2\}$, and similarly for $ {\cal{O}}_{quH\phi^2}$. For our EFT analysis, we refer to a single parameter $\wcqdh$ , or $\wcquh$ in the plots we present. These are to be understood as corresponding  to having taken the WCs for all four pairs of flavour indices to be the same. 
	
\begin{figure}[t]
		\centering
		\includegraphics[width=4.9 cm]{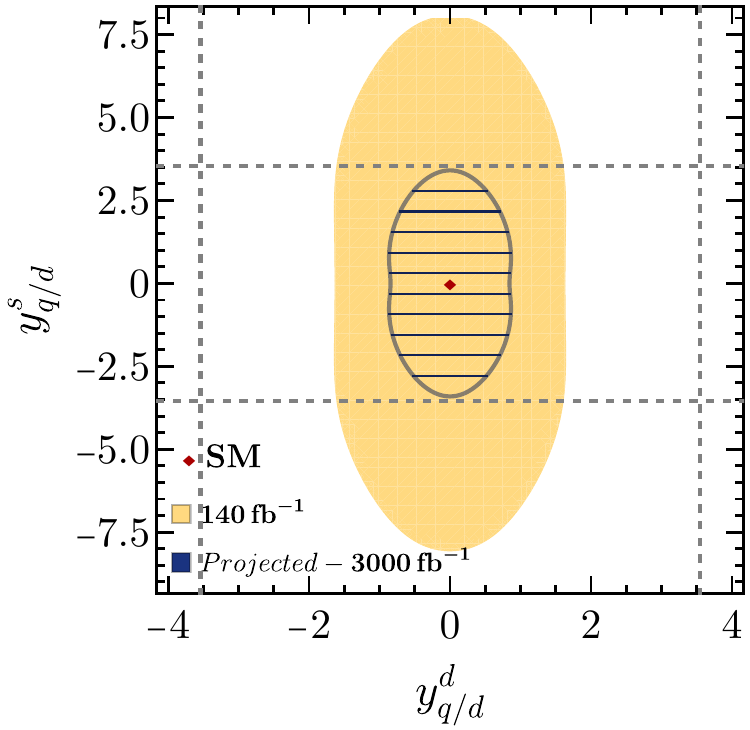}
		\includegraphics[width= 4.9 cm]{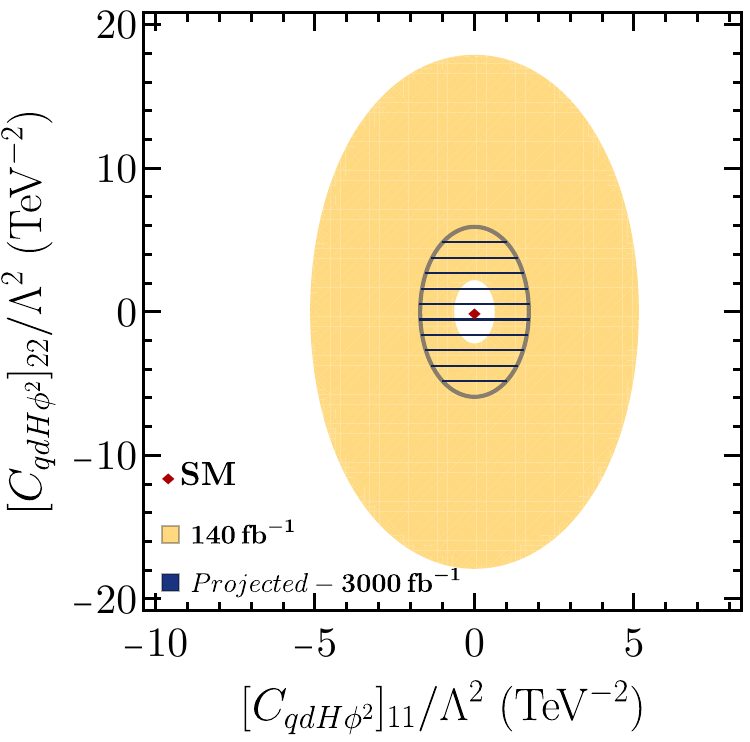}
		\includegraphics[width= 4.9 cm]{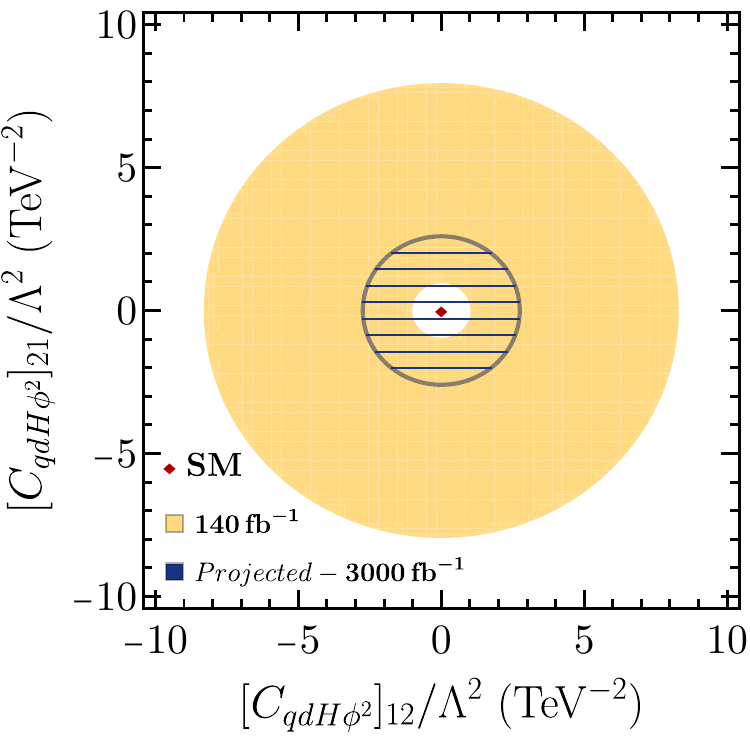}
		\includegraphics[width= 4.9 cm]{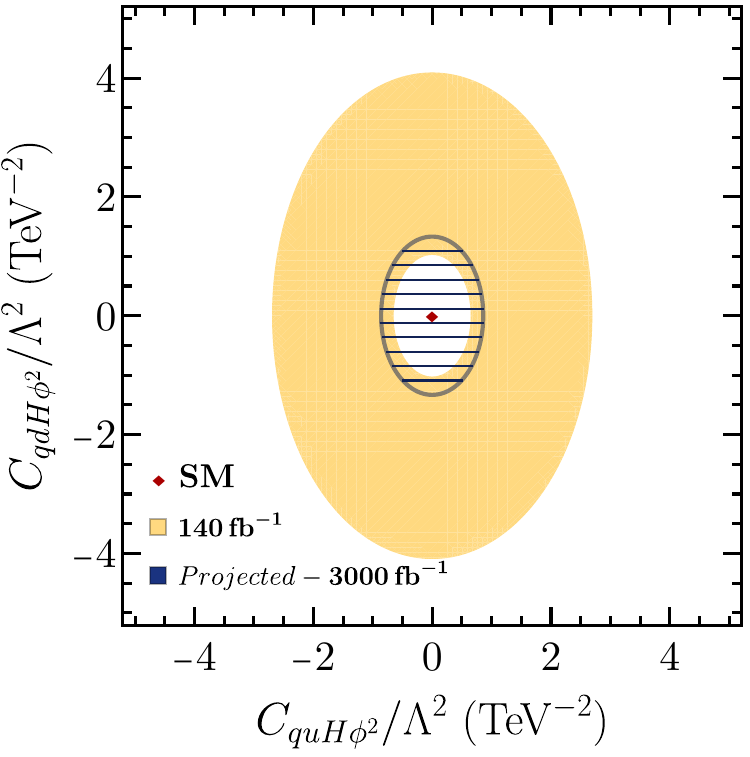}
		\includegraphics[width= 4.9 cm]{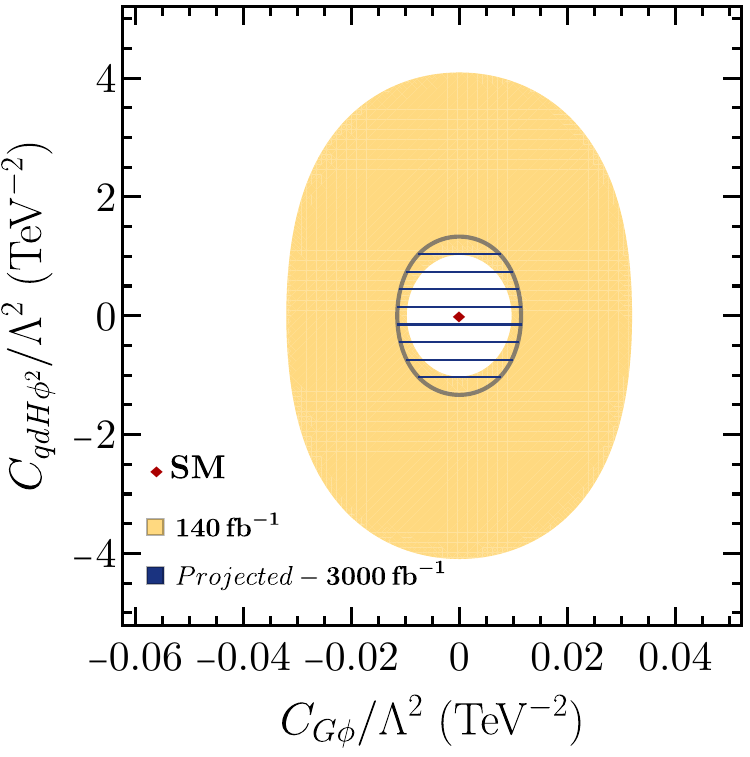}
		\caption{\small Allowed (solid) and projected (hatched) 95$\%$ CL regions at luminosities $\mathcal{L}=140$ and $3000~\rm fb^{-1}$ respectively for VLQS parameters ($y_{q/d}^d, y_{q/d}^s$) at the chosen benchmark, $y_1=1,  \kappa=\lambda=0, M_{Q/D}=3~\rm TeV$ (top left), and selected $\rm \phi SMEFT$ WCs (remaining panels) shown in yellow. The DM mass is chosen to be $m_\phi=1$ GeV. The dashed lines in the top-left panel indicate perturbativity limits.
		}
		\label{fig:region95CL1}
	\end{figure}

	The parameter regions allowed at the 95$\%$ CL from the monojet measurements on both the UV-model and the $\rm \phi SMEFT$ WCs, are shown in Figure~\ref{fig:region95CL1}. The yellow regions correspond to LHC Run-II data with an integrated luminosity of $\mathcal{L} = 140\; \mathrm{fb^{-1}}$. These regions do not include the SM (0,0) for $\rm \phi SMEFT$. This peculiar result can be traced back mainly to the ATLAS data differing from the SM by more than 5$\sigma$ for the EM10 bin ($1000{\rm ~GeV}<\MET<1100{\rm ~GeV}$), as can be seen in the bottom panel of Figure~\ref{fig:mphiphi_monojet}, where we also show the pull from the different exclusive ATLAS bins. On the other hand, for the VLQS, the effective number of events in the higher $p_T$ bins is significantly larger. Although the point $(0,0)$ does not correspond to the minimum of the total $\chi^2$, the minimum lies close enough that $(0,0)$ (the SM) remains within the $95\%$ CL region. This indicates that, at high $\MET$, the UV completion can more easily accommodate a deviation than the $\rm \phi SMEFT$.
	
	To estimate the future projection at the LHC with $\mathcal{L} = 3000\; \mathrm{fb^{-1}}$, we adopt an Asimov dataset approach \cite{Cowan:2010js} under the background-only hypothesis, i.e., assuming that no signal excess is observed (i.e., $\Delta\mathcal{N}^{exp}=0$) and that the number of observed events equals the expected SM background scaled by luminosity. The uncertainty in SM background prediction is assumed to scale as $\delta_{sys} \to \sqrt{3000/140}\,\delta_{sys}$ (hashed blue).\footnote{The ATLAS result quotes only a combination of systematic and statistical uncertainty for the SM background estimation, and since the study is partly data driven, it is difficult to know how the systematic uncertainty may be reduced in the future.} The projections always contain the SM point by construction, and only estimate future sensitivity. 
	
	A rough estimate of the improvement in sensitivity at the HL-LHC, assuming no excess ($\Delta\mathcal{N}^{exp}=0$ in both cases), suggests that, 
	\begin{itemize}
		\item For the VLQS, in terms of a generic coupling $y$ and associated sensitivity $\delta_y$, we have $\sigma(y) \propto y^4$, which combined with  $\chi^2 \propto \frac{\sigma(y)^2 \mathcal{L}^2}{ \mathcal{L}}$, results in $(\delta_y)_{HL}/(\delta_y)_{\rm current}\sim  \left(\frac{\mathcal{L}_1}{\mathcal{L}_2}\right)^{1/8}\sim 0.68$. 
		\item For the EFT, in terms of a generic coupling $\mathcal{C}$ and associated sensitivity $\delta_{\mathcal{C}}$, we have $\sigma(\mathcal{C}) \propto \mathcal{C}^2$, which combined with $\chi^2 \propto \frac{\sigma(\mathcal{C})^2 \mathcal{L}^2}{\Delta^2 \mathcal{L}} \propto \mathcal{C}^4 \times \mathcal{L}$, results in $(\delta_{\mathcal{C}})_{HL}/(\delta_{\mathcal{C}})_{\rm current}\sim\left(\frac{\mathcal{L}_1}{\mathcal{L}_2}\right)^{1/4}\sim 0.46$.
	\end{itemize}
	These estimates are in rough agreement with the contours shown in Figure~\ref{fig:region95CL1}, with small differences arising from the different treatment of $\Delta\mathcal{N}^{\text{exp}}$ in the current versus projected luminosity estimates.
	
	The 95\% CL regions displayed in Figure~\ref{fig:region95CL1} are obtained as follows. The VLQS plot (top left), is based on a two-parameter minimisation of the $\chi^2$ with the assumption $y_q^d = y_d^d$ and $y_q^s = y_d^s$. For the first two EFT plots, also in the top panel, we minimise a $\chi^2$ with four independent parameters $[\wcqdh]_{ij}, ij \in {11, 22, 12, 21}$, and show two-dimensional slices passing through the origin. For the two EFT plots in the bottom panel, there are only two independent parameters since 
	$\wcqdh$ and $\wcquh$ refer to a scenario where all four combinations of flavour indices are taken to be the same. Together with $\wcgh$, this results in a three-dimensional parameter space for which we show 
	two-dimensional slices through the origin. The centre and right panels at the top display a hierarchy of constraints with the tightest occurring for $[\wcqdh]_{11}$, followed by $[\wcqdh]_{21}$ and lastly $[\wcqdh]_{22}$. This can be understood as arising from the parton distribution functions for a $pp$ collider where $f_u>f_d>f_s$.
	
	\subsection{EFT-UV comparison and discussion}
	
	If the EFT provides a valid low-energy description of the high-scale UV model, we expect the bounds presented in Figure~\ref{fig:region95CL1} on the UV and EFT parameters to be \textit{compatible}. This means that translating the bounds on the WCs into the VLQS parameter space using the matching equations should yield approximately the same allowed region as that obtained by applying the bounds directly to the VLQS model. 
	
\begin{figure}[tb!]
		\centering
		\includegraphics[width=0.31\linewidth]{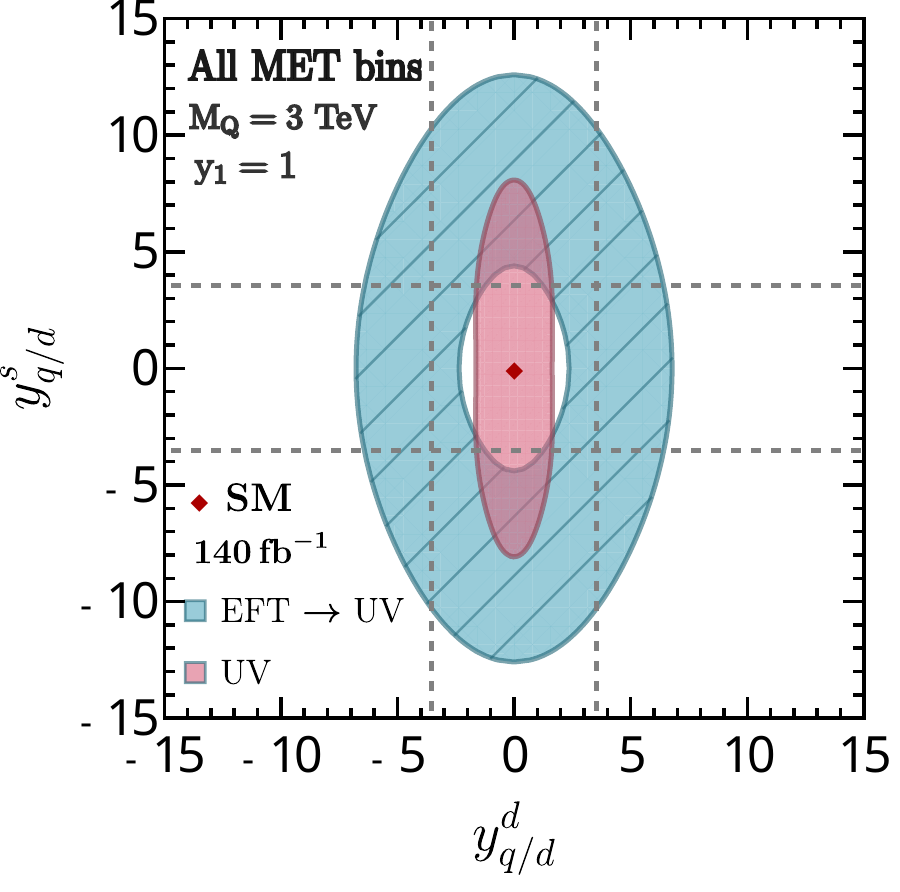}
		\includegraphics[width=0.31\linewidth]{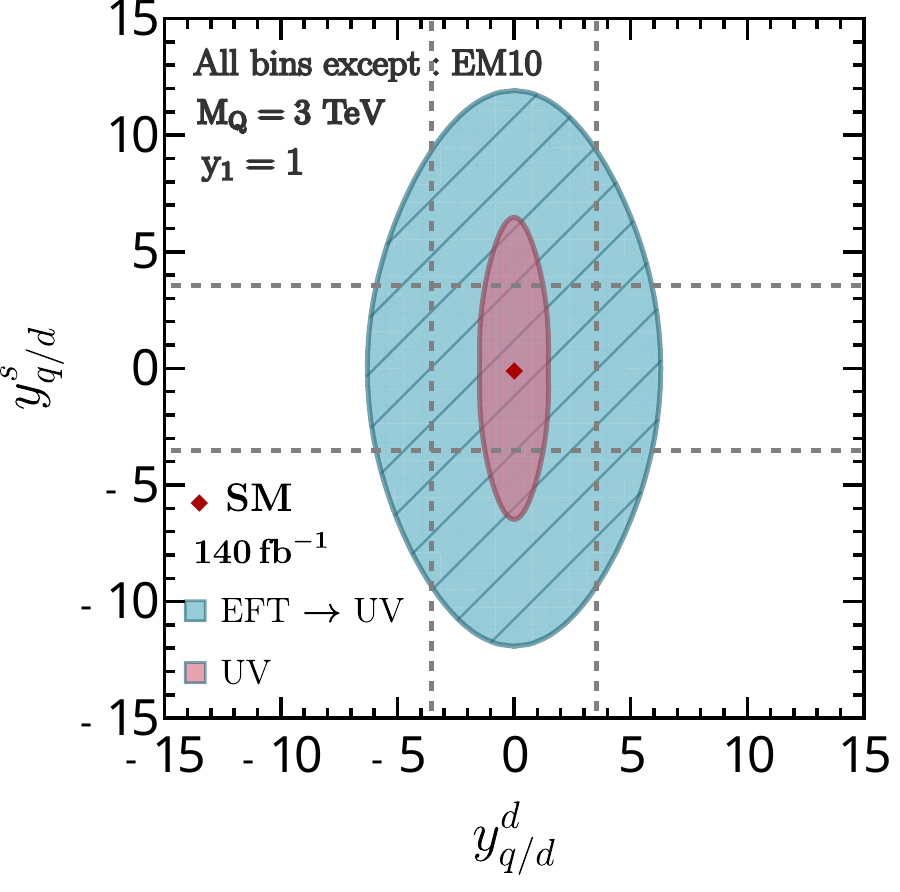}
		\includegraphics[width=0.31\linewidth]{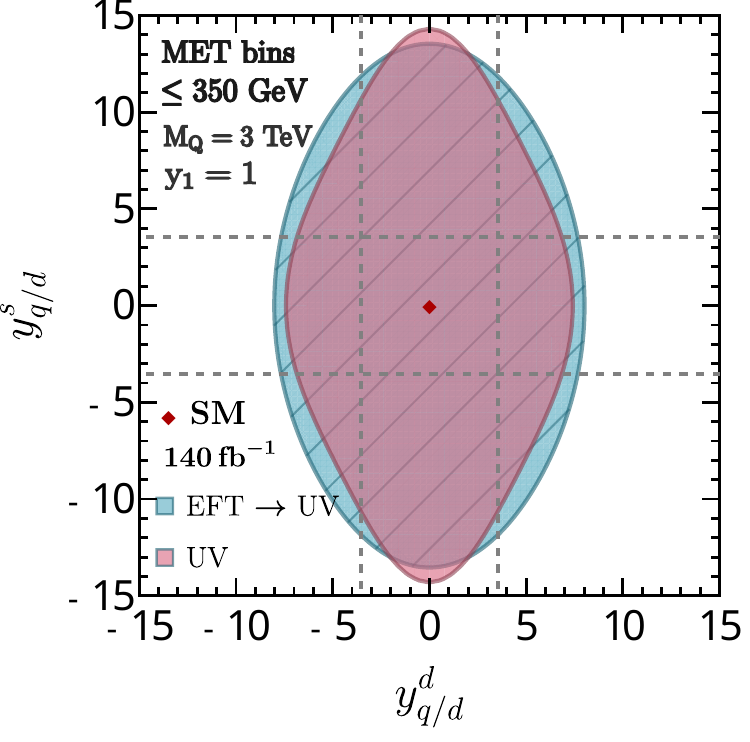}
		\caption{Comparison of the allowed parameter region for a UV complete model obtained in two different manners:  directly constraining the UV completion (VLQS) (purple), translating constraints placed on the EFT ($\rm \phi SMEFT$) (blue) via the matching equations \eqref{eq:matching1} for $M_{Q}=3$ TeV and $m_{\phi}=1$ GeV. The left plot shows the result when using all the exclusive $\MET$ bins, and the centre panel excludes the EM10 $\MET$ bin. The right panel uses only bins with $\MET<350$ GeV. The gray dashed lines indicate the perturbativity limit for $y_{q/d}^{d/s}$.
		}
		\label{fig:EFT_UV_comparison}
	\end{figure}

	Using Eq.~\eqref{eq:matching1}, we transform the allowed EFT parameter region to the corresponding region in the $y_{q/d}^{1}-y_{q/d}^{2}$ parameter space, for the benchmark $y_{q}^i=y_{d}^i$. The result is presented in Figure~\ref{fig:EFT_UV_comparison} as obtained in two different manners:
	\begin{enumerate}
		\item by directly constraining the VLQS (purple).
		\item by constraining the $C_{qdH\phi^2}$ WCs in the $\rm \phi SMEFT$ and then mapping those constraints onto the parameters of the UV completion via the matching equations \eqref{eq:matching1}  (blue).
	\end{enumerate}
	
	This figure illustrates how adopting constraints on EFT WCs as indicative of constraints on UV-complete theories via matching conditions can go wrong. 
	In the left panel of the top row, we use all MET bins for the comparison. There are two salient features: the EFT region does not contain the SM, and it is significantly larger than the VLQS region. The first issue can be traced to the EM10 bin, as already mentioned. To further verify this, the center panel shows the same comparison with EM10 excluded. In the EM10 range of $\MET$ the UV completion can more easily accommodate a deviation than the EFT when the relevant parameters are away from zero. This explains the relative size of the excluded region around the SM in Figure~\ref{fig:EFT_UV_comparison}. 
	
	The rightmost panel is obtained using only bins with $\MET<350$ GeV, a region where the EFT and its UV completion have very similar $\MET$ and $M (\phi,\phi,j)$ spectra as seen in Figure~\ref{fig:mphiphi_monojet}, and where the ATLAS data does not show any large deviations from the SM. One expects this case to result in very similar allowed parameter regions from the two procedures, and this can be seen in the Figure. As can also be seen in this panel, the bounds in this case are not significant as they are outside the perturbativity limits.
	Below, we discuss in some detail the origin of the discrepancy in this example.

	\subsection{Direct comparison of distributions between the EFT and its UV completion}
	
	\begin{figure}[tb!]
		\centering
		\includegraphics[width=6.5 cm]{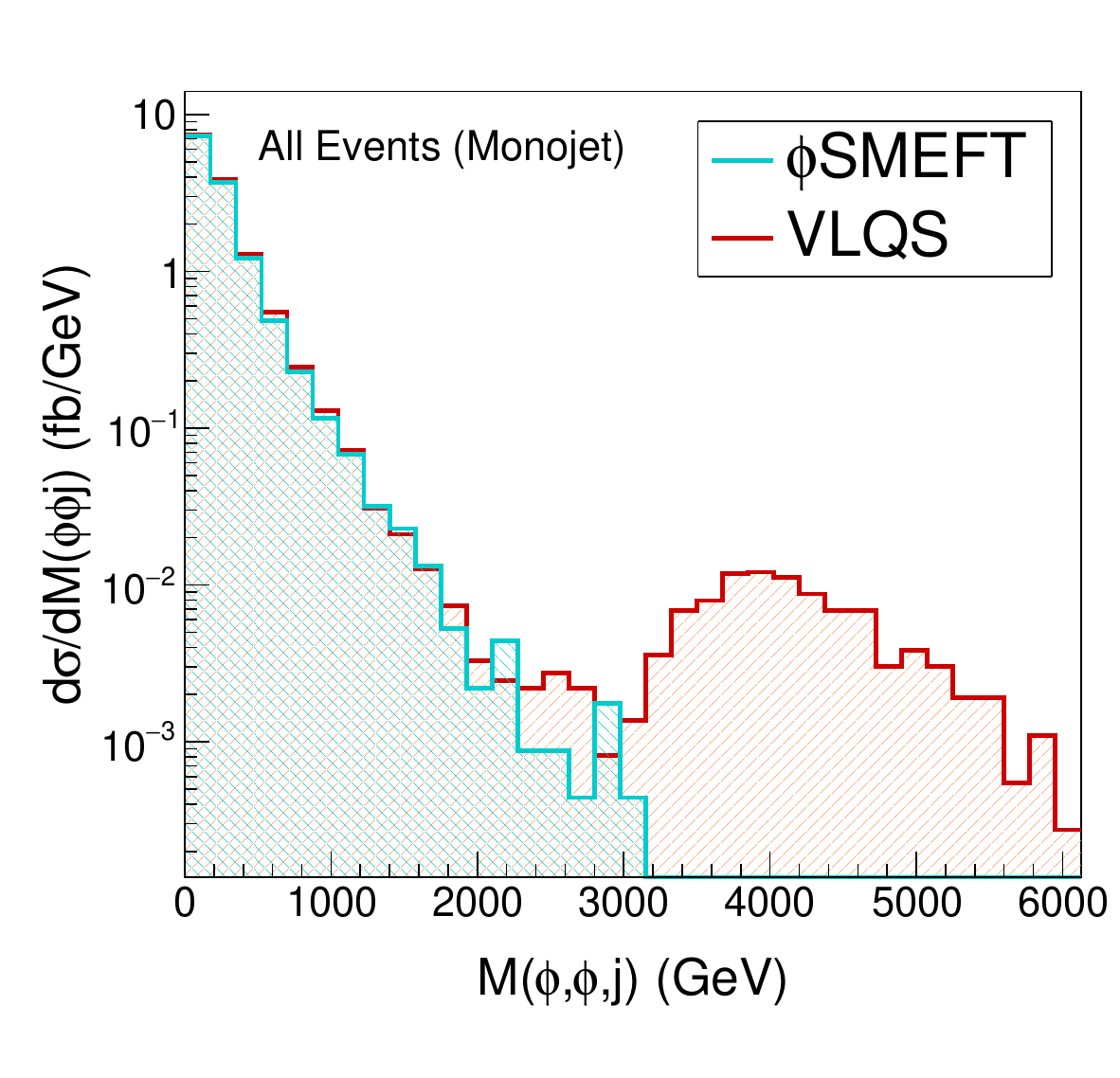}
		\includegraphics[width=6.5 cm]{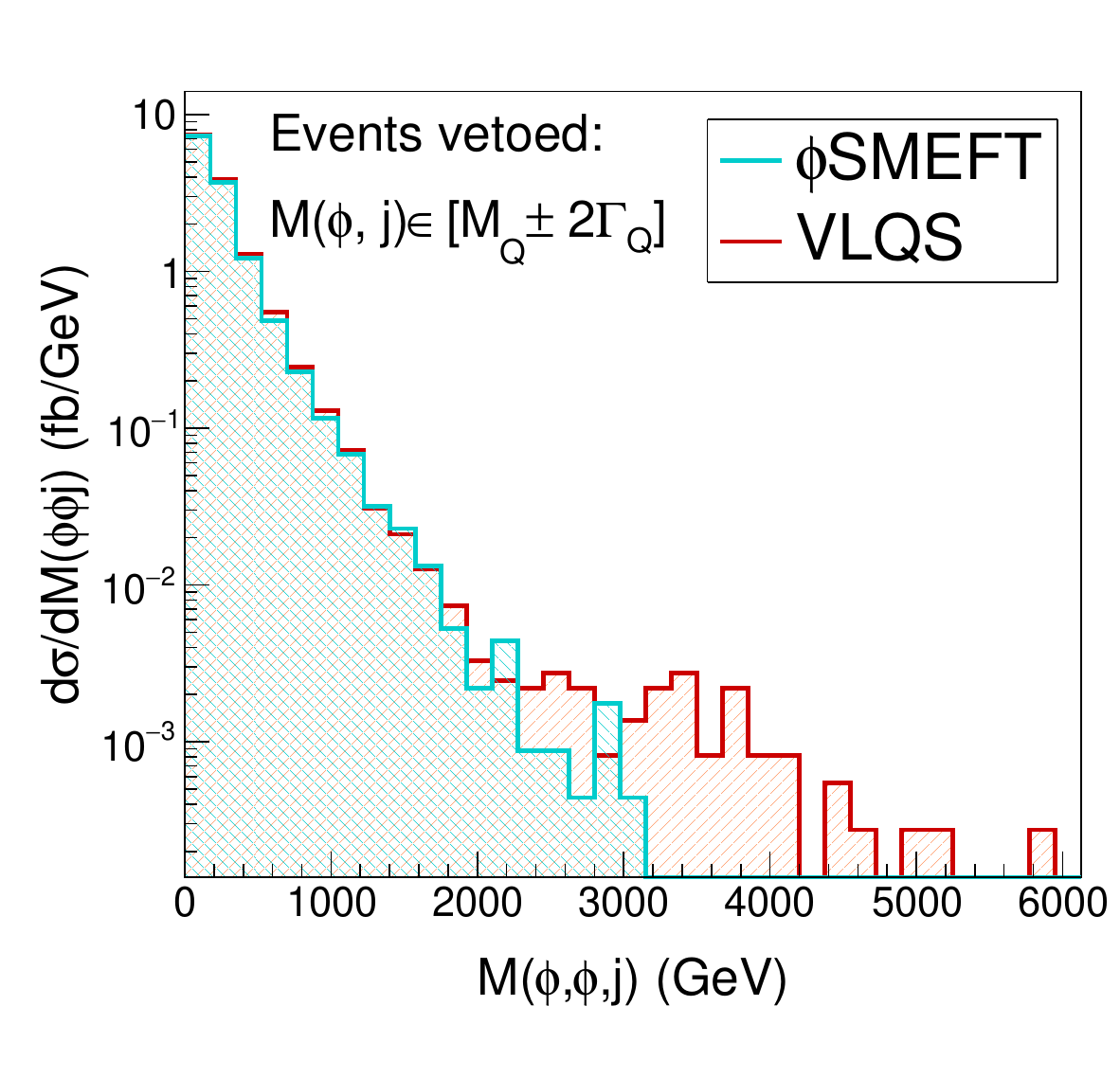}
		\includegraphics[width=6.5 cm]{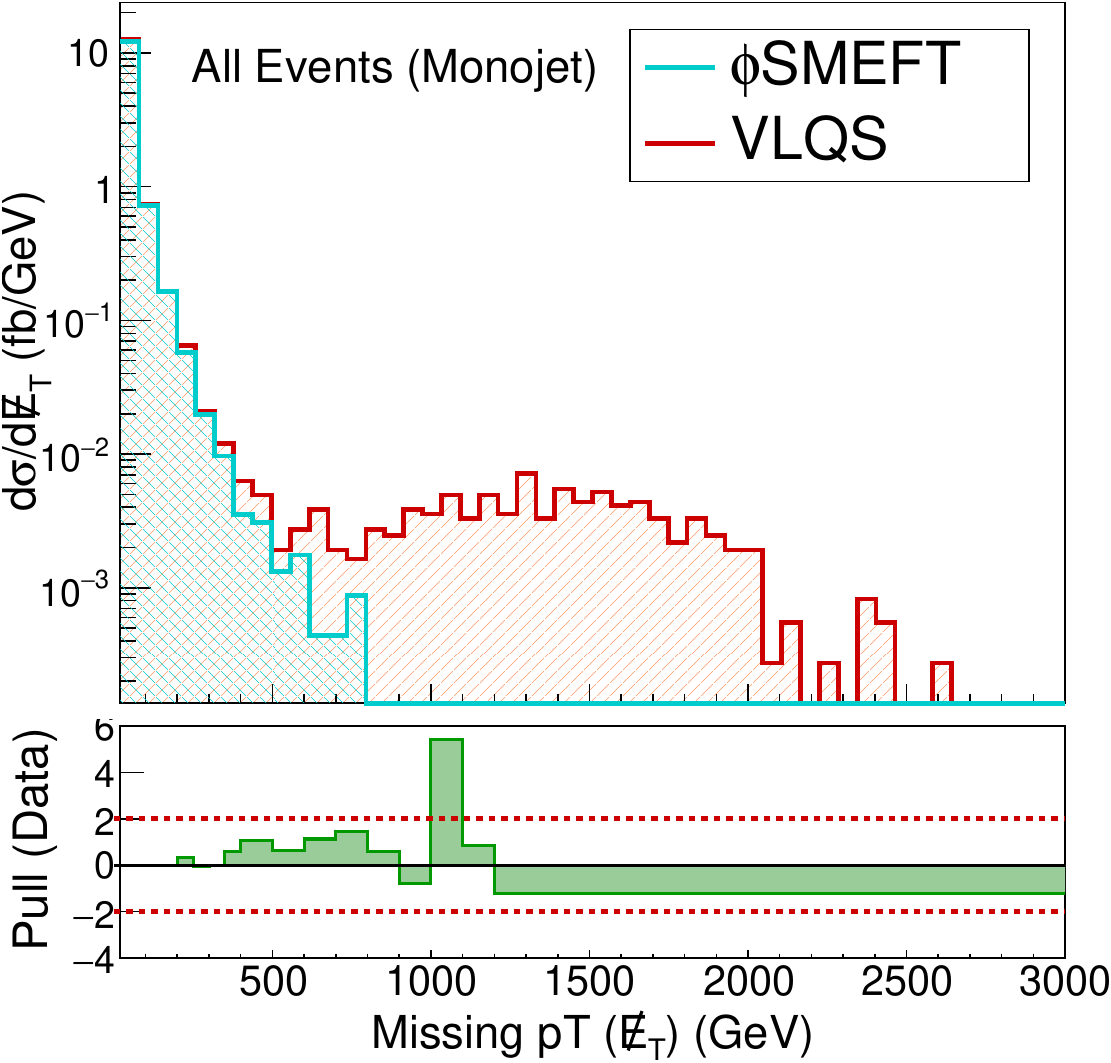}
		\includegraphics[width=6.5 cm]{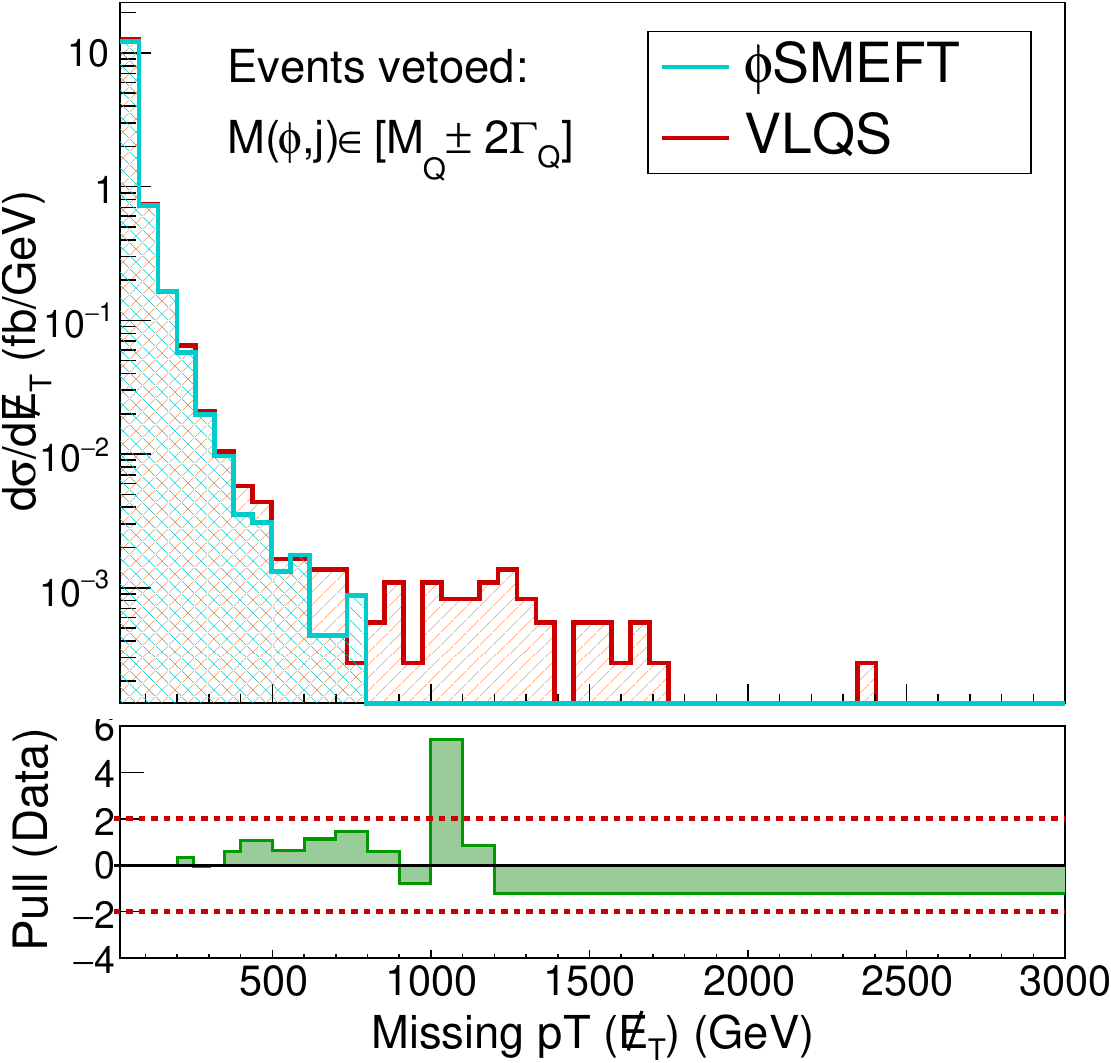}
		\caption{\small Distribution of the truth-level invariant mass of two $\phi s$ and the jet ($M(\phi,\phi,j)$, top panel) and Missing pT ($\MET$, bottom panel) for the monojet process $pp\to \phi\phi j$ at $\sqrt{s}=13$ TeV LHC. The left figures consist of the full set of events, whereas in the right figures, events with at least one $\rm M(\phi,j)\in [M_Q-2\Gamma_{Q}, M_Q+2\Gamma_{Q}]$ are vetoed. The NP-scale ($M_{Q/D}$ and $\Lambda$) is set to 3 TeV and $\rm m_{\phi}=1$ GeV. The parameters for the VLQS $y_{q/d}^d = y_{q/d}^s  = y_1 = 1$ and all the relevant $\rm \phi SMEFT$ $\wc = 1$. The subplots in the bottom panel present the deviation of the observed ATLAS events from the standard model prediction, quantified by pull.}
		\label{fig:mphiphi_monojet}
	\end{figure}
	
	In Figure~\ref{fig:mphiphi_monojet} we directly compare the predictions of the EFT and its UV completion for two distributions:  the invariant mass of the final state particles $\rm M(\phi,\phi, j)$ (top panel) and the missing-$\rm p_T$ ($\MET$, equivalent to $\rm p_T$(jet)) (bottom panel), noting that only the latter is observable. To assist our discussion, we also show with the $\rm \MET$ distribution the pull of the different exclusive bins in the ATLAS analysis, defined as ${\rm Pull}_i=(N_{\rm observed}-N_{\rm predicted})/\sigma_{\rm predicted}$ using the numbers reported in Table~VIII of \cite{ATLAS:2021kxv}. For the NP predictions, we choose the benchmark point 
	$M_{Q/D}=3$ TeV, $y_{q/d}^d=y_{q/d}^s=1$, and $y_1=1$ for the UV model, whereas for the EFT, we use $\wcqdh/\Lambda^2=\wcquh/\Lambda^2=1/(3~\mathrm{TeV})^2$ to match the magnitude of the UV model benchmark point. 
	
	The $\rm M(\phi,\phi, j)$ differential cross-section in the left panel shows the EFT tracking its UV completion until around 2.5~TeV, after which the EFT keeps falling as $\rm M(\phi,\phi, j)$ increases, whereas the UV completion rises above it.  This rise in cross-section near the scale $\Lambda$ is due to on-shell production of a vector-like quark followed by its decay, as in the top-right Feynman diagram of Fig.~\ref{fig:monojet_feynman}, and it is also reflected in the $\MET$ distribution.
	This feature is responsible for allowing the VLQS to match the up-fluctuation of events ATLAS sees in bin EM10 with smaller values of the new Yukawa couplings than would be necessary with the EFT. The right panels present the distributions resulting from removing events with at least one $\mathrm{M}(\phi,j) \in [M_Q - 2\Gamma_{Q}, M_Q + 2\Gamma_{Q}]$, which eliminates a significant fraction of these peaks, confirming the above statements. To understand these effects across different $\MET$ bins in the ATLAS analysis, we observe in the bottom panel of Figure~\ref{fig:mphiphi_monojet} that the EFT closely tracks the UV model up to around $\MET = 350$~GeV, and that significant deviations occur beyond this point. A lower mass for the heavy mediators (i.e., $M_Q=M_D<3$ TeV) lowers this breakdown scale, whereas a larger mass extends it. An important observation is that the discrepancy between EFT and VLQS seen in the $\rm \MET$ distribution begins at a much lower scale than one would expect from the invariant mass distribution.
	
	\subsubsection{Underlying \texorpdfstring{$2\to2$}{2->2} quark-level sub-process }
	
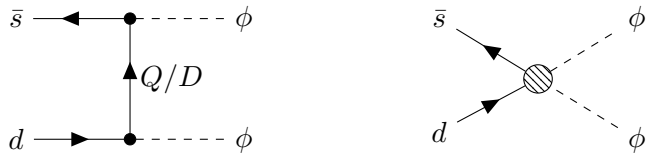
\begin{figure}
		\centering
		\begin{subfigure}[b]{0.49\textwidth}
			\centering
			\begin{tikzpicture}
			\begin{feynman}
			\vertex (a) at (-1.5, 0.8) {$\bar{s}$};
			\vertex (b) at (-1.5, -0.8) {$d$};
			\vertex [dot](m1) at (0, 0.8) {};
			\vertex [dot](m2) at (0, -0.8) {};
			\vertex (c) at (1.5, 0.8) {$\phi$};
			\vertex (d) at (1.5, -0.8) {$\phi$};
			\diagram* {
				(a) -- [anti fermion] (m1) -- [scalar] (c),
				(b) -- [fermion] (m2) -- [scalar] (d),
				(m1) -- [anti fermion, edge label=$Q/D$] (m2),
			};
			\end{feynman}
			\end{tikzpicture}
		\end{subfigure}
		\begin{subfigure}[b]{0.49\textwidth}
			\begin{tikzpicture}
			\centering
			\begin{feynman}
			\vertex[blob, scale=0.5] (m) at ( 0, 0) {\contour{white}{}};
			\vertex (a) at (-1.3,-0.7) {$d$};
			\vertex (b) at ( 1.3,-0.8) {$\phi$};
			\vertex (c) at (-1.3, 0.8) {$\bar{s}$};
			\vertex (d) at ( 1.3, 0.8) {$\phi$};
			\diagram* {
				(a) -- [fermion] (m) -- [fermion] (c),
				(b) -- [scalar] (m) -- [scalar] (d),
			};
			\end{feynman}
			\end{tikzpicture}
		\end{subfigure}
		\caption{Feynman diagrams of $d\bar{s}\to \phi\phi$ in VLQS (left) and $\rm \phi SMEFT$(right) .}
		\label{fig:dspp_feynman}
	\end{figure}

\begin{figure}
		\centering
		\includegraphics[width=0.6\linewidth]{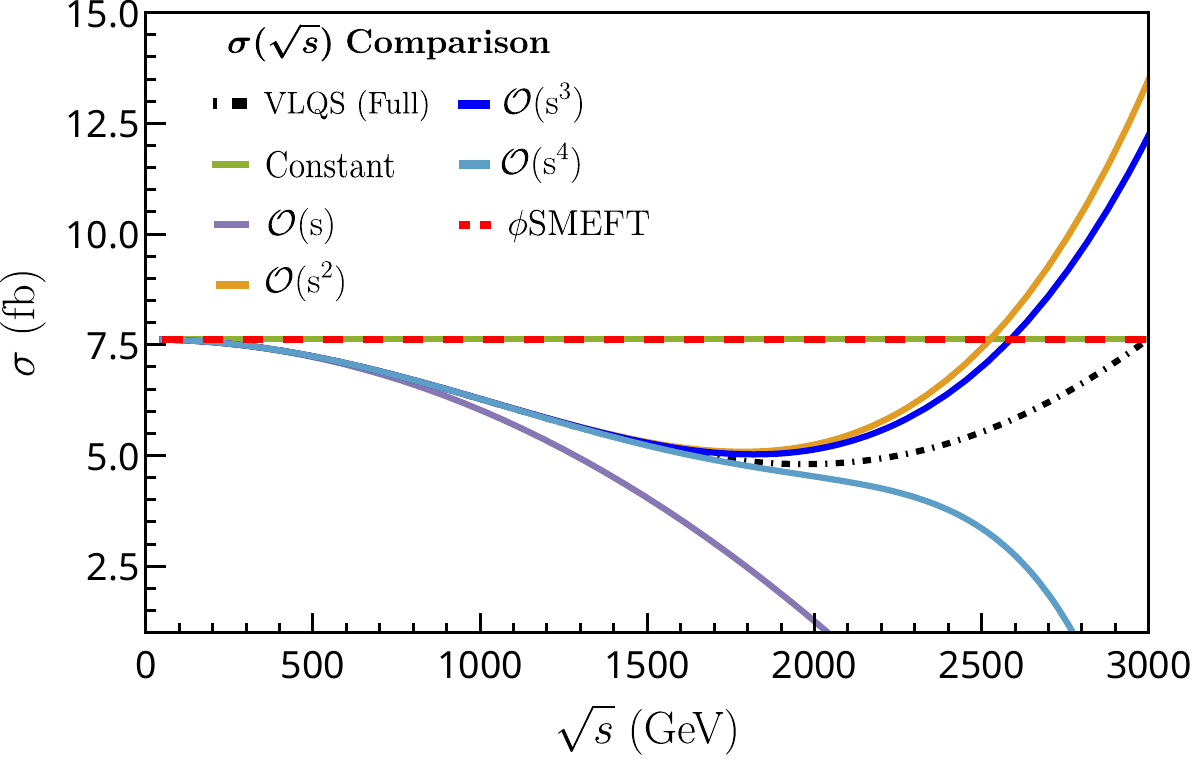}
		\caption{\small Variation of cross-section with $s$, computed analytically for the $2\to 2$ subprocess $d\bar{s}\to \phi\phi$. The different lines mark different orders in an energy expansion of the VLQS cross-section (black dashed line) as well as the $s$-independent $\phi$SMEFT cross-section (red dashed line). The NP scale $M_{\rm Q/D}=\Lambda$ is set to $3$ TeV.
		}
		\label{fig:s_dependence_analytic}
	\end{figure}

	To understand the origin of the differences seen in Figure~\ref{fig:mphiphi_monojet}, we performed an analytical comparison of the cross-sections for the underlying $2 \to 2$ subprocess $d\bar{s} \to \phi\phi$. The relevant Feynman diagrams are presented in Figure~\ref{fig:dspp_feynman}. We first look at the dependence of the cross-section on the centre-of-mass energy $\sqrt{{s}}$ for the benchmark choice $M_{Q/D} = \Lambda = 3$ TeV, and an appropriate choice of parameters ($y_1= y_{q/d}^d=y_{q/d}^s=1$ for VLQS and $[\wcqdh]_{12} = [\wcqdh]_{21} = 1$ for $\rm \phi SMEFT$), guided by  Eq.~\eqref{eq:matching1}, so that the two descriptions agree when the VLQS result is expanded to leading order. In the EFT, the invariant matrix element squared for the scattering process grows with ${s}$, resulting in a cross-section that remains constant as a function of ${s}$ (dashed red line in Figure~\ref{fig:s_dependence_analytic}). In contrast, the VLQS cross-section has a complicated dependence on ${s}$. Whereas the first order expansion (a constant term) matches the EFT by construction, Figure~\ref{fig:s_dependence_analytic} shows that one needs to go to higher order in the expansion (we show terms up to $s^4$) for it to track the VLQS cross-section up to around $\sqrt{{s}}\sim 2$~TeV or 60\% of the scale $\Lambda$.  The leading (constant) term, however, is a very poor approximation to the full result for $\sqrt{{s}}\gtrsim 400$~GeV. 
	
	\subsubsection{Underlying parton-level sub-process }
	
\begin{figure}
		\centering
		\includegraphics[width=0.49\linewidth]{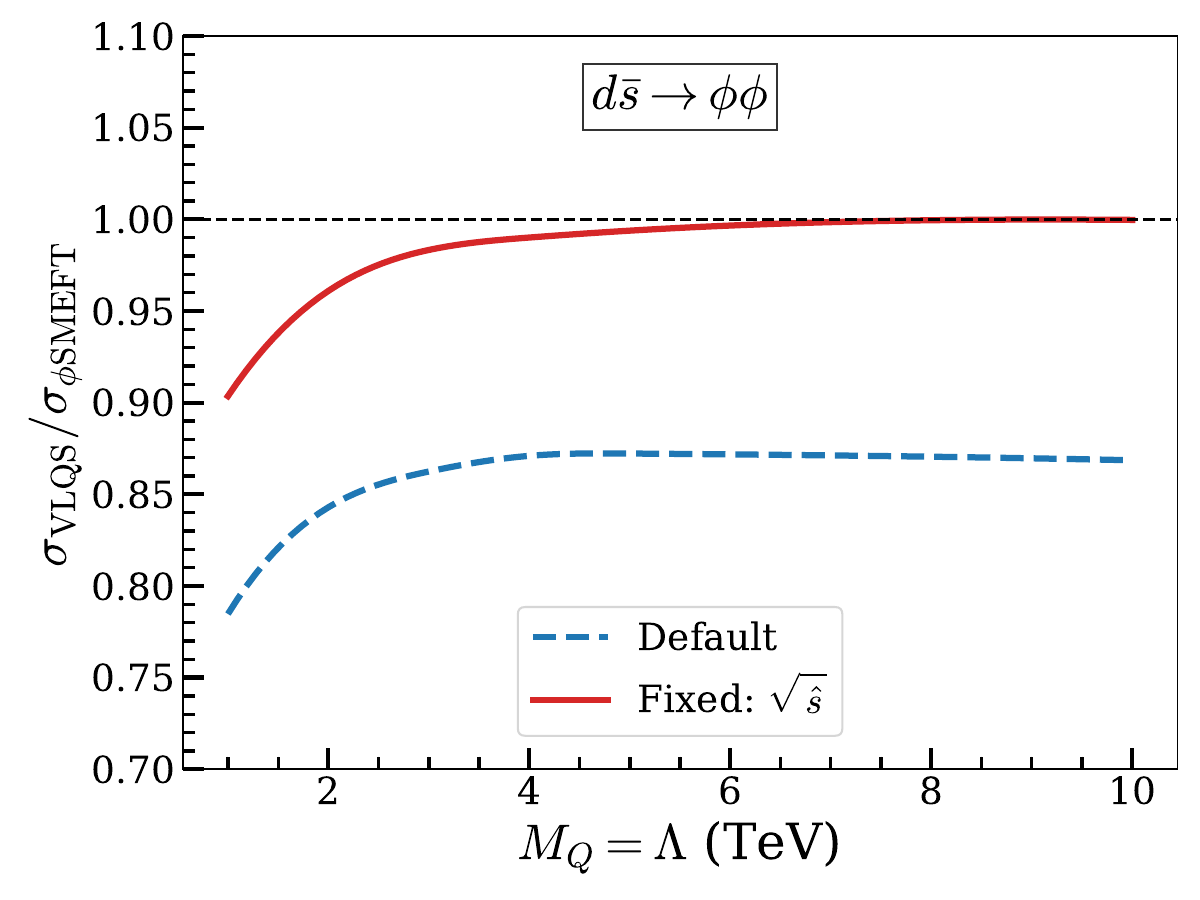}
		\includegraphics[width=0.49\linewidth]{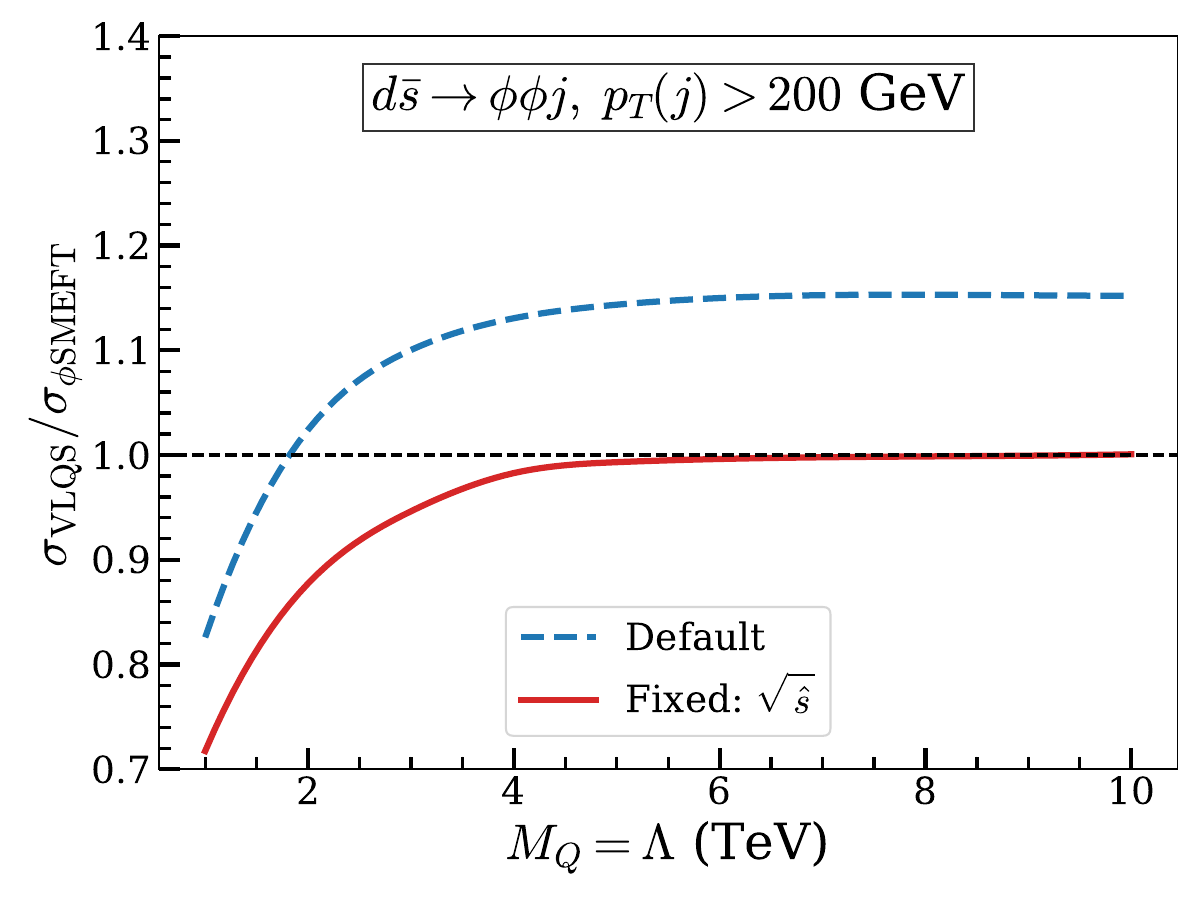}
		\includegraphics[width=0.49\linewidth]{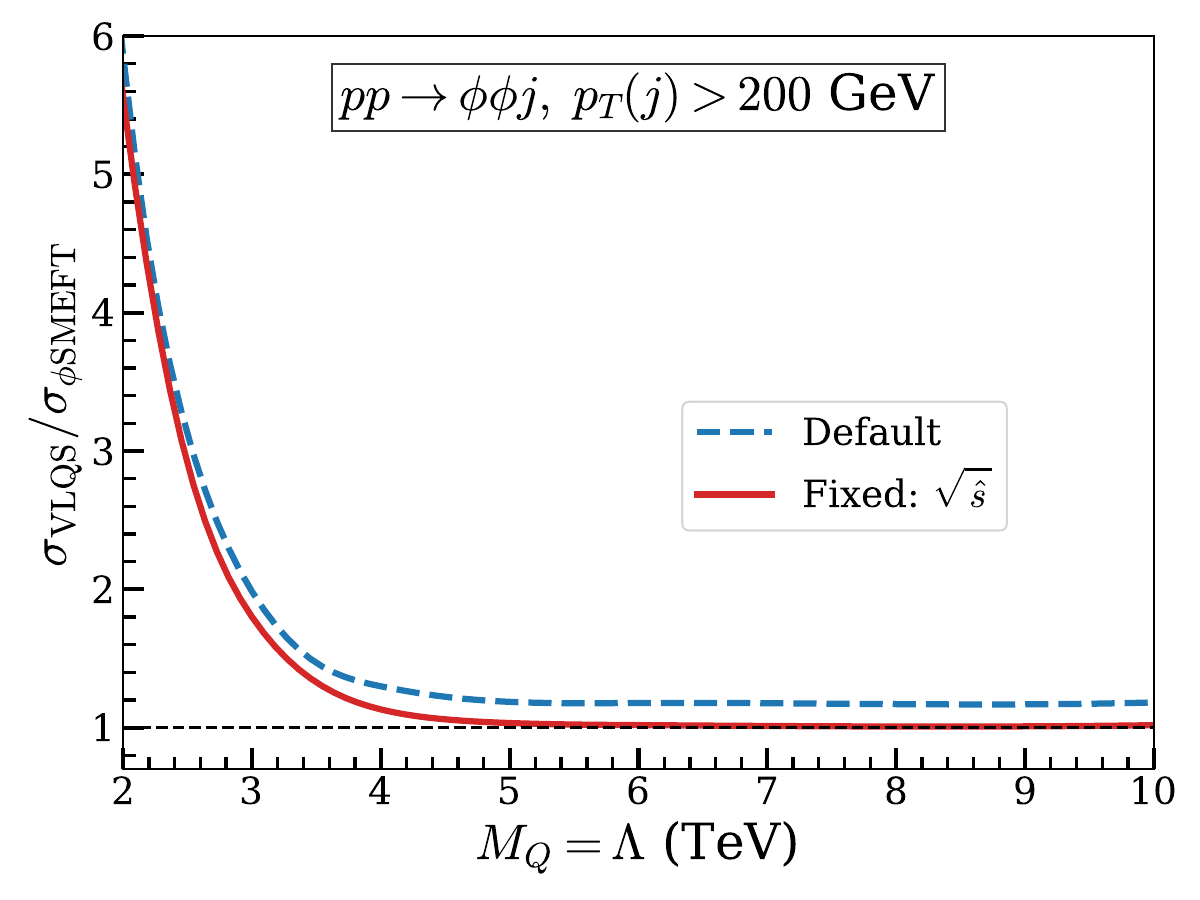}
		\caption{\small Ratio of partonic cross-sections ($\sigma_{\text{VLQS}}/\sigma_{\phi\text{SMEFT}}$) for the $2\to 2$ subprocess $d\bar{s} \to \phi\phi$ with $\sqrt{\hat{s}}>100~\rm GeV$ (top left), monojet $d\bar{s} \to \phi\phi j$ (top right), and $pp\to \phi\phi j$ (bottom)  with $ p_T(j)>200$ GeV at $\sqrt{s} = 13$ TeV for two benchmark scenarios, VLQS: $y_{q/d}^d = y_{q/d}^s  = y_1 = 1$, $\rm \phi SMEFT$: $[\wcqdh]_{12} = [\wcqdh]_{21} = 1$ (for top panel),  $[\wcqdh]_{12} = [\wcqdh]_{21} = [\wcqdh]_{11} = [\wcqdh]_{22} = 1$ (for bottom panel) and all other $\wc = 0$. $\Lambda$ is chosen to analytically match EFT and UV rates and $m_\phi=1$ GeV. 
			The blue dashed line corresponds to MG5's default scale, while the red solid line shows the same for the fixed scale choice $\sqrt{\hat{s}}$. }
		\label{fig:cross-sec_dsphiphi}
	\end{figure}

\begin{figure}[tbp!]
		\centering
		\includegraphics[width=0.5\linewidth]{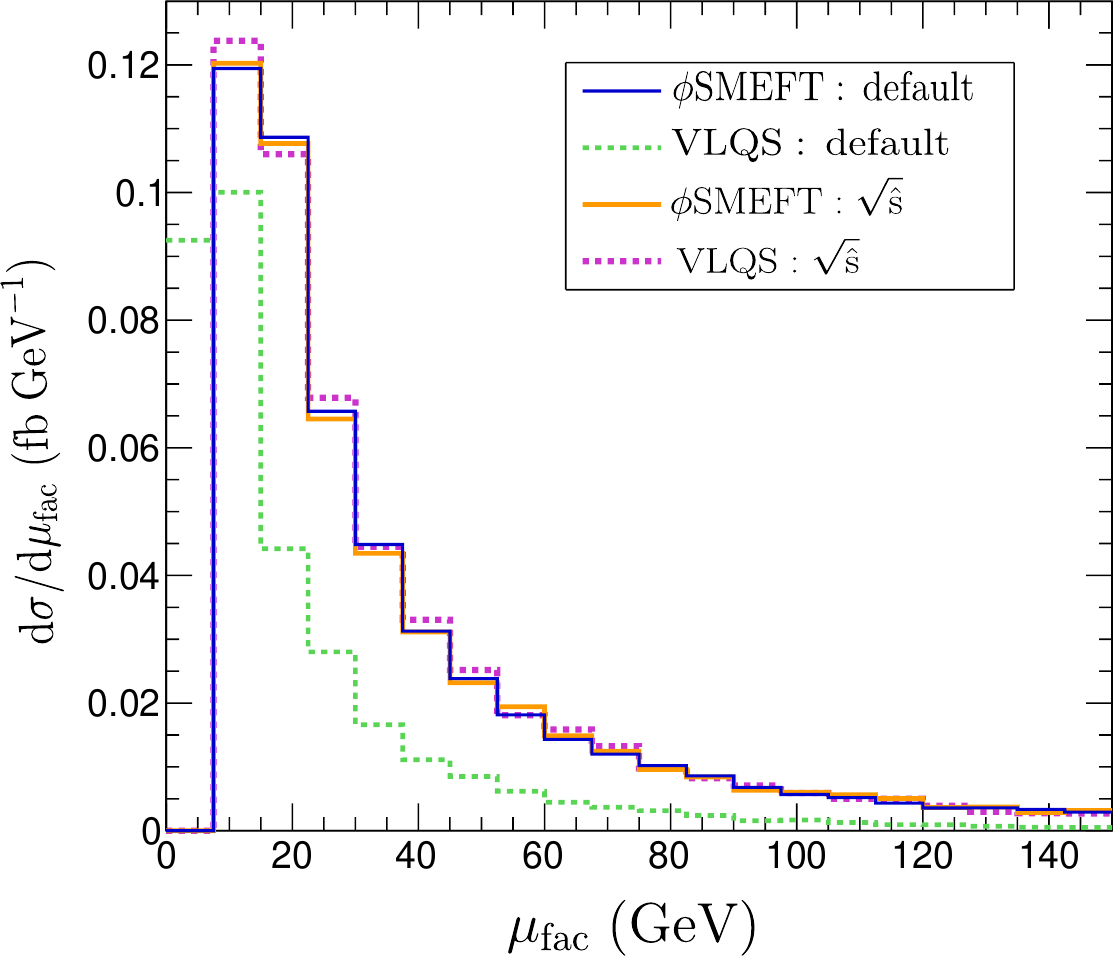}
		\caption{\small Differential cross-sections for the subprocess $d\bar{s} \to \phi\phi$ at $\sqrt{s} = 13$ TeV for different renormalisation and factorisation scale choices in MG5. 
			The NP-scale is fixed at $M_{Q/D}=\Lambda = 8$ TeV, with all WCs set to 1 in the EFT, and couplings $y_{q/d}^d = y_{q/d}^s  = y_1 = 1$ in VLQS, ensuring the EFT and UV amplitudes match as explained in the text. In this plot, \textit{default} refers to the `default dynamic scale' and $\rm\sqrt{\hat{s}}$ is another scale choice in MG5.
		}
		\label{fig:differential_scale}
	\end{figure}

	Additional complications appear when the quark-level cross-sections are convoluted with the parton distribution functions to obtain LHC cross-sections, and we illustrate one of them in Figure~\ref{fig:cross-sec_dsphiphi} (top left). The figure shows the ratio of  cross-sections, $\sigma_{\text{VLQS}}/\sigma_{\phi\text{SMEFT}}$,  for the $2\to 2$ subprocess $d\bar{s} \to \phi\phi$ at $\sqrt{s} = 13$ TeV with benchmark parameters 
	$[\wcqdh]_{12} = [\wcqdh]_{21} = 1$, $y_{q/d}^d = y_{q/d}^s  = y_1 = 1$, and $\Lambda$ chosen to analytically match the leading expansion of VLQS and the EFT rates. As already illustrated in Figure~\ref{fig:s_dependence_analytic}, the EFT should then track VLQS up to an energy scale that is some fraction of the new physics scale. We then expect that as the new physics scale gets very large compared to typical LHC energies, the ratio $\sigma_{\text{VLQS}}/\sigma_{\phi\text{SMEFT}}\to 1$. Figure~\ref{fig:cross-sec_dsphiphi} shows that this is not the case when we use the \textit{default dynamic} renormalization and factorization scale ($\rm \mu_{fac}$)\footnote{The default dynamical scale is defined as the transverse mass of the $2\to2$ system that results from a $kT$ clustering~\cite{Hirschi:2015iia}.} in {\tt Madgraph5} (MG5). The same can be seen in the other two panels for $d\bar{s} \to \phi\phi~j$ and $pp \to \phi\phi~j$. The figure also shows that this is resolved by selecting a fixed scale instead, for illustration $\rm \mu_{fac}=\sqrt{\hat s}$
	is used in Figure~\ref{fig:cross-sec_dsphiphi}, but we have checked that other fixed scales, such as $\rm \sum E_T$ (sum over the transverse energy of the final-state particles), also work.
	To understand the discrepancy observed with the default MG5 scale, we examine the differential distribution as a function of $\mu_{\rm fac}$ in Figure~\ref{fig:differential_scale}. 
	While the distributions are comparable for fixed choices of scale (solid orange and dashed purple curves), they differ significantly when using the \textit{default} MG5 scale (solid blue and dashed green curves). Since the default scale varies considerably on an event-by-event basis, this leads to a mismatch in the predicted cross-sections between the EFT and the UV model. It is important to note that the results shown in Figures.~\ref{fig:region95CL1}–\ref{fig:mphiphi_monojet} use a fixed factorization and renormalization scale, $\sqrt{\hat{s}}$,  for this reason.
	
	\subsubsection{ Process: \texorpdfstring{$pp\to \phi\phi j$}{pp->phi phi j}} 
	
\begin{figure}[tpb!]
		\centering
		\includegraphics[width=0.49\linewidth]{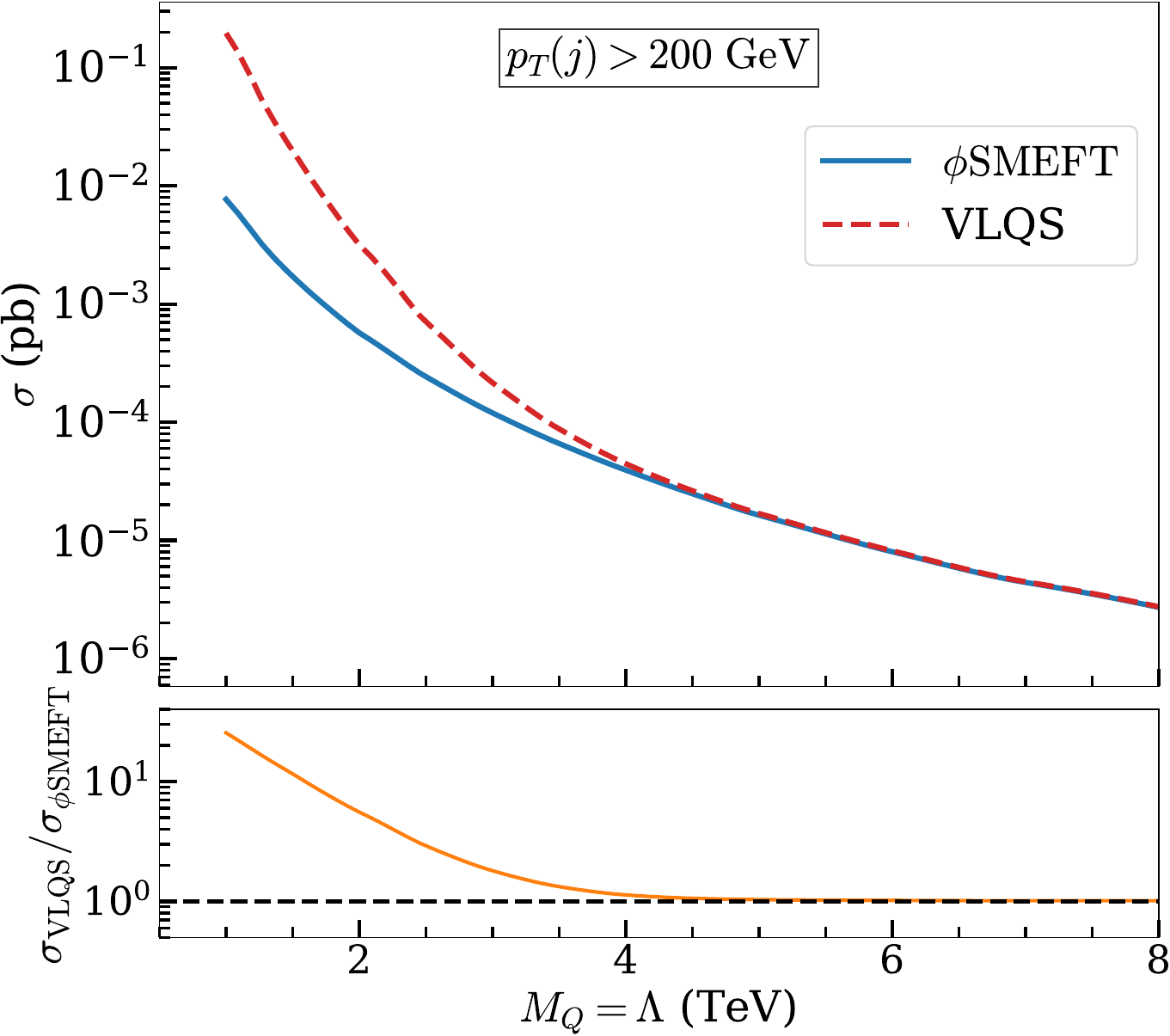}
		\includegraphics[width=0.49\linewidth]{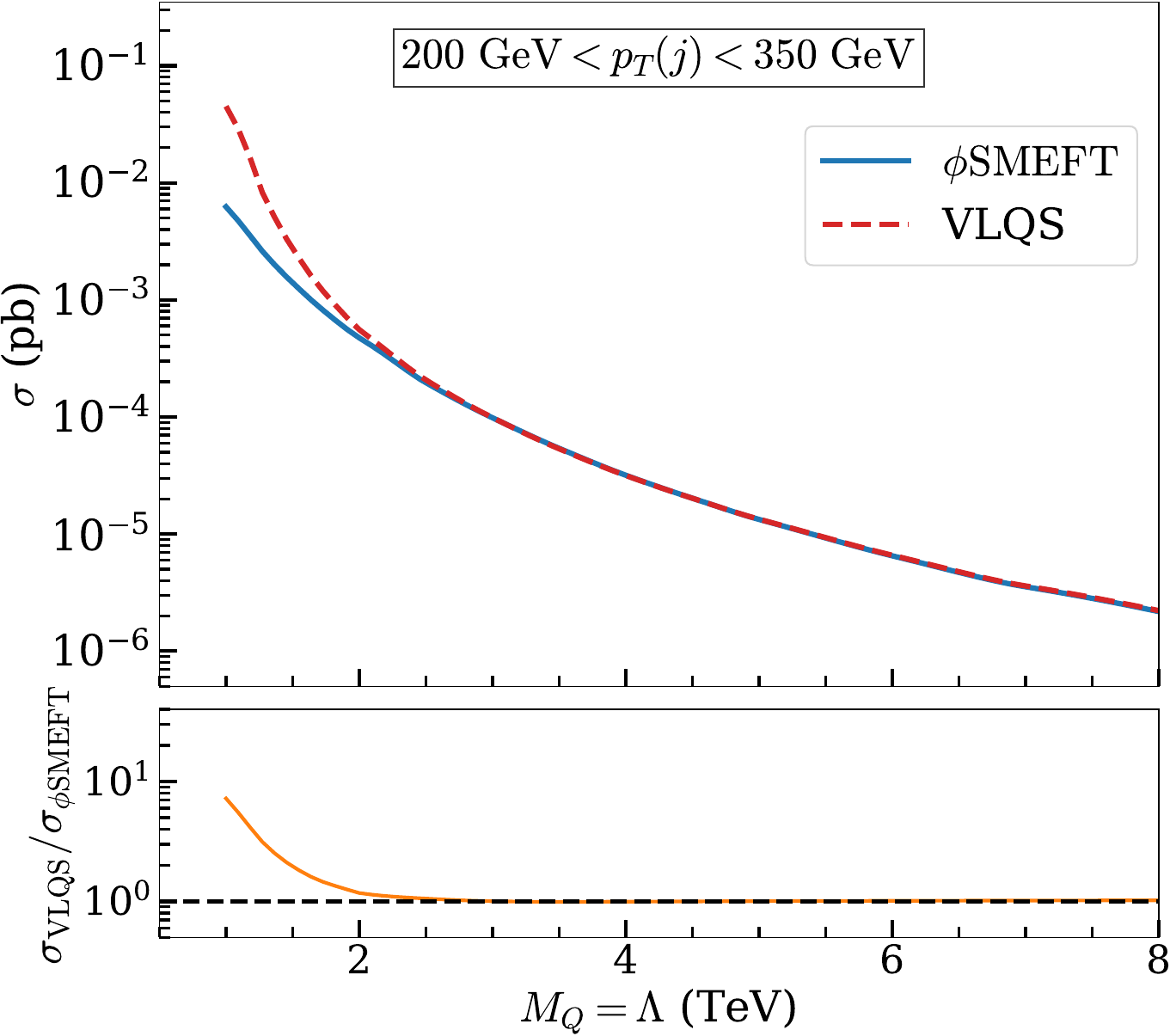}
		\caption{\small Variation of cross-sections for the process $pp \to \phi\phi j$ (monojet) as a function of the NP-scale $\Lambda$ at $\sqrt{s} = 13$ TeV, using the scale choice $\sqrt{\hat{s}}$. for the two benchmark scenarios: 
			$\rm \phi SMEFT:$ $[\wcqdh]_{12} = [\wcqdh]_{21} = [\wcqdh]_{11} = [\wcqdh]_{22} = 1$, all other $\wc = 0$, 
			VLQS: $y_{q/d}^d = y_{q/d}^s  = y_1 = 1$; $\rm M_{Q/D}=\Lambda$ is chosen to ensure matched cross-sections, and $m_{\phi}= 1$ GeV. The left panel shows the results for events with $\rm p_T(j) > 200~GeV$. The right panel shows the same comparison for  $200~{\rm GeV}<\rm p_T(j) < 350~GeV$.
		}
		\label{fig:cross-sec_monojet}
	\end{figure}

	Additional insight can be gained from the monojet process by first setting the factorization scale to the partonic center-of-mass energy $\sqrt{\hat{s}}$ and then varying the accepted ranges of the $\rm p_T$ of jet (equivalent to $\rm \MET$ of the event).  Figure~\ref{fig:cross-sec_monojet} illustrates the difference between cross-sections calculated using the EFT and the VLQS  model as a function of the NP-scale $\Lambda$ at $\sqrt{s} = 13$ TeV. On the left panel, with a minimal phase-space cut $\rm p_T(j) > 200~GeV$, the two differ by $80\%$ for $\Lambda = 3$~TeV and agree to within 4\% at $\Lambda \geq 5$~TeV. The right panel uses instead the phase-space cut  $200~\rm GeV<\rm p_T(j) < 350~GeV$. 
	Comparing these two panels illustrates why the high $\MET$ ATLAS bins, as well as the inclusive bins,  lead to misleading constraints on the EFT as discussed before. Quantitatively, the constraints on the EFT WCs derived from the ATLAS monojet measurement -- which probes event yields for $\MET > 200$ GeV --  using only $\MET$ bins up to 350~GeV, should give a good approximation to those for the VLQS for scales above $\Lambda \gtrsim 3$~TeV, whereas those that use all the $\MET$ bins would not.  This can be seen in the right panel, as well as in the right panel of Figure~\ref{fig:EFT_UV_comparison}.

	\subsection{Results}
	\begin{figure}[tbp]
		\centering
		\includegraphics[width=0.45\linewidth]{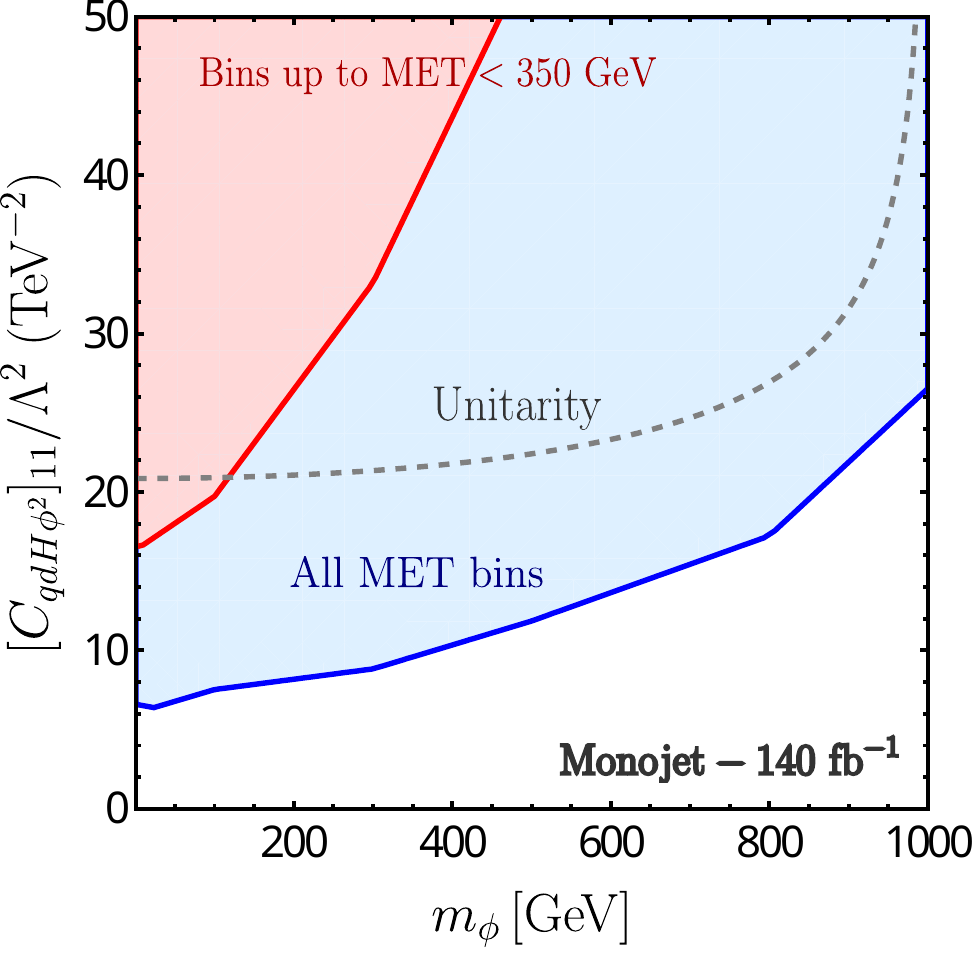}
		\includegraphics[width=0.45\linewidth]{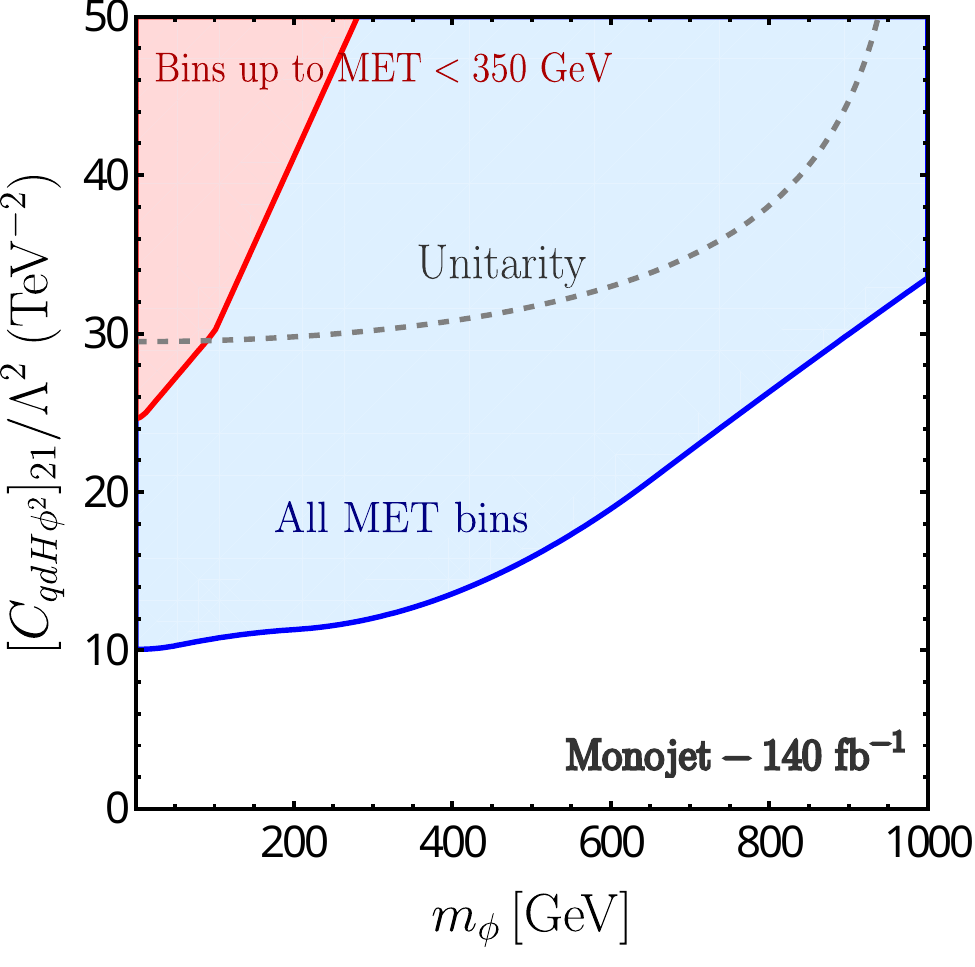}
		\includegraphics[width=0.45\linewidth]{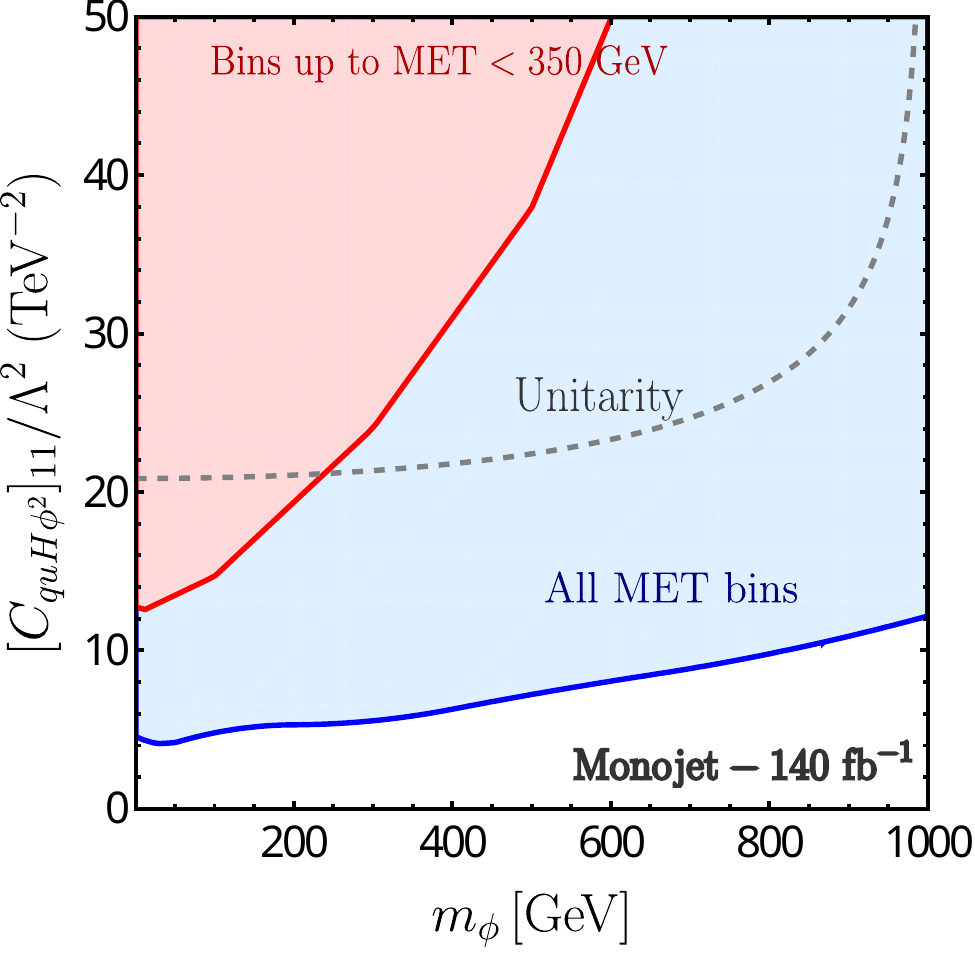}
		\includegraphics[width=0.45\linewidth]{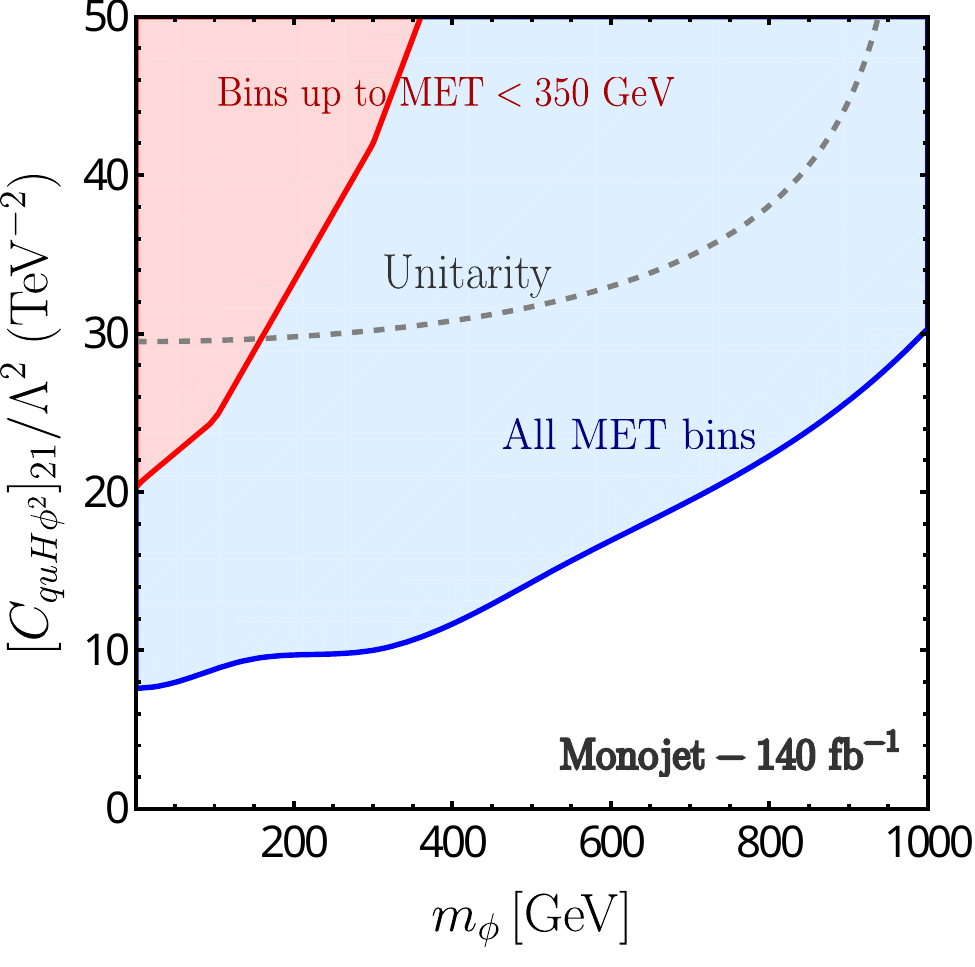}
		\includegraphics[width=0.45\linewidth]{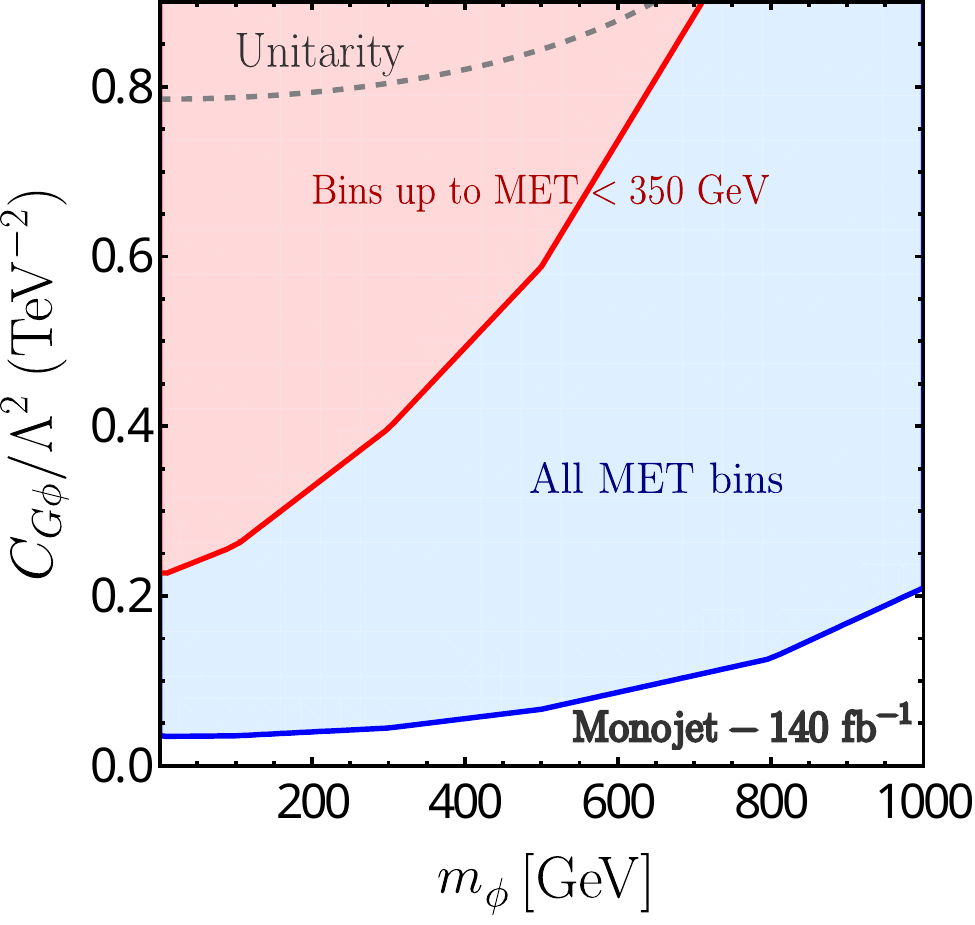}
		\includegraphics[width=0.46\linewidth]{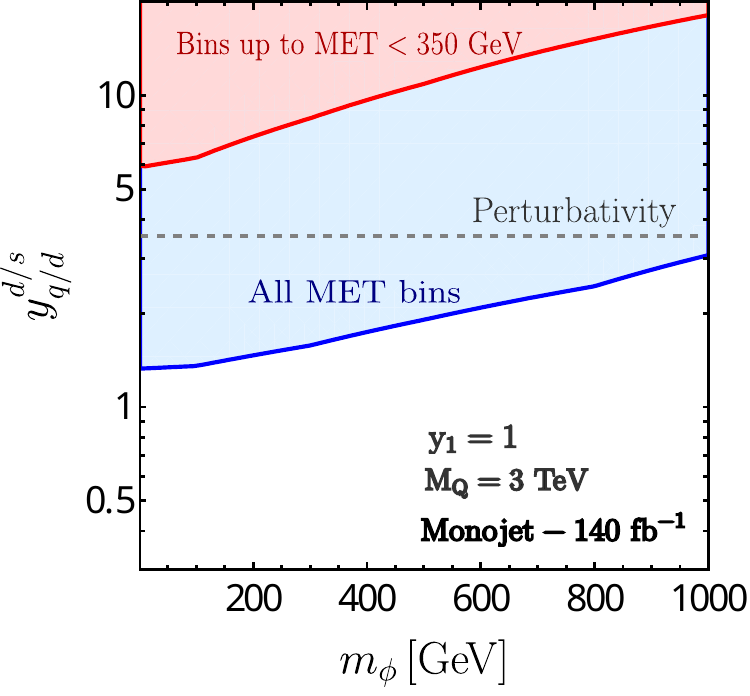}
		\caption{\small Regions excluded at $95\%$ CL using monojet measurements~\cite{ATLAS:2021kxv} for individual WCs $[\wcqdh]_{11,21}$ (top panel), $[\wcquh]_{11,21}$ (center panel), $\wcgh$ (bottom left), with $\Lambda=1$ TeV, and VLQS model parameters (bottom right), with assumption for the VLQS model: $y_{q/d}^d=y_{q/d}^s=y_{q/d}^{d/s}$ and $ y_1=1, M_{Q/D}=3~\rm TeV$. The blue regions show exclusions obtained using all $\rm \MET$ bins, while the red regions use only bins up to 350 GeV -- ensuring compatibility with the UV completion. The region above the dashed gray line violates perturbative unitarity or perturbativity.
		}
		\label{fig:dm_exclusion_monojet}
	\end{figure}
	
	The $\phi$SMEFT WCs are also constrained by perturbative unitarity considerations, and the VLQS Yukawa couplings by perturbativity. Comparison with these theoretical constraints enables us to assess the significance of the monojet bounds. For definiteness, we use the following theoretical constraints, with detailed derivations presented in Appendix~\ref{sec:Unitarity}:
	\begin{align}
	\rm EFT-quark: ~~~~ & \frac{|\mathcal{C}_{q_iq_j}|}{\Lambda^2}<
	\frac{8\pi}{\sqrt{3 (1+\delta_{ij})}v\sqrt{\hat{s}}}\left(\frac{\hat{s}-4m_{\phi}^2}{\hat{s}}\right)^{-1/4}, 
	\\
	\rm EFT-gluon: ~~~~ & \frac{\wcgh}{\Lambda^2}<\frac{\pi}{\hat{s}}\left(\frac{\hat{s}-4m_{\phi}^2}{\hat{s}}\right)^{-1/4},\label{eq:unitarity_CG}\\
	\rm VLQS: ~~~~ & y<\sqrt{4\pi},
	\end{align}
	where $q_{i}$ denotes the quark of flavour $i$.
	For multiple non-zero quark WCs, the constraint applies to the Euclidean norm of the WCs.
	
	\begin{figure}[tb]
		\centering
		\includegraphics[width=0.7\linewidth]{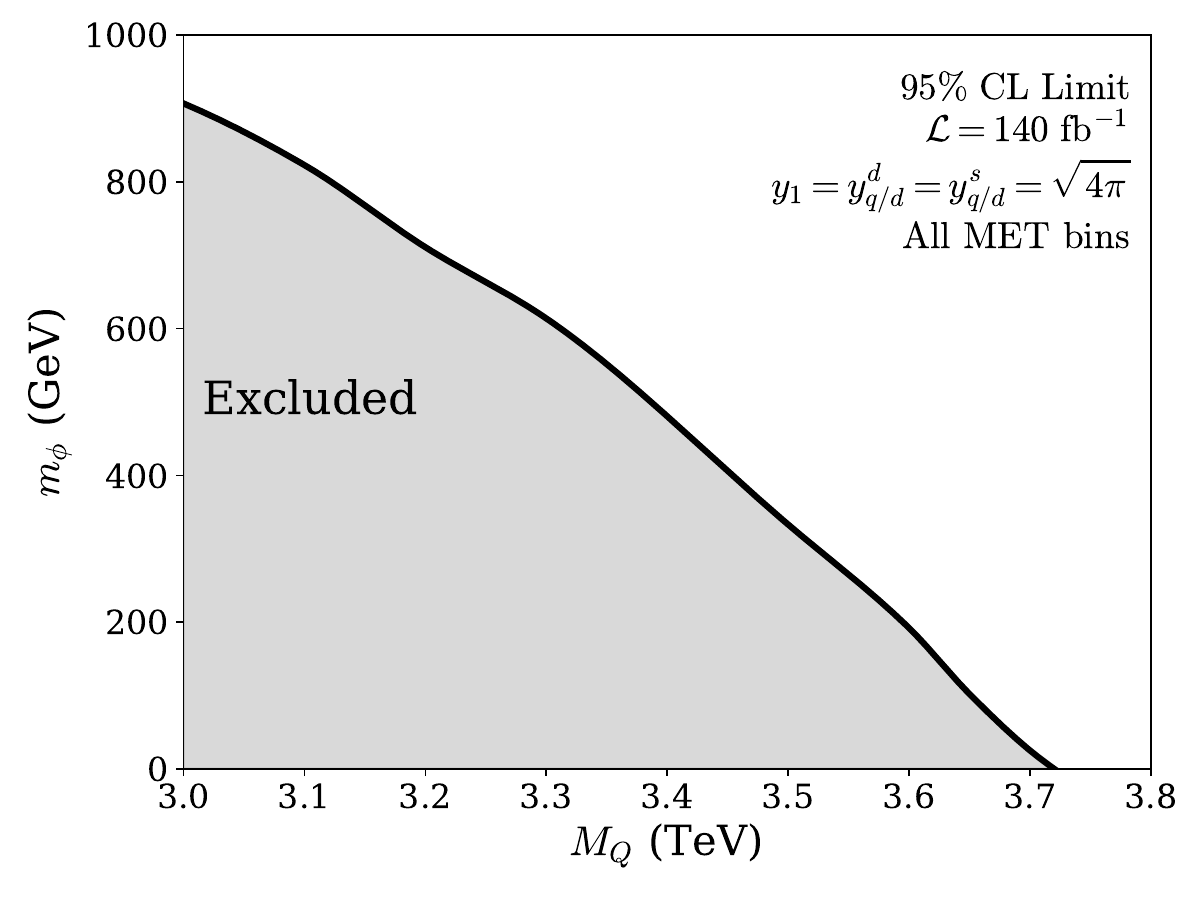}
		\caption{\small Region excluded at $95\%$ CL in the $M_Q$-$m_\phi$ plane of the VLQS model in gray. The Yukawa parameters are set to their  perturbativity limit $y_1=y_{q/d}^{d}=y_{q/d}^{s}=\sqrt{4\pi}$ and $M_D\gg M_Q$.}
		\label{fig:MQ_MPHI_EXC}
	\end{figure}

	In Figure~\ref{fig:dm_exclusion_monojet}, we present the exclusion limits on the scalar $\phi$ as a function of its mass $m_\phi$, considering two scenarios: (a) the EFT is assumed to be valid across the entire phase space as indicated by the blue shaded region, and (b) the EFT description is restricted to the lower $\MET$ bins of the ATLAS measurement, up to 350 GeV (red shaded region). We observe that the exclusion limits are significantly weaker in the second scenario. Although gluon operators are better constrained by monojet searches, their WCs are suppressed by a loop factor $(4\pi)^{2}$ and thus monojet searches probe similar scales for quark and gluon operators. We also show, with a dashed gray line, the upper bound from tree-level perturbative unitarity (or perturbativity for VLQS), indicating where the LHC constraints supersede the simple theoretical ones. Following our discussion for Figure~\ref{fig:region95CL1}, we show constraints for $[\wcqdh]_{11}$ (the strongest) and for $[\wcqdh]_{21}$, a flavour off-diagonal case that cannot be directly constrained by DD experiments.
	Monojet constraints on the parameters of the VLQS model are presented in Figure~\ref{fig:dm_exclusion_monojet} (bottom right), using a suitable benchmark $ y_1=1, M_{Q/D}=3~\rm TeV$, and an assumption $y_{q/d}^d=y_{q/d}^s=y_{q/d}$. This panel indicates that, in this case, using only the lower $\MET$ bins does not constrain the parameter region where the model is perturbative.
	
	To end this section, we quantify the reach of the LHC in searching for a vector-like quark that decays into a jet and missing energy in the $(M_Q, m_\phi)$ plane. We consider the VLQS model in the limit in which one of the vector-like quarks is very heavy and decouples. We further set the values of the Yukawa parameters to their perturbativity limit   $y_1 = y_{q/d}^d = y_{q/d}^s = \sqrt{4\pi}$ to find the largest cross-section for a given mass while checking that the width of the remaining VLQ remains within 25\% of its mass. To this end, we construct a combined $\chi^2$ likelihood for each point in VLQS parameter space using all MET bins ($\MET \in 200 \text{--}1200$ GeV),  and its SM counterpart $\chi^2_{0} = \sum_{i} \frac{(\Delta \mathcal{N}^{exp})^2}{[\delta(\Delta \mathcal{N})]^2}$, and find the region where they differ by less than $\chi^2-\chi_0^2\leq 5.99$ , equivalent to a 95\% CL region. The resulting exclusion limit is presented in Figure~\ref{fig:MQ_MPHI_EXC} in gray. This provides an idea of the optimal LHC reach of the VLQS model from monojet searches.

	\section{Constraints from DM direct detection and DM relic density} \label{sec:directdetection}
	
	The operators $\opquh$, $\opqdh$, and $\opgh$ also contribute to the DM annihilation cross-section into quarks, which determines the DM relic abundance, and the DM-nucleon scattering cross-section, 
	which is relevant for the theoretical prediction of event rates in DM direct detection experiments. 
	We identify the region of parameter space consistent with the observed relic abundance by comparing the Planck measurement, $\Omega h^2 = 0.120 \pm 0.001$~\cite{Planck:2018vyg}, with the relic density computed numerically using \texttt{MadDM}~\cite{Backovic:2013dpa,Backovic:2015cra} for different values of ${\wc_i, m_{\phi}}$. Similarly, we evaluate the DM-nucleon scattering cross-section for different combinations of ${\wc_i, m_{\phi}}$, and compare the results with experimental upper limits to put constraints on the $\wc_i$–$m_{\phi}$ plane. The spin-independent (SI) scattering cross-section is~\cite{DelNobile:2021wmp}
	\begin{align}
	\sigma_{\text{SI}}^N &= \frac{\mu_{\phi N}^2}{4\pi m_{\phi}^2} \left[\frac{\sqrt{2}v}{\Lambda^2}\sum_{q=u,d,s}\left( f_{T_q}^{(p)} \frac{m_N}{m_q} \frac{Z}{A} \,\wcqq + f_{T_q}^{(n)} \frac{m_N}{m_q} \frac{A - Z}{A} \, \wcqq\right)- \frac{16 \pi m_N f_{T G}^{(N)}}{9 \alpha_s} \frac{\wcgh}{\Lambda^2}\right]^2,
	\label{eq:DD_SI_EFT}
	\end{align}
	in terms of the DM-gluon WC $\wcgh$ and the DM-quark WCs $\wcquh$ and $\wcqdh$, which are collectively denoted as $\wcqq$. All WCs, running masses $m_q\in m_{u,d,s}$ , and the strong coupling are evaluated at the hadronic scale $\mu=2$ GeV, and are related to the WCs at the NP scale $\Lambda=1$ TeV
	by\footnote{This has been obtained using the RG invariance of $\frac{\wcqq}{m_q}$ and $\frac{\wcgh}{\alpha_s}$ together with the numerical values of the running masses and strong gauge coupling, which we calculated using RunDec~\cite{Chetyrkin:2000yt,Herren:2017osy} with $\alpha_s(m_Z)=0.1181$ and the running quark masses at the top quark mass scale in~\cite{Huang:2020hdv} as input.
	} 
	$\wcqq(2~\rm GeV)=2.04\,\wcqq(1~\rm TeV)$ and $\wcgh(2~\rm GeV) = 3.33\, \wcgh(1~\rm TeV)$.
	The nucleon form factors associated with the scalar quark current at zero momentum transfer are denoted by $f_{T_{u,d,s}}^{(N)}$ and the nucleon form factor for the gluon operator is $f_{TG}^{(N)}$.\footnote{We use the latest numerical values as given in~\cite{FlavourLatticeAveragingGroupFLAG:2024oxs} and \cite{DelNobile:2021wmp}:
		$
		f_{Tu}^p  =0.017 (2),\;\;
		f_{Tu}^n  =0.012 (1),\;\;
		f_{Td}^p  =0.025 (1),\;\; 
		f_{Td}^n  =0.036 (4),\;\;
		f_{Ts}^{(p,n)}  =0.048 (7),\;\text{and}\;
		f_{TG}^{(p,n)}  =0.910 (7).
		$
	} 
	The reduced mass of the DM-nucleon system is $\mu_{\phi N}$ and $m_N$ is the nucleon mass. $Z$ is the charge of the nucleus and $A$ the total number of nucleons. 
	
	For multi-TeV vector-like quark masses, the DM-nucleon scattering is well described by $\phi\rm SMEFT$ and the SI scattering cross-section in VLQS can be obtained from Eq.~\eqref{eq:DD_SI_EFT} using the matching equations \eqref{eq:matching1}. 
	We choose $m_{Q/D}=3$ TeV and the WCs at the low scale are $\wcqq(2\, \mathrm{GeV}) = 2.18\, \wcqq(3 \,\mathrm{TeV})$ and $\wcgh(2\, \mathrm{GeV})=3.70\, \wcgh(3\, \mathrm{TeV})$. The contribution from the gluon operator is suppressed compared to the quark operators and can be neglected.
	

	\begin{figure}[tbp!]
		\centering
		\includegraphics[width=0.45\linewidth]{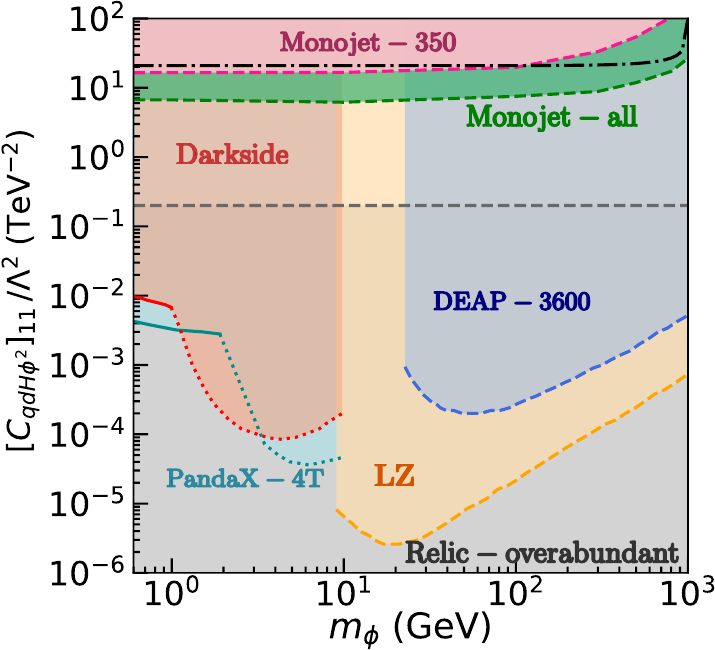}
		\includegraphics[width=0.45\linewidth]{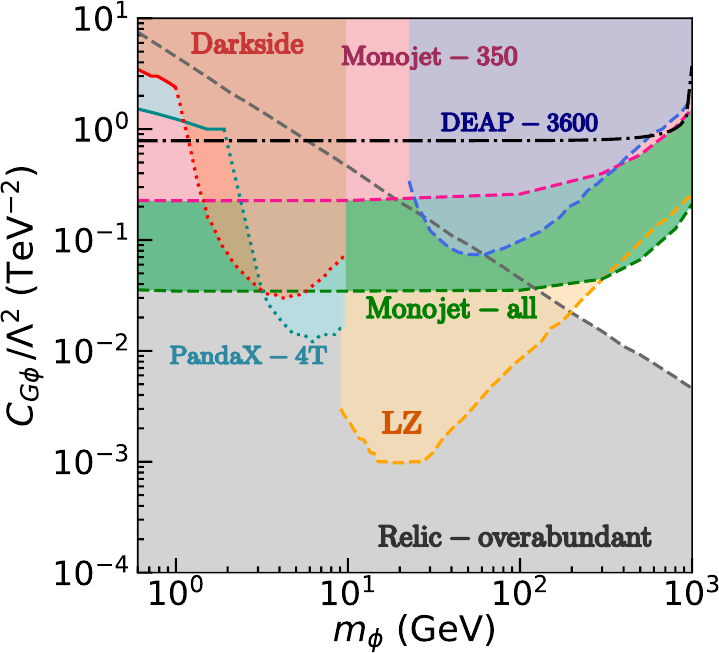}
		\includegraphics[width=0.45\linewidth]{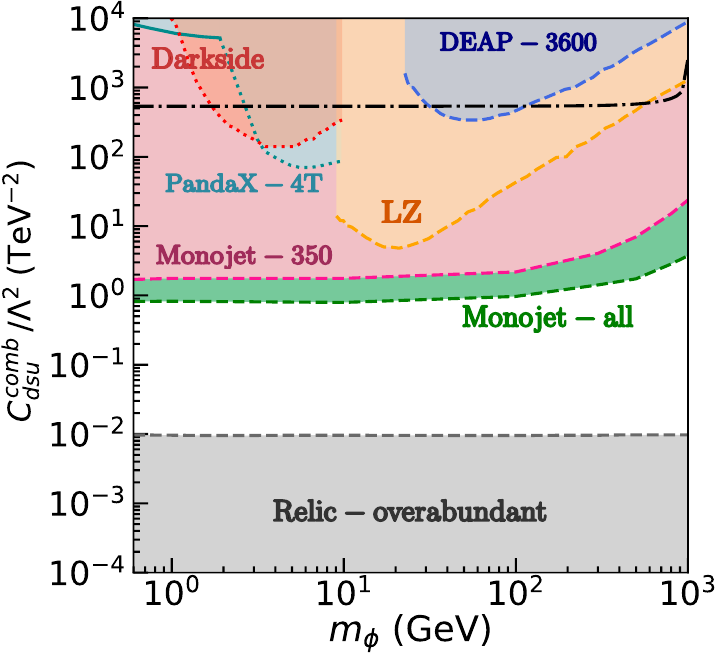}
		\includegraphics[width=0.45\linewidth]{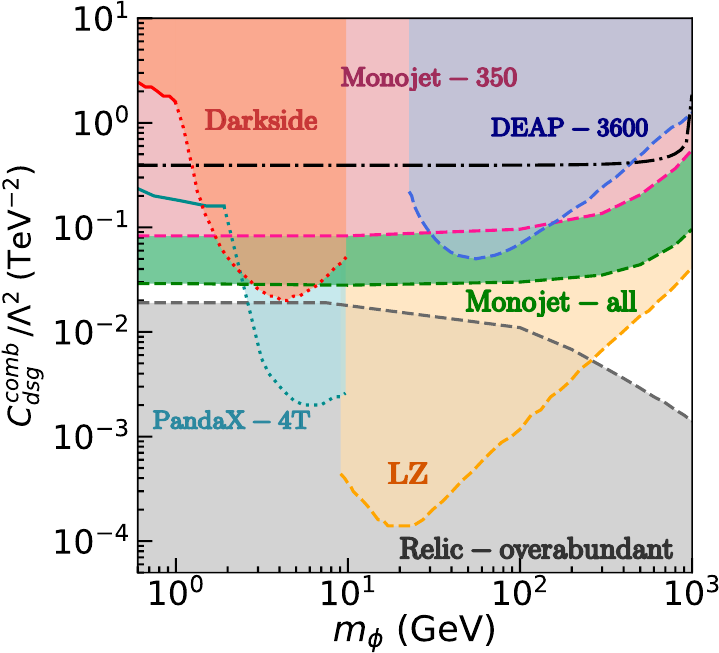}
		\includegraphics[width=0.45\linewidth]{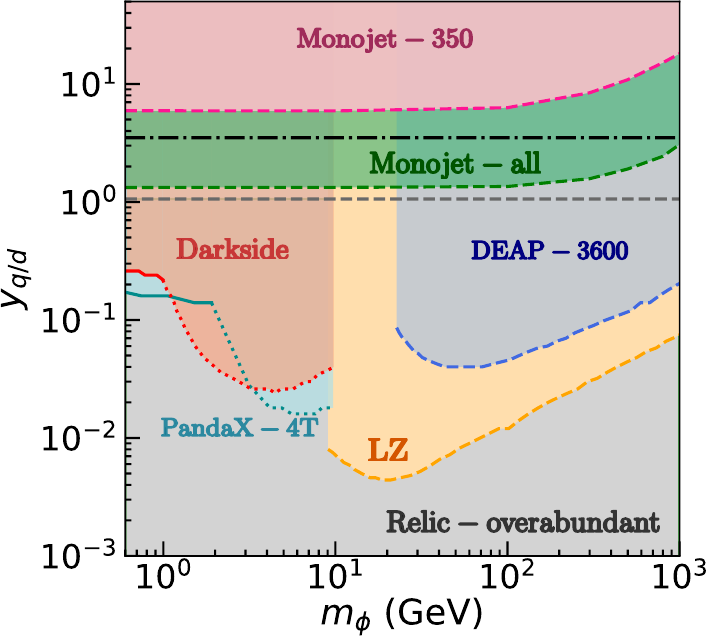}
		\includegraphics[width=0.45\linewidth]{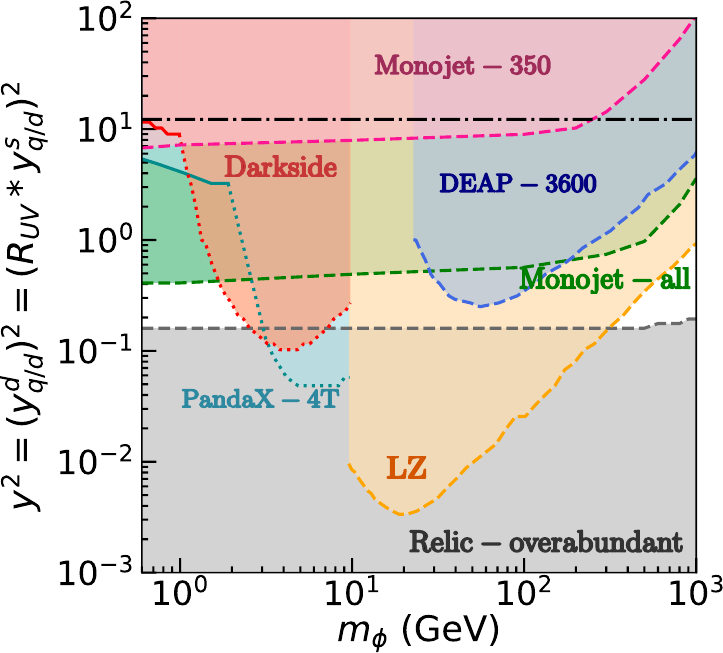}
		\caption{\small Limits on individual $\phi$SMEFT operators (first panel), their combinations (second panel), and different VLQS benchmark choices (third panel).
			The region above the black dash-dot line is not allowed by either perturbative unitarity or by perturbativity.
			The combinations in the second panel are
			$C_{dsu}^{comb}\equiv [\wcqdh]_{11}-R_{ds}\; [\wcqdh]_{22} -R_{du}\; [\wcquh]_{11}$ (left) and $C_{dsg}^{comb}\equiv [\wcqdh]_{11}-R_{ds}\; [\wcqdh]_{22} -R_{dg} \;\wcgh$ (right).
			VLQS benchmark assumptions are: $y_1 = 1$, $M_{Q/D} = 3~\rm TeV$, with 
			$y_{q/d}^d = y_{q/d}^s  = y_{q/d}$ (left), and $y = y_{q/d}^d = R_{UV} \,y_{q/d}^s$ (right). The ratios $R_{ds}, R_{du}, R_{dg}$, $R_{UV}$ minimize the DD scattering cross-sections. Coefficients absent in a plot are set to zero. Excluded regions are denoted by darker shades: Monojet-all (all $\MET$ bins, green), Monojet-350 (up to $\MET = 350$ GeV bin, pink), LZ (yellow), PandaX-4T (blue), DarkSide (red), DEAP (violet). All exclusions extend to the upper edge of the plot; overlaps are implicit. The region disfavored by DM relic density constraints (assuming thermal freeze-out) is shown in gray.}
		\label{fig:single_param_bounds}
	\end{figure}
	
	Figure~\ref{fig:single_param_bounds} summarizes DM constraints on individual and combinations of EFT and UV parameters from monojet searches and DM direct detection search experiments, together with the relic density requirement for thermal DM. The PandaX-4T~\cite{PandaX:2022xqx,PandaX:2023xgl} and DarkSide-50~\cite{DarkSide:2022dhx} experiments present the exclusion of the DM-nucleon SI scattering cross-section for different values of the DM mass, considering the Migdal effect~\cite{Migdal:1941,Ibe:2017yqa} for light DM with $m_{\phi}\lesssim 2$ GeV. Above $m_{\phi}\simeq 9~\rm GeV$, the LZ experiment~\cite{LZ:2024zvo} places the strongest limits. While PandaX and LZ use Xenon targets, DarkSide is an Argon target experiment, and the SI cross-sections for combinations of the EFT WCs are sensitive to the $Z/A$ ratio, which varies across different targets. For completeness, we also include the DEAP-3600 \cite{DEAP:2019yzn} experiment, which uses an Argon target and puts constraints on masses $m_\phi\gtrsim 20$ GeV. 
	In each case, we compute the effective $Z/A$ ratio in Eq.~\eqref{eq:DD_SI_EFT} by taking a weighted average over all naturally occurring isotopes ($\langle Z/A \rangle = Z \cdot \sum_i \frac{f_i}{A_i}$), where $f_i$ is the natural fractional abundance of the isotope with mass number $A_i$ \cite{DelNobile:2021wmp}. This ensures that the nuclear response is accurately normalized for xenon and argon targets.
	DM is overabundant below the gray dashed line. It is independent of the DM mass for the dimension-6 $\phi$SMEFT operator $\opqdh$ (see Figure~\ref{fig:single_param_bounds} (top left)), because the DM annihilation cross-section does not depend on the DM mass to leading order and scales like $\sigma\sim {v^2}|\wcqdh|^2 / {\Lambda^4}$. For the gluon operator, the explanation of the observed DM abundance requires a smaller WC for larger DM masses $\wcgh \propto m_\phi^{-1}$ (see Figure~\ref{fig:single_param_bounds} (top right)), since the DM annihilation cross-section scales like $\sigma\sim {m_\phi^2 |\wcgh|^2}/{\Lambda^4}$.

	A typical constraint on the operator $\opqdh$ is shown in Figure~\ref{fig:single_param_bounds} (top left), indicating that the DD constraints are significantly better than those coming from the monojet studies. For the gluon operator $\opgh$, the DD is comparatively less sensitive, and monojet searches put the strongest bounds for the DM mass $m_{\phi}\lesssim 2$ GeV, see Figure~\ref{fig:single_param_bounds} (top right). 
	
	The complementarity of the two sets of constraints can be appreciated in the centre panel, where the constraints on the combinations $C_{dsu}^{comb}\equiv [\wcqdh]_{11}-R_{ds}\; [\wcqdh]_{22} -R_{du}\; [\wcquh]_{11}$ and $C_{dsg}^{comb}\equiv [\wcqdh]_{11}-R_{ds}\; [\wcqdh]_{22} -R_{dg} \;\wcgh$ are shown.\footnote{The coefficients $R_{ds}$, $R_{du}$, and $R_{dg}$ are chosen such that the event rates in DD experiments are suppressed, with the precise values depending on the detector type and the nucleon form factors. For $C_{dsu}^{comb}$, we use $R_{ds}=-25.72, R_{du}=1.05$, while for $C_{dsg}^{comb}$ we take $R_{ds}=-12.77$ and $R_{dg}=3.14$. Note that with three degrees of freedom, these minima for the DD cross section are not unique, but instead lie on a plane, and the quoted values are two of the many possibilities.}  The unitarity constraints on the parameter combination $C_{dsu}^{comb}$ is,
	\begin{align}
	\frac{C_{dsu}}{\Lambda^2} < \frac{8\pi \sqrt{1+R_{ds}^2+R_{du}^2} }{ \sqrt{6} v \sqrt{\hat{s}}}\left(\frac{\hat{s}-4m_\phi^2}{\hat{s}}\right)^{-1/4}.
	\end{align}
	Whereas, for $C_{dsg}^{comb}$, the scattering amplitude is dominated by the $\wcgh$, so we use the same limit as in Eq.~\eqref{eq:unitarity_CG}.
	In this case, constraints from the monojet process turn out to be the strongest across the entire DM mass range for $C_{dsu}^{comb}$. In contrast, for $C_{dsg}^{comb}$, monojet constraints are primarily effective for DM masses below a few GeV or above a few hundred GeV. In both cases, certain regions remain viable for a DM candidate. 
	
	The bottom panel shows the VLQS scenario for $y_{q/d}^d = y_{q/d}^s = y_{q/d}$. For the full range of DM masses, the DD constraints are stronger than those of the monojet, as shown in the bottom left panel. The global picture indicates that this benchmark scenario is excluded as a viable thermal DM explanation. However, under the assumption $y = y_{q/d}^d = R_{UV}\, y_{q/d}^s$, an appropriate choice of $R_{UV}$  ($R_{UV}=2.18$) suppresses the cross-section of the DD, allowing monojet bounds to dominate around the GeV scale (bottom right). In this case, certain regions below a few GeV and above a few hundred GeV are still allowed by the DM requirements.
	
	\section{Summary and Conclusions} \label{sec:conclusions}
	
	We have studied the constraints on real scalar dark matter from monojet studies at the LHC as well as from direct detection experiments. We contrasted two frameworks introducing the DM scalars: a dimension-six effective field theory, and a simple UV completion involving two new vector-like quarks. There are two aspects to this study: determining the constraints on this type of dark matter and comparing the results obtained in the EFT with those from its UV completion.
	
	The parameter space that is allowed by current LHC measurements of monojets is shown in Figure~\ref{fig:region95CL1}. The bounds on Yukawa couplings of the VLQ model are near their perturbativity limit, $\sqrt{4\pi}$. While monojet searches constrain $|y_{q/d}^d|\lesssim 1.8$ for $M_{Q/D}=3$ TeV, the Yukawa couplings $y_{q/d}^s$ are more strongly constrained by perturbative unitarity. This difference is understood as it follows from the smaller strange-quark parton distribution function.
	The results are compared to a sensitivity projection for HL-LHC, for which it is assumed that statistical uncertainties will dominate. At the EFT level, the tightest constraint is on the gluon operator, but one should bear in mind that this operator is typically induced at one-loop level within specific models. The hole seen in all the currently allowed regions that suggests exclusion of the SM is due to the large fluctuation seen in the data for one $\MET$ exclusive bin, EM10.
	
	A more direct comparison between the results from the two approaches is shown in Figure~\ref{fig:EFT_UV_comparison}. When all available exclusive bins are used, the constraints imposed on the EFT are much weaker than the corresponding ones in its UV completion. However, most of this difference occurs in regions where the couplings exceed their perturbativity limit. Exclusion of bin EM10 confirms that it is responsible for the apparent exclusion of the SM in the current data. Finally, when only the low $\MET$ bins are taken into consideration, the two descriptions result in very similar allowed parameter regions. Restricting the study to the  low $\MET$ bins results in constraints that do not improve what is known from simple theoretical arguments, in this case perturbativity.
	
	Figure~\ref{fig:mphiphi_monojet} illustrates the kinematic regimes where the two descriptions overlap. As the EFT represents the first term in an energy expansion, it is useful to first compare the $\sqrt{\hat s}=M(\phi\phi j)$ distributions, which are known at the simulation level. As expected, the two distributions disagree at higher values, specifically, this occurs around 70\% of the NP scale. The $\MET$ distribution is an observable proxy for $M(\phi\phi j)$ and shows the two descriptions deviating at a lower point, near 20\% of the NP scale. This is understandable as $\MET$ represents only a fraction of the available event energy. The figure reveals an unexpected second peak at higher energies, above the NP scale. This feature can be traced to contributions originating from diagrams such as the top-right one in Figure~\ref{fig:monojet_feynman}. At higher partonic centre-of-mass energies, it is possible to produce an on-shell vector-like quark, with the resulting effective $2\to2$ process enhancing the cross-section. Removal of these contributions at the simulation level confirms them as responsible for the second peak in the $M(\phi\phi j)$ distribution.
	
	To further study the differences between the two descriptions, we show in Figure~\ref{fig:s_dependence_analytic} the parton-level cross-section for $q\bar{q}\to \phi\phi$ as a function of $\sqrt{\hat{s}}$. Surprisingly, the two deviate already at about 15\% of the NP scale, even though there are no s-channel resonances in this UV completion. The hadron-level cross-section integrates this curve after weighing it with the PDFs, resulting in a much smaller difference between the two. We confirm this in Figure~\ref{fig:cross-sec_dsphiphi}, where we see how with a fixed factorisation scale, $\mu=\sqrt{\hat{s}}$, the EFT and its UV completion differ by less than 5\% for NP scales as low as 2~TeV (60\% of the NP scale) for both $pp\to \phi\phi$ and $pp \to \phi\phi j$. The figure also reveals a surprising result, that the matching between the two descriptions can be spoiled by using the default {\tt MadGraph5} factorisation scale. Evidently, the event topologies in the two cases are sufficiently different to cause a mismatch in this scale. We confirm this observation with the distributions shown in Figure~\ref{fig:differential_scale}, showing that it is systematically higher for the EFT. This issue is resolved by switching to a fixed factorization scale $\rm \mu_{fac}$, as illustrated for $\rm \mu_{fac}=\sqrt{\hat s}$ in Figures~\ref{fig:s_dependence_analytic} and \ref{fig:cross-sec_dsphiphi}.
	
	The influence of the higher energy events on the comparison between the two descriptions can be seen in Figure~\ref{fig:cross-sec_monojet}. When only the lower $\MET$ bins are used, the two agree for NP scales as low as 2~TeV, whereas when all $\MET$ bins are included, the agreement is reached only for NP scales near 4~TeV.
	
	We find that for monojet studies at the LHC, using inclusive bins does not produce reliable parameter constraints for EFTs. Using exclusive bins allows us to fine-tune the best $\MET$ range to constrain such theories: whereas the highest $\MET$ bins correspond to kinematic regions where the EFT is not valid, the lowest $\MET$ bins do not probe the EFT beyond what is known from  perturbative unitarity. With our t-channel UV completion model, we have identified on-shell production of narrow, heavy mediators as another factor affecting the fidelity of an EFT. 
	
	Figure~\ref{fig:dm_exclusion_monojet} illustrates the regions of parameter space allowed by monojet searches as a function of the dark scalar mass. The perturbative unitarity (or perturbativity) limits superimposed on these figures are always improved by the constraints obtained with all $\MET$ bins. The same is not true of those obtained using only low $\MET$ bins. 
	
	In Figure~\ref{fig:MQ_MPHI_EXC}, we quantify the reach of the LHC for the VLQS model in the $(M_Q, m_\phi)$ plane. For light scalar masses $m_\phi\lesssim 10$ GeV with Yukawa couplings near the perturbativity limit, monojet searches constrain the VLQ mediator to be lighter than $M_Q\lesssim 3.7$ TeV.   
	
	Finally, we compare the monojet constraints to those obtained from DD experiments. DD experiments pose strong constraints on dark scalars, excluding most of the parameter space. Monojet constraints are complementary and are seen to close some of the windows still allowed by DD. These windows occur with certain relations between different Yukawa couplings that lead to cancellations in the DD cross-sections. The cancellation points differ for different targets, and we illustrate some of these conditions. The monojet constraints also close the gap for flavour off-diagonal couplings to scalars, which are not probed by DD experiments.

	\vspace{1 cm}

	\acknowledgments
	
	This work is supported by an Australian Research Council Discovery Project. We thank Paul Jackson for clarification on \cite{ATLAS:2021kxv}, and Xiao-Dong Ma for useful discussions.
	
	\appendix
	
	\section{Perturbative Unitarity Bounds}
	\label{sec:Unitarity}
	We impose perturbative unitarity on $2\to2$ amplitudes,
	\begin{align}
	|{\rm Re}\,a_j| \le \tfrac12 \, ,
	\end{align}
	where $a_j$ is the partial wave amplitude with total angular momentum $J^2=j(j+1)$. The dominant constraint comes from the $j=0$ partial wave
	\begin{align}
	a_0
	= \frac{1}{32\pi}\int_{-1}^{+1} d\!\cos\theta\, \mathcal{M}(\cos\theta) \, .
	\label{eq:PWintegral}
	\end{align}
	For two identical particles in the initial or final states, the scattering amplitude includes a symmetry factor $1/\sqrt{2}$~\cite{Logan:2022uus}.
	For massive initial and final states, the phase space factors $\beta_{i,f}$ are important. In our case only final-state DM ($\phi$) is massive, so we use a factor:
	\begin{align}
	\beta_f\equiv\beta= \left(\tfrac{\hat{s}-4m_\phi^2}{\hat{s}}\right)^{1/4},
	\end{align}
	which multiplies the partial waves~\cite{Endo:2014mja,Cohen:2021gdw}.
	Perturbative unitarity then implies 
	\begin{equation}
	|\mathrm{Re}(a_0)| \leq \frac12 \,.
	\end{equation}

	\subsection{Quark operators}
	For the operators $\opqdh,\opquh$, the partonic amplitude for $d(p_2)\, \bar s(p_1) \to \phi\phi$ is
	\begin{align}
	i\mathcal{M}(d \bar s \to \phi\phi)
	= i\frac{\sqrt{2}\,v}{\Lambda^2}\,\bar v_s(p_1)\left([\wcqdh]_{12} P_L+[\wcqdh]_{21}P_R\right) u_d(p_2)\,,
	\end{align}
	and similar for other initial states $d\bar d,s\bar s$ with different WCs.
	For massless external quarks in the centre-of-mass frame, the only non-vanishing helicity configurations are
	\begin{align}
	\mathcal{M}_{--} &= \frac{\sqrt{2}\,v}{\Lambda^2}\, [\wcqdh]_{12}\,\sqrt{\hat{s}}\,,\qquad
	\mathcal{M}_{++} = \frac{\sqrt{2}\,v}{\Lambda^2}\, [\wcqdh]_{21}\,\sqrt{\hat{s}}\,,
	\end{align}
	for the colour-singlet combination with $\mathcal{M}_{+-}=\mathcal{M}_{-+}=0$.
	Including the phase-space factor $\beta$ and the set of initial and final states of two particles $(\ket{\phi\phi},\ket{\bar s d_L}, \ket{\bar s d_R}, \ket{\bar d s_L}, \ket{\bar d s_R})$, and a symmetry factor $1/\sqrt{2}$ for the identical final state $\phi\phi$~\cite{Logan:2022uus}, we obtain 
	\begin{align}
	\frac{|[\wcqdh]_{ij}|}{\Lambda^2}
	&\leq \frac{8\pi}{\sqrt{3}\,v\sqrt{\hat{s}}}\,
	\left(\frac{\hat{s}-4m_\phi^2}{\hat{s}}\right)^{-1/4}
	\end{align}
	for real $\wcqdh$ with $i\neq j$. For multiple non-zero WCs, the constraint applies to the Euclidean norm of the WCs, e.g. for $[\wcqdh]_{12}$ and $[\wcqdh]_{21}$, it applies to the combination  $\sqrt{|[\wcqdh]_{21}|^2+ |[\wcqdh]_{12}|^2}$.
	Similarly, for a single non-vanishing real WC $[\wcqdh]_{11}$ perturbative unitarity implies
	\begin{align}
	\frac{|[\wcqdh]_{11}|}{\Lambda^2}
	&\leq \frac{8\pi}{\sqrt{6}\,v\sqrt{\hat{s}}}\,
	\left(\frac{\hat{s}-4m_\phi^2}{\hat{s}}\right)^{-1/4}\;.
	\end{align}
	The same constraints apply to WCs $\wcquh$.
	
	\subsection{Gluon operator}
	For the gluon operator $\mathcal{L} = \frac{\wcgh}{\Lambda^2} G^A_{\mu\nu} G^{A\mu\nu} \phi^2$ the scattering matrix element for $gg\to \phi\phi$ reads
	\begin{align} 
	i\mathcal{M} &= \langle \phi\phi| i \int d^4x \mathcal{L} | \epsilon_1^A(\lambda_1) ,\epsilon_2^B(\lambda_2)\rangle 
	\\
	& = \frac{8 i \wcgh \delta^{AB}}{\Lambda^2} \left(p_1 \cdot \epsilon_2(\lambda_2) p_2\cdot\epsilon_1(\lambda_1) - p_1 \cdot p_2 \epsilon_1(\lambda_1) \cdot \epsilon(\lambda_2)\right).
	\end{align}
	which includes a factor 2 from the multiplicity of $\phi$ and factor 4 from the multiplicity of the gauge fields in the field strength tensor $G^{\mu\nu} G_{\mu\nu}$. Using the explicit form of the momenta $p^\mu_{1,2}=(E,0,0,\pm E)^T$ and the polarisation vectors $\epsilon^\mu_{\pm} = \frac{1}{\sqrt{2}}(0,1,\pm i,0)^T$ this leads to
	\begin{align} 
	\mathcal{M}_{++}=\mathcal{M}_{--} 
	& = - \frac{4  \wcgh \,\hat{s} }{\Lambda^2} \delta^{AB}
	\end{align}
	for the two matrix elements. Taking into account the colour factor and a symmetry factor $1/2$ for the multiplicity of particles in identical initial and final states $(\ket{\phi\phi},\ket{gg}_{\pm\pm})$, the perturbative unitarity condition for the $j=0$ helicity amplitude implies
	\begin{align}
	\frac{ |\wcgh|}{\Lambda^2}\leq \frac{\pi}{\hat{s}} \left(\frac{\hat{s}-4m_{\phi}^2}{\hat{s}}\right)^{-1/4}\;. 
	\end{align}
	
	\subsection{Operator combinations}
	We start with non-zero $[\wcqdh]_{11,22}$ and $[\wcquh]_{11}$. 
	\begin{align}
	\mathcal{L} = \frac{v}{\sqrt{\Lambda^2}} \left(
	[\wcqdh]_{11}  \bar d d \phi^2
	+[\wcqdh]_{22}  \bar s s \phi^2
	+[\wcquh]_{11}  \bar u u \phi^2
	\right)
	\end{align}
	Note that this assumes that the WCs are real, so that there are no projection operators since all operators are hermitian. The scattering matrix element from the two-quark state $\ket{\psi_a\psi_b}\equiv \ket{d\bar d} + R_{ds} \ket{s\bar s} + R_{du} \ket{u\bar u})/\sqrt{1+R_{ds}^2+R_{du}^2}$ to the two-DM state $\ket{\phi\phi}$ is thus
	\begin{align}
	\langle \phi \phi |i \int d^4x \mathcal{L}_{\rm int} | \bar \psi_a \psi_b \rangle 
	&= \frac{\sqrt{2} v C_{dsu}}{\Lambda^2 \sqrt{1+R_{ds}^2 + R_{du}^2}} \bar v_a (P_L + P_R )u_b\,.
	\end{align}
	We thus find for the $j=0$ partial waves
	\begin{align}
	[a_0]_{\pm\pm} = \frac{\sqrt{3}  v \sqrt{\hat{s}} \,C_{dsu}}{16\pi \Lambda^2 \sqrt{1+R_{ds}^2 + R_{du}^2}}\,. 
	\end{align}
	Imposing the perturbative unitarity condition on the largest eigenvalue of the $3\times 3$ $T$ matrix results in
	\begin{align}
	\frac{|C_{dsu}|}{\Lambda^2} < \frac{8\pi \sqrt{1+R_{ds}^2+R_{du}^2} }{ \sqrt{6} v \sqrt{\hat{s}}}\left(\frac{\hat{s}-4m_\phi^2}{\hat{s}}\right)^{-1/4} . 
	\end{align}
	
	As the perturbative unitarity constraint for the gluon operator is much stronger than that for the combinations of the quark operators, we impose the gluon operator constraint on gluon fraction of $C_{dsg}^{comb}$ and neglect the additional contributions of the quark operators.
	
	\section{Fitted coefficients for the number of NP events}\label{sec:fitted_coefficients}
\begin{figure}[t!]
		\centering
		\includegraphics[width=0.5\linewidth]{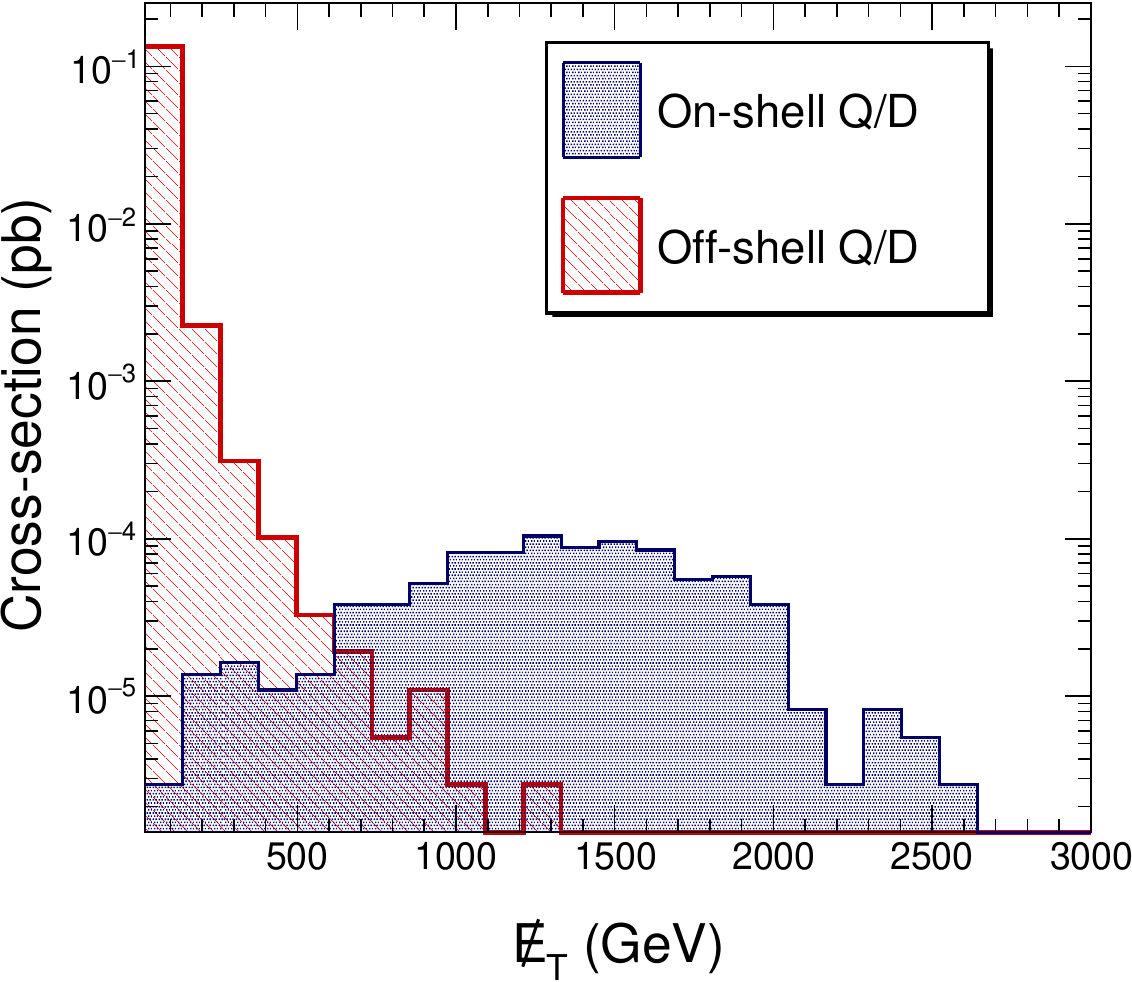}
		\caption{Distribution of the truth-level $\MET$ in the monojet process $pp \to \phi\phi j$ at $\sqrt{s}=13$ TeV LHC, comparing events with an on-shell $Q/D$ contribution to those with off-shell $Q/D$ contributions, for $M_{Q/D}=3$ TeV, $m_{\phi}=1$ GeV, and $y_{q/d}^d = y_{q/d}^s = y_1 = 1$.}
		\label{fig:on_shell_effect}
	\end{figure}

	We use the exclusive monojet event yields in different phase-space regions reported by ATLAS analysis~\cite{ATLAS:2021kxv}. The corresponding values of the coefficients, parameterized as $\gamma_{ij}$, $\beta_{mn}$, and $\kappa_{mn}$ 
	in Eq.~\eqref{eq:fit_polynomial}, are presented for the various phase-space regions in Tables \ref{tab:coefficients_EFT} and \ref{tab:coefficients_VLQS} for $\rm \phi SMEFT$ and VLQS, respectively. The uncertainties in the fitted coefficients are negligible compared to the experimental ones, so they are omitted in our numerical estimates.
	\begin{table}[btp]
		\centering
		\resizebox{1.0\textwidth}{!}{
			\begin{tabular}{c|ccccccccccccc}
				\hline\hline
				WC, Bin & 200 & 250 & 300 & 350 & 400 & 500 & 600 & 700 & 800 & 900 & 1000 & 1100 & 1200 \\
				\hline
				$([\wcqdh]_{11})^2$ & 328 & 257 & 157 & 58 & 31 & 31 & 13 & 6.3 & 2.3 & 1.5 & 0.87 & 0.50 & 0.73 \\
				$([\wcqdh]_{12})^2$ & 141 & 107 & 66 & 39 & 22 & 12 & 5.3 & 2.3 & 0.81 & 0.55 & 0.31 & 0.18 & 0.26 \\
				$([\wcqdh]_{21})^2$ & 155 & 120 & 72 & 41 & 25 & 13 & 5.7 & 2.7 & 0.97 & 0.62 & 0.36 & 0.20 & 0.29 \\
				$([\wcqdh]_{22})^2$ & 49 & 33 & 22 & 6.0 & 2.6 & 2.6 & 1.6 & 0.49 & 0.16 & 0.081 & 0.062 & 0.033 & 0.045 \\
				$(\wcquh)^2$ & 1000 & 760 & 490 & 300 & 190 & 95 & 45 & 21 & 8.1 & 5.3 & 3.2 & 1.7 & 2.8 \\
				$(\wcqdh)^2$ & 530 & 410 & 250 & 150 & 87 & 46 & 20 & 8.5 & 3.4 & 2.1 & 1.1 & 0.73 & 0.97 \\
				$(\wcgh)^2/1000$  & 990 & 930 & 780 & 560 & 390 & 270 & 150 & 92 & 35 & 27 & 19 & 12 & 25 \\
				\hline\hline
			\end{tabular}
		}
		\caption{Fitted $\rm \phi SMEFT$ coefficients $\gamma_{ij}$ (Eq.~\eqref{eq:fit_polynomial}) for each $\MET$ bin (in $\rm TeV^4$). The columns present different exclusive bins in the ATLAS analysis~\cite{ATLAS:2021kxv}. $\wcqdh$ indicates all four $[\wcqdh]_{ij}$ , $i,j\in\{1,2\}$ are set to the same value, and similar for $\wcquh$.}
		\label{tab:coefficients_EFT}
	\end{table}
	\begin{table}[tbp]
		\centering
		\resizebox{1.0\textwidth}{!}{
			\begin{tabular}{c|cccccc|ccccccc}
				\hline\hline
				& \multicolumn{6}{c|}{Off-shell fitting} & \multicolumn{7}{|c}{On-shell fitting} \\
				\hline\hline
				Paramater, Bins & 200 & 250 & 300 & 350 & 400 & 500 & 600 & 700 & 800 & 900 & 1000 & 1100 & 1200 \\
				\hline
				$\kappa_{dd}$      & 4.47 & 3.04 & 1.89 & 1.10 & 0.92 & 0.48 & 0.29 & 0.16 & 0.11 & 0.09 & 0.06 & 0.04 & 0.09 \\
				$\kappa_{ss}$      & 0.46 & 0.28 & 0.14 & 0.065 & - & - & - & - & - & - & - & - & - \\
				$\kappa_{ds}$ & 5.80 & 4.00 & 2.53& 1.58 & 1.48 & 0.84 & 0.68 & 0.37 & 0.22 & 0.14 & 0.09 & 0.07 & 0.10  \\
				$\beta_{dd}$      &  &  &  &  &  &  & - & - & 0.28 & 0.42 & 0.71 & 0.88 &  6.44 \\
				$\beta_{ss}$      &  &  &  &  &  &  & 0.43 & 0.20 & 0.18 & 0.14 & 0.10  & 0.10 & 0.36 \\
				$\beta_{ds}$  &  &  &  &  &  &  & -9.65 & -4.77 & -2.05 & -0.63 & 0.25 & 0.31 & 5.41 \\
				\hline\hline
		\end{tabular}}
		\caption{Fitted VLQS coefficients $\beta_{mn},~\kappa_{mn}$ for Eq.~\eqref{eq:fit_polynomial} for each $\MET$ bin. Other relevant parameters are $y_1=1, M_{Q/D}=3$ TeV, $y_2=0$. The columns present different exclusive bins in the ATLAS analysis~\cite{ATLAS:2021kxv}. Entries marked with ``-'' correspond to coefficients whose fitted values are not statistically significant at the $2\sigma$ level. }
		\label{tab:coefficients_VLQS}
	\end{table}
	We also tested terms of the form $(y_{q/d}^d)^3 y_{q/d}^s$ and $y_{q/d}^d (y_{q/d}^s)^3$ in this fit, finding negligible coefficients that modify the total number of events by $\lesssim 1\%$. The `off-shell fitting’ corresponds to using only the coefficients $\kappa_{jk}$ in the fit in the region where the $Q/D$ states are dominantly off-shell, as guided by Fig.~\ref{fig:on_shell_effect}. In all other regions, both $\beta_{mn}$ and $\kappa_{mn}$ are included, where, $\beta_{mn}$ takes into account the width of Q/D for $M_{Q/D}=3$ TeV, $\Gamma\simeq 35.8~\text{GeV} \big[(y_{q/d}^d)^2+(y_{q/d}^s)^2\big]$.

	\bibliographystyle{JHEP}
	\bibliography{ref.bib}

@article{Dawson:2025dmi,
    author = "Dawson, Sally and Roy, Arnab and Valencia, German",
    title = "{Semi-visible higgs decay as a probe for new invisible particles}",
    eprint = "2511.09778",
    archivePrefix = "arXiv",
    primaryClass = "hep-ph",
    month = "11",
    year = "2025"
}

@article{Fuentes-Martin:2022jrf,
    author = {Fuentes-Mart\'\i{}n, Javier and K\"onig, Matthias and Pag\`es, Julie and Thomsen, Anders Eller and Wilsch, Felix},
    title = "{A proof of concept for matchete: an automated tool for matching effective theories}",
    eprint = "2212.04510",
    archivePrefix = "arXiv",
    primaryClass = "hep-ph",
    reportNumber = "MITP-22-105, TUM-HEP-1443/22, ZU-TH-58/22",
    doi = "10.1140/epjc/s10052-023-11726-1",
    journal = "Eur. Phys. J. C",
    volume = "83",
    number = "7",
    pages = "662",
    year = "2023"
}

@article{Belle-II:2023esi,
    author = "Adachi, I. and others",
    collaboration = "Belle-II",
    title = "{Evidence for $B^+ \to K^+ \nu\bar\nu$ decays}",
    eprint = "2311.14647",
    archivePrefix = "arXiv",
    primaryClass = "hep-ex",
    reportNumber = "Belle II Preprint 2023-017, KEK Preprint 2023-35",
    doi = "10.1103/PhysRevD.109.112006",
    journal = "Phys. Rev. D",
    volume = "109",
    number = "11",
    pages = "112006",
    year = "2024"
}

@article{Criado:2021trs,
    author = "Criado, Juan Carlos and Djouadi, Abdelhak and Perez-Victoria, Manuel and Santiago, Jose",
    title = "{A complete effective field theory for dark matter}",
    eprint = "2104.14443",
    archivePrefix = "arXiv",
    primaryClass = "hep-ph",
    doi = "10.1007/JHEP07(2021)081",
    journal = "JHEP",
    volume = "07",
    pages = "081",
    year = "2021"
}

@article{Aebischer:2022wnl,
    author = "Aebischer, Jason and Altmannshofer, Wolfgang and Jenkins, Elizabeth E. and Manohar, Aneesh V.",
    title = "{Dark matter effective field theory and an application to vector dark matter}",
    eprint = "2202.06968",
    archivePrefix = "arXiv",
    primaryClass = "hep-ph",
    doi = "10.1007/JHEP06(2022)086",
    journal = "JHEP",
    volume = "06",
    pages = "086",
    year = "2022"
}

@article{Bell:2016obu,
    author = "Bell, Nicole and Busoni, Giorgio and Kobakhidze, Archil and Long, David M. and Schmidt, Michael A.",
    title = "{Unitarisation of EFT Amplitudes for Dark Matter Searches at the LHC}",
    eprint = "1606.02722",
    archivePrefix = "arXiv",
    primaryClass = "hep-ph",
    doi = "10.1007/JHEP08(2016)125",
    journal = "JHEP",
    volume = "08",
    pages = "125",
    year = "2016"
}

@article{Busoni:2014haa,
    author = "Busoni, Giorgio and De Simone, Andrea and Jacques, Thomas and Morgante, Enrico and Riotto, Antonio",
    title = "{On the Validity of the Effective Field Theory for Dark Matter Searches at the LHC Part III: Analysis for the $t$-channel}",
    eprint = "1405.3101",
    archivePrefix = "arXiv",
    primaryClass = "hep-ph",
    doi = "10.1088/1475-7516/2014/09/022",
    journal = "JCAP",
    volume = "09",
    pages = "022",
    year = "2014"
}

@article{Herren:2017osy,
    author = "Herren, Florian and Steinhauser, Matthias",
    title = "{Version 3 of RunDec and CRunDec}",
    eprint = "1703.03751",
    archivePrefix = "arXiv",
    primaryClass = "hep-ph",
    reportNumber = "TTP17-011",
    doi = "10.1016/j.cpc.2017.11.014",
    journal = "Comput. Phys. Commun.",
    volume = "224",
    pages = "333--345",
    year = "2018"
}

@article{Chetyrkin:2000yt,
    author = "Chetyrkin, K. G. and Kuhn, Johann H. and Steinhauser, M.",
    title = "{RunDec: A Mathematica package for running and decoupling of the strong coupling and quark masses}",
    eprint = "hep-ph/0004189",
    archivePrefix = "arXiv",
    reportNumber = "DESY-00-034, TTP-00-05",
    doi = "10.1016/S0010-4655(00)00155-7",
    journal = "Comput. Phys. Commun.",
    volume = "133",
    pages = "43--65",
    year = "2000"
}

@article{Busoni:2014sya,
    author = "Busoni, Giorgio and De Simone, Andrea and Gramling, Johanna and Morgante, Enrico and Riotto, Antonio",
    title = "{On the Validity of the Effective Field Theory for Dark Matter Searches at the LHC, Part II: Complete Analysis for the $s$-channel}",
    eprint = "1402.1275",
    archivePrefix = "arXiv",
    primaryClass = "hep-ph",
    reportNumber = "SISSA-04-2014-FISI",
    doi = "10.1088/1475-7516/2014/06/060",
    journal = "JCAP",
    volume = "06",
    pages = "060",
    year = "2014"
}

@article{Busoni:2013lha,
    author = "Busoni, Giorgio and De Simone, Andrea and Morgante, Enrico and Riotto, Antonio",
    title = "{On the Validity of the Effective Field Theory for Dark Matter Searches at the LHC}",
    eprint = "1307.2253",
    archivePrefix = "arXiv",
    primaryClass = "hep-ph",
    reportNumber = "CERN-PH-TH-2013-151, SISSA-29-2013-FISI",
    doi = "10.1016/j.physletb.2013.11.069",
    journal = "Phys. Lett. B",
    volume = "728",
    pages = "412--421",
    year = "2014"
}

@article{Huang:2020hdv,
    author = "Huang, Guo-yuan and Zhou, Shun",
    title = "{Precise Values of Running Quark and Lepton Masses in the Standard Model}",
    eprint = "2009.04851",
    archivePrefix = "arXiv",
    primaryClass = "hep-ph",
    doi = "10.1103/PhysRevD.103.016010",
    journal = "Phys. Rev. D",
    volume = "103",
    number = "1",
    pages = "016010",
    year = "2021"
}

@article{LZ:2024zvo,
    author = "Aalbers, J. and others",
    collaboration = "LZ",
    title = "{Dark Matter Search Results from 4.2 Tonne-Years of Exposure of the LUX-ZEPLIN (LZ) Experiment}",
    eprint = "2410.17036",
    archivePrefix = "arXiv",
    primaryClass = "hep-ex",
    reportNumber = "FERMILAB-PUB-24-0796-V",
    month = "10",
    year = "2024"
}

@article{He:2025jfc,
    author = "He, Xiao-Gang and Ma, Xiao-Dong and Tandean, Jusak and Valencia, German",
    title = "{B {\textrightarrow} K+ invisible, dark matter, and CP violation in hyperon decays}",
    eprint = "2502.09603",
    archivePrefix = "arXiv",
    primaryClass = "hep-ph",
    doi = "10.1007/JHEP07(2025)078",
    journal = "JHEP",
    volume = "07",
    pages = "078",
    year = "2025"
}

@article{He:2024iju,
    author = "He, Xiao-Gang and Ma, Xiao-Dong and Schmidt, Michael A. and Valencia, German and Volkas, Raymond R.",
    title = "{Scalar dark matter explanation of the excess in the Belle II B$^{+}$\textrightarrow{} K$^{+}$+ invisible measurement}",
    eprint = "2403.12485",
    archivePrefix = "arXiv",
    primaryClass = "hep-ph",
    reportNumber = "CPPC-2024-04",
    doi = "10.1007/JHEP07(2024)168",
    journal = "JHEP",
    volume = "07",
    pages = "168",
    year = "2024"
}

@article{ATLAS:2021kxv,
    author = "Aad, Georges and others",
    collaboration = "ATLAS",
    title = "{Search for new phenomena in events with an energetic jet and missing transverse momentum in $pp$ collisions at $\sqrt {s}$ =13  TeV with the ATLAS detector}",
    eprint = "2102.10874",
    archivePrefix = "arXiv",
    primaryClass = "hep-ex",
    reportNumber = "CERN-EP-2020-238",
    doi = "10.1103/PhysRevD.103.112006",
    journal = "Phys. Rev. D",
    volume = "103",
    number = "11",
    pages = "112006",
    year = "2021"
}

@article{DarkSide:2022dhx,
    author = "Agnes, P. and others",
    collaboration = "DarkSide",
    title = "{Search for Dark-Matter\textendash{}Nucleon Interactions via Migdal Effect with DarkSide-50}",
    eprint = "2207.11967",
    archivePrefix = "arXiv",
    primaryClass = "hep-ex",
    doi = "10.1103/PhysRevLett.130.101001",
    journal = "Phys. Rev. Lett.",
    volume = "130",
    number = "10",
    pages = "101001",
    year = "2023"
}

@article{Hirschi:2015iia,
    author = "Hirschi, Valentin and Mattelaer, Olivier",
    title = "{Automated event generation for loop-induced processes}",
    eprint = "1507.00020",
    archivePrefix = "arXiv",
    primaryClass = "hep-ph",
    reportNumber = "IPPP-15-35, DCPT-15-70, MCNET-15-14",
    doi = "10.1007/JHEP10(2015)146",
    journal = "JHEP",
    volume = "10",
    pages = "146",
    year = "2015"
}

@article{Papucci:2014iwa,
    author = "Papucci, Michele and Vichi, Alessandro and Zurek, Kathryn M.",
    title = "{Monojet versus the rest of the world I: t-channel models}",
    eprint = "1402.2285",
    archivePrefix = "arXiv",
    primaryClass = "hep-ph",
    doi = "10.1007/JHEP11(2014)024",
    journal = "JHEP",
    volume = "11",
    pages = "024",
    year = "2014"
}

@article{Jacques:2016dqz,
    author = "Jacques, Thomas and Katz, Andrey and Morgante, Enrico and Racco, Davide and Rameez, Mohamed and Riotto, Antonio",
    title = "{Complementarity of DM searches in a consistent simplified model: the case of $Z^{\prime}$}",
    eprint = "1605.06513",
    archivePrefix = "arXiv",
    primaryClass = "hep-ph",
    reportNumber = "SISSA-22-2016-FISI, CERN-TH-2016-099, SISSA 22/2016/FISI",
    doi = "10.1007/JHEP10(2016)071",
    journal = "JHEP",
    volume = "10",
    pages = "071",
    year = "2016",
    note = "[Erratum: JHEP 01, 127 (2019)]"
}

@article{Chala:2015ama,
    author = "Chala, Mikael and Kahlhoefer, Felix and McCullough, Matthew and Nardini, Germano and Schmidt-Hoberg, Kai",
    title = "{Constraining Dark Sectors with Monojets and Dijets}",
    eprint = "1503.05916",
    archivePrefix = "arXiv",
    primaryClass = "hep-ph",
    reportNumber = "DESY-15-036, CERN-PH-TH-2015-057",
    doi = "10.1007/JHEP07(2015)089",
    journal = "JHEP",
    volume = "07",
    pages = "089",
    year = "2015"
}

@article{Liew:2016oon,
    author = "Liew, Seng Pei and Papucci, Michele and Vichi, Alessandro and Zurek, Kathryn M.",
    title = "{Mono-X Versus Direct Searches: Simplified Models for Dark Matter at the LHC}",
    eprint = "1612.00219",
    archivePrefix = "arXiv",
    primaryClass = "hep-ph",
    doi = "10.1007/JHEP06(2017)082",
    journal = "JHEP",
    volume = "06",
    pages = "082",
    year = "2017"
}

@article{Belyaev:2018pqr,
    author = "Belyaev, Alexander and Bertuzzo, Enrico and Caniu Barros, Cristian and Eboli, Oscar and Grilli Di Cortona, Giovanni and Iocco, Fabio and Pukhov, Alexander",
    title = "{Interplay of the LHC and non-LHC Dark Matter searches in the Effective Field Theory approach}",
    eprint = "1807.03817",
    archivePrefix = "arXiv",
    primaryClass = "hep-ph",
    doi = "10.1103/PhysRevD.99.015006",
    journal = "Phys. Rev. D",
    volume = "99",
    number = "1",
    pages = "015006",
    year = "2019"
}

@article{Bruggisser:2016nzw,
    author = "Bruggisser, Sebastian and Riva, Francesco and Urbano, Alfredo",
    title = "{The Last Gasp of Dark Matter Effective Theory}",
    eprint = "1607.02475",
    archivePrefix = "arXiv",
    primaryClass = "hep-ph",
    reportNumber = "DESY-16-123, CERN-TH-2016-154",
    doi = "10.1007/JHEP11(2016)069",
    journal = "JHEP",
    volume = "11",
    pages = "069",
    year = "2016"
}

@article{Pobbe:2017wrj,
    author = "Pobbe, F. and Wulzer, A. and Zanetti, M.",
    title = "{Setting limits on Effective Field Theories: the case of Dark Matter}",
    eprint = "1704.00736",
    archivePrefix = "arXiv",
    primaryClass = "hep-ph",
    reportNumber = "CERN-TH-2017-074",
    doi = "10.1007/JHEP08(2017)074",
    journal = "JHEP",
    volume = "08",
    pages = "074",
    year = "2017"
}

@article{Belwal:2017nkw,
    author = "Belwal, Swasti and Drees, Manuel and Kim, Jong Soo",
    title = "{Analysis of the Bounds on Dark Matter Models from Monojet Searches at the LHC}",
    eprint = "1709.08545",
    archivePrefix = "arXiv",
    primaryClass = "hep-ph",
    doi = "10.1103/PhysRevD.98.055017",
    journal = "Phys. Rev. D",
    volume = "98",
    number = "5",
    pages = "055017",
    year = "2018"
}

@article{Racco:2015dxa,
    author = "Racco, Davide and Wulzer, Andrea and Zwirner, Fabio",
    title = "{Robust collider limits on heavy-mediator Dark Matter}",
    eprint = "1502.04701",
    archivePrefix = "arXiv",
    primaryClass = "hep-ph",
    reportNumber = "CERN-PH-TH-2015-014, DFPD-2015-TH-05",
    doi = "10.1007/JHEP05(2015)009",
    journal = "JHEP",
    volume = "05",
    pages = "009",
    year = "2015"
}

@article{DeSimone:2016fbz,
    author = "De Simone, Andrea and Jacques, Thomas",
    title = "{Simplified models vs. effective field theory approaches in dark matter searches}",
    eprint = "1603.08002",
    archivePrefix = "arXiv",
    primaryClass = "hep-ph",
    reportNumber = "SISSA-21-2016-FISI",
    doi = "10.1140/epjc/s10052-016-4208-4",
    journal = "Eur. Phys. J. C",
    volume = "76",
    number = "7",
    pages = "367",
    year = "2016"
}

@article{Bauer:2016pug,
    author = "Bauer, Martin and Butter, Anja and Desai, Nishita and Gonzalez-Fraile, Juan and Plehn, Tilman",
    title = "{Validity of dark matter effective theory}",
    eprint = "1611.09908",
    archivePrefix = "arXiv",
    primaryClass = "hep-ph",
    doi = "10.1103/PhysRevD.95.075036",
    journal = "Phys. Rev. D",
    volume = "95",
    number = "7",
    pages = "075036",
    year = "2017"
}

@article{Butterworth:2015oua,
    author = "Butterworth, Jon and others",
    title = "{PDF4LHC recommendations for LHC Run II}",
    eprint = "1510.03865",
    archivePrefix = "arXiv",
    primaryClass = "hep-ph",
    reportNumber = "OUTP-15-17P, SMU-HEP-15-12, TIF-UNIMI-2015-14, LCTS-2015-27, CERN-PH-TH-2015-249",
    doi = "10.1088/0954-3899/43/2/023001",
    journal = "J. Phys. G",
    volume = "43",
    pages = "023001",
    year = "2016"
}

@article{Sjostrand:2006za,
    author = "Sjostrand, Torbjorn and Mrenna, Stephen and Skands, Peter Z.",
    title = "{PYTHIA 6.4 Physics and Manual}",
    eprint = "hep-ph/0603175",
    archivePrefix = "arXiv",
    reportNumber = "FERMILAB-PUB-06-052-CD-T, LU-TP-06-13",
    doi = "10.1088/1126-6708/2006/05/026",
    journal = "JHEP",
    volume = "05",
    pages = "026",
    year = "2006"
}

@article{Sjostrand:2007gs,
    author = "Sjostrand, Torbjorn and Mrenna, Stephen and Skands, Peter Z.",
    title = "{A Brief Introduction to PYTHIA 8.1}",
    eprint = "0710.3820",
    archivePrefix = "arXiv",
    primaryClass = "hep-ph",
    reportNumber = "CERN-LCGAPP-2007-04, LU-TP-07-28, FERMILAB-PUB-07-512-CD-T",
    doi = "10.1016/j.cpc.2008.01.036",
    journal = "Comput. Phys. Commun.",
    volume = "178",
    pages = "852--867",
    year = "2008"
}

@article{deFavereau:2013fsa,
    author = "de Favereau, J. and Delaere, C. and Demin, P. and Giammanco, A. and Lema\^\i{}tre, V. and Mertens, A. and Selvaggi, M.",
    collaboration = "DELPHES 3",
    title = "{DELPHES 3, A modular framework for fast simulation of a generic collider experiment}",
    eprint = "1307.6346",
    archivePrefix = "arXiv",
    primaryClass = "hep-ex",
    doi = "10.1007/JHEP02(2014)057",
    journal = "JHEP",
    volume = "02",
    pages = "057",
    year = "2014"
}

@article{Alwall:2014hca,
    author = "Alwall, J. and Frederix, R. and Frixione, S. and Hirschi, V. and Maltoni, F. and Mattelaer, O. and Shao, H. -S. and Stelzer, T. and Torrielli, P. and Zaro, M.",
    title = "{The automated computation of tree-level and next-to-leading order differential cross sections, and their matching to parton shower simulations}",
    eprint = "1405.0301",
    archivePrefix = "arXiv",
    primaryClass = "hep-ph",
    reportNumber = "CERN-PH-TH-2014-064, CP3-14-18, LPN14-066, MCNET-14-09, ZU-TH-14-14",
    doi = "10.1007/JHEP07(2014)079",
    journal = "JHEP",
    volume = "07",
    pages = "079",
    year = "2014"
}

@article{Roy:2024avj,
    author = "Roy, Arnab and Valencia, German",
    title = "{High-p$_{T}$ LHC constraints on SMEFT operators affecting rare kaon and hyperon decays}",
    eprint = "2410.05859",
    archivePrefix = "arXiv",
    primaryClass = "hep-ph",
    doi = "10.1007/JHEP05(2025)088",
    journal = "JHEP",
    volume = "05",
    pages = "088",
    year = "2025"
}

@article{Guchait:2022ktz,
    author = "Guchait, Monoranjan and Roy, Arnab",
    title = "{Exploring SMEFT operators in the tHq production at the LHC}",
    eprint = "2210.05503",
    archivePrefix = "arXiv",
    primaryClass = "hep-ph",
    doi = "10.1007/JHEP10(2023)064",
    journal = "JHEP",
    volume = "10",
    pages = "064",
    year = "2023"
}

@article{Backovic:2013dpa,
    author = "Backovic, Mihailo and Kong, Kyoungchul and McCaskey, Mathew",
    title = "{MadDM v.1.0: Computation of Dark Matter Relic Abundance Using MadGraph5}",
    eprint = "1308.4955",
    archivePrefix = "arXiv",
    primaryClass = "hep-ph",
    doi = "10.1016/j.dark.2014.04.001",
    journal = "Physics of the Dark Universe",
    volume = "5-6",
    pages = "18--28",
    year = "2014"
}

@article{Backovic:2015cra,
    author = "Backovi\'c, Mihailo and Martini, Antony and Mattelaer, Olivier and Kong, Kyoungchul and Mohlabeng, Gopolang",
    title = "{Direct Detection of Dark Matter with MadDM v.2.0}",
    eprint = "1505.04190",
    archivePrefix = "arXiv",
    primaryClass = "hep-ph",
    doi = "10.1016/j.dark.2015.09.001",
    journal = "Phys. Dark Univ.",
    volume = "9-10",
    pages = "37--50",
    month = "5",
    year = "2015"
}

@article{PandaX:2022xqx,
    author = "Li, Shuaijie and others",
    collaboration = "PandaX",
    title = "{Search for Light Dark Matter with Ionization Signals in the PandaX-4T Experiment}",
    eprint = "2212.10067",
    archivePrefix = "arXiv",
    primaryClass = "hep-ex",
    doi = "10.1103/PhysRevLett.130.261001",
    journal = "Phys. Rev. Lett.",
    volume = "130",
    number = "26",
    pages = "261001",
    year = "2023"
}

@article{PandaX:2023xgl,
    author = "Huang, Di and others",
    collaboration = "PandaX",
    title = "{Search for Dark-Matter\textendash{}Nucleon Interactions with a Dark Mediator in PandaX-4T}",
    eprint = "2308.01540",
    archivePrefix = "arXiv",
    primaryClass = "hep-ex",
    doi = "10.1103/PhysRevLett.131.191002",
    journal = "Phys. Rev. Lett.",
    volume = "131",
    number = "19",
    pages = "191002",
    year = "2023"
}

@article{Migdal:1941,
    author = "A. B. Migdal",
    title = "{Ionization of atoms accompanying $\alpha$-and $\beta$-decay}",
    journal = "J. Phys. 4, 449 (1941)",
    year = "1941"
}

@article{Ibe:2017yqa,
    author = "Ibe, Masahiro and Nakano, Wakutaka and Shoji, Yutaro and Suzuki, Kazumine",
    title = "{Migdal Effect in Dark Matter Direct Detection Experiments}",
    eprint = "1707.07258",
    archivePrefix = "arXiv",
    primaryClass = "hep-ph",
    reportNumber = "IPMU17-0100",
    doi = "10.1007/JHEP03(2018)194",
    journal = "JHEP",
    volume = "03",
    pages = "194",
    year = "2018"
}

@article{CMS:2022qva,
    author = "Tumasyan, Armen and others",
    collaboration = "CMS",
    title = "{Search for invisible decays of the Higgs boson produced via vector boson fusion in proton-proton collisions at $\sqrt{s}=13$ TeV}",
    eprint = "2201.11585",
    archivePrefix = "arXiv",
    primaryClass = "hep-ex",
    reportNumber = "CMS-HIG-20-003, CERN-EP-2021-273",
    doi = "10.1103/PhysRevD.105.092007",
    journal = "Phys. Rev. D",
    volume = "105",
    number = "9",
    pages = "092007",
    year = "2022"
}

@article{ATLAS:2022yvh,
    author = "Aad, Georges and others",
    collaboration = "ATLAS",
    title = "{Search for invisible Higgs-boson decays in events with vector-boson fusion signatures using 139 fb$^{-1}$ of proton-proton data recorded by the ATLAS experiment}",
    eprint = "2202.07953",
    archivePrefix = "arXiv",
    primaryClass = "hep-ex",
    reportNumber = "CERN-EP-2021-258",
    doi = "10.1007/JHEP08(2022)104",
    journal = "JHEP",
    volume = "08",
    pages = "104",
    year = "2022"
}

@article{Strigari:2009bq,
    author = "Strigari, Louis E.",
    title = "{Neutrino Coherent Scattering Rates at Direct Dark Matter Detectors}",
    eprint = "0903.3630",
    archivePrefix = "arXiv",
    primaryClass = "astro-ph.CO",
    doi = "10.1088/1367-2630/11/10/105011",
    journal = "New J. Phys.",
    volume = "11",
    pages = "105011",
    year = "2009"
}

@article{Billard:2013qya,
    author = "Billard, J. and Strigari, L. and Figueroa-Feliciano, E.",
    title = "{Implication of neutrino backgrounds on the reach of next generation dark matter direct detection experiments}",
    eprint = "1307.5458",
    archivePrefix = "arXiv",
    primaryClass = "hep-ph",
    doi = "10.1103/PhysRevD.89.023524",
    journal = "Phys. Rev. D",
    volume = "89",
    number = "2",
    pages = "023524",
    year = "2014"
}

@article{Monroe:2007xp,
    author = "Monroe, Jocelyn and Fisher, Peter",
    title = "{Neutrino Backgrounds to Dark Matter Searches}",
    eprint = "0706.3019",
    archivePrefix = "arXiv",
    primaryClass = "astro-ph",
    doi = "10.1103/PhysRevD.76.033007",
    journal = "Phys. Rev. D",
    volume = "76",
    pages = "033007",
    year = "2007"
}

@article{Arcadi:2021mag,
    author = "Arcadi, Giorgio and Djouadi, Abdelhak and Kado, Marumi",
    title = "{The Higgs-portal for dark matter: effective field theories versus concrete realizations}",
    eprint = "2101.02507",
    archivePrefix = "arXiv",
    primaryClass = "hep-ph",
    doi = "10.1140/epjc/s10052-021-09411-2",
    journal = "Eur. Phys. J. C",
    volume = "81",
    number = "7",
    pages = "653",
    year = "2021"
}

@article{DelNobile:2021wmp,
    author = "Del Nobile, Eugenio",
    title = "{The Theory of Direct Dark Matter Detection: A Guide to Computations}",
    eprint = "2104.12785",
    archivePrefix = "arXiv",
    primaryClass = "hep-ph",
    doi = "10.1007/978-3-030-95228-0",
    month = "4",
    year = "2021"
}

@article{Cowan:2010js,
    author = "Cowan, Glen and Cranmer, Kyle and Gross, Eilam and Vitells, Ofer",
    title = "{Asymptotic formulae for likelihood-based tests of new physics}",
    eprint = "1007.1727",
    archivePrefix = "arXiv",
    primaryClass = "physics.data-an",
    doi = "10.1140/epjc/s10052-011-1554-0",
    journal = "Eur. Phys. J. C",
    volume = "71",
    pages = "1554",
    year = "2011",
    note = "[Erratum: Eur.Phys.J.C 73, 2501 (2013)]"
}

@article{DEAP:2019yzn,
    author = "Ajaj, R. and others",
    collaboration = "DEAP",
    title = "{Search for dark matter with a 231-day exposure of liquid argon using DEAP-3600 at SNOLAB}",
    eprint = "1902.04048",
    archivePrefix = "arXiv",
    primaryClass = "astro-ph.CO",
    doi = "10.1103/PhysRevD.100.022004",
    journal = "Phys. Rev. D",
    volume = "100",
    number = "2",
    pages = "022004",
    year = "2019"
}

@article{Planck:2018vyg,
    author = "Aghanim, N. and others",
    collaboration = "Planck",
    title = "{Planck 2018 results. VI. Cosmological parameters}",
    eprint = "1807.06209",
    archivePrefix = "arXiv",
    primaryClass = "astro-ph.CO",
    doi = "10.1051/0004-6361/201833910",
    journal = "Astron. Astrophys.",
    volume = "641",
    pages = "A6",
    year = "2020",
    note = "[Erratum: Astron.Astrophys. 652, C4 (2021)]"
}

@article{FlavourLatticeAveragingGroupFLAG:2024oxs,
    author = "Aoki, Y. and others",
    collaboration = "Flavour Lattice Averaging Group (FLAG)",
    title = "{FLAG Review 2024}",
    eprint = "2411.04268",
    archivePrefix = "arXiv",
    primaryClass = "hep-lat",
    reportNumber = "CERN-TH-2024-192, FERMILAB-PUB-24-0785-T",
    month = "11",
    year = "2024"
}

@article{Endo:2014mja,
    author = "Endo, Motoi and Yamamoto, Yasuhiro",
    title = "{Unitarity Bounds on Dark Matter Effective Interactions at LHC}",
    eprint = "1403.6610",
    archivePrefix = "arXiv",
    primaryClass = "hep-ph",
    reportNumber = "UT-14-11, UG-FT-309-14, CAFPE-179-14",
    doi = "10.1007/JHEP06(2014)126",
    journal = "JHEP",
    volume = "06",
    pages = "126",
    year = "2014"
}

@article{Cohen:2021gdw,
    author = "Cohen, Timothy and Doss, Joel and Lu, Xiaochuan",
    title = "{Unitarity bounds on effective field theories at the LHC}",
    eprint = "2111.09895",
    archivePrefix = "arXiv",
    primaryClass = "hep-ph",
    doi = "10.1007/JHEP04(2022)155",
    journal = "JHEP",
    volume = "04",
    pages = "155",
    year = "2022"
}

@article{Logan:2022uus,
    author = "Logan, Heather E.",
    title = "{Lectures on perturbative unitarity and decoupling in Higgs physics}",
    eprint = "2207.01064",
    archivePrefix = "arXiv",
    primaryClass = "hep-ph",
    month = "7",
    year = "2022"
}

@article{Benbrik:2024fku,
    author = "Benbrik, Rachid and Boukidi, Mohammed and Ech-chaouy, Mohamed and Moretti, Stefano and Salime, Khawla and Yan, Qi-Shu",
    title = "{Vector-Like Quarks at the LHC: A unified perspective from ATLAS and CMS exclusion limits}",
    eprint = "2412.01761",
    archivePrefix = "arXiv",
    primaryClass = "hep-ph",
    doi = "10.1007/JHEP03(2025)020",
    journal = "JHEP",
    volume = "03",
    pages = "020",
    year = "2025"
}

@article{Roy:2024ear,
    author = "Roy, Arnab and Dasgupta, Basudeb and Guchait, Monoranjan",
    title = "{Constraining Asymmetric Dark Matter using colliders and direct detection}",
    eprint = "2402.17265",
    archivePrefix = "arXiv",
    primaryClass = "hep-ph",
    doi = "10.1007/JHEP08(2024)095",
    journal = "JHEP",
    volume = "08",
    pages = "095",
    year = "2024"
}

@article{Ghosh:2025agw,
    author = "Ghosh, Anupam",
    title = "{Unveiling a natural multicomponent dark sector: an inert doublet guided by Peccei{\textendash}Quinn}",
    doi = "10.1140/epjp/s13360-025-06621-5",
    journal = "Eur. Phys. J. Plus",
    volume = "140",
    number = "7",
    pages = "688",
    year = "2025"
}

@article{Ghosh:2023xhs,
    author = "Ghosh, Anupam and Konar, Partha",
    title = "{Precision prediction of a democratic up-family philic KSVZ axion model at the LHC}",
    eprint = "2305.08662",
    archivePrefix = "arXiv",
    primaryClass = "hep-ph",
    doi = "10.1016/j.dark.2024.101746",
    journal = "Phys. Dark Univ.",
    volume = "47",
    pages = "101746",
    year = "2025"
}

@article{Ghosh:2022rta,
    author = "Ghosh, Anupam and Konar, Partha and Roshan, Rishav",
    title = "{Top-philic dark matter in a hybrid KSVZ axion framework}",
    eprint = "2207.00487",
    archivePrefix = "arXiv",
    primaryClass = "hep-ph",
    doi = "10.1007/JHEP12(2022)167",
    journal = "JHEP",
    volume = "12",
    pages = "167",
    year = "2022"
}

@article{Ghosh:2024nkj,
    author = "Ghosh, Anupam and Konar, Partha and Show, Sudipta",
    title = "{Collider fingerprints of freeze-in dark matter produced during the fast expansion phase of Universe}",
    eprint = "2411.09464",
    archivePrefix = "arXiv",
    primaryClass = "hep-ph",
    doi = "10.1103/pj7s-zhcr",
    journal = "Phys. Rev. D",
    volume = "112",
    number = "5",
    pages = "055012",
    year = "2025"
}

@article{Das:2024xle,
    author = "Das, Prasanta Kumar and Dey, Shyamashish and Kundu, Saumyen and Rai, Santosh Kumar",
    title = "{Revisiting the inert scalar dark matter with vector-like quarks}",
    eprint = "2412.17719",
    archivePrefix = "arXiv",
    primaryClass = "hep-ph",
    reportNumber = "HRI-RECAPP-2024-07",
    doi = "10.1007/JHEP05(2025)027",
    journal = "JHEP",
    volume = "05",
    pages = "027",
    year = "2025"
}

@article{Bhattacharya:2025mlg,
    author = "Bhattacharya, Subhaditya and Kolay, Lipika and Pradhan, Dipankar and Sarkar, Abhik",
    title = "{Up-type FCNC in presence of Dark Matter}",
    eprint = "2504.20045",
    archivePrefix = "arXiv",
    primaryClass = "hep-ph",
    month = "4",
    year = "2025"
}

@article{Babu:2021hef,
    author = "Babu, K. S. and Jana, Sudip and Thapa, Anil",
    title = "{Vector boson dark matter from trinification}",
    eprint = "2112.12771",
    archivePrefix = "arXiv",
    primaryClass = "hep-ph",
    doi = "10.1007/JHEP02(2022)051",
    journal = "JHEP",
    volume = "02",
    pages = "051",
    year = "2022"
}

@article{Borah:2020nsz,
    author = "Borah, Debasish and Roshan, Rishav and Sil, Arunansu",
    title = "{Sub-TeV singlet scalar dark matter and electroweak vacuum stability with vectorlike fermions}",
    eprint = "2007.14904",
    archivePrefix = "arXiv",
    primaryClass = "hep-ph",
    doi = "10.1103/PhysRevD.102.075034",
    journal = "Phys. Rev. D",
    volume = "102",
    number = "7",
    pages = "075034",
    year = "2020"
}

@article{Moretti:2017qby,
    author = "Moretti, Stefano and O'Brien, Dermot and Panizzi, Luca and Prager, Hugo",
    title = "{Production of extra quarks decaying to Dark Matter beyond the Narrow Width Approximation at the LHC}",
    eprint = "1705.07675",
    archivePrefix = "arXiv",
    primaryClass = "hep-ph",
    doi = "10.1103/PhysRevD.96.035033",
    journal = "Phys. Rev. D",
    volume = "96",
    number = "3",
    pages = "035033",
    year = "2017"
}

@article{Olgoso:2025jot,
    author = "Olgoso, Pablo and Paradisi, Paride and Selimovic, Nudzeim",
    title = "{The Dark Side of a Tera-Z Factory}",
    eprint = "2507.17803",
    archivePrefix = "arXiv",
    primaryClass = "hep-ph",
    month = "7",
    year = "2025"
}

@article{Ruhdorfer:2024dgz,
    author = "Ruhdorfer, Maximilian and Salvioni, Ennio and Wulzer, Andrea",
    title = "{Building the case for forward muon detection at a muon collider}",
    eprint = "2411.00096",
    archivePrefix = "arXiv",
    primaryClass = "hep-ph",
    doi = "10.1103/PhysRevD.111.053010",
    journal = "Phys. Rev. D",
    volume = "111",
    number = "5",
    pages = "053010",
    year = "2025"
}

@article{Arina:2025zpi,
    author = "Arina, Chiara and others",
    title = "{t-channel dark matter models {\textendash} a whitepaper}",
    eprint = "2504.10597",
    archivePrefix = "arXiv",
    primaryClass = "hep-ph",
    reportNumber = "CERN-LPCC-2025-001, IRMP-CP3-25-07, TTK-25-07",
    doi = "10.1140/epjc/s10052-025-14635-7",
    journal = "Eur. Phys. J. C",
    volume = "85",
    number = "9",
    pages = "975",
    year = "2025"
}

@article{Alloul:2013bka,
    author = "Alloul, Adam and Christensen, Neil D. and Degrande, C{\'e}line and Duhr, Claude and Fuks, Benjamin",
    title = "{FeynRules  2.0 - A complete toolbox for tree-level phenomenology}",
    eprint = "1310.1921",
    archivePrefix = "arXiv",
    primaryClass = "hep-ph",
    reportNumber = "CERN-PH-TH-2013-239, MCNET-13-14, IPPP-13-71, DCPT-13-142, PITT-PACC-1308",
    doi = "10.1016/j.cpc.2014.04.012",
    journal = "Comput. Phys. Commun.",
    volume = "185",
    pages = "2250--2300",
    year = "2014"
}

@article{Arina:2020udz,
    author = "Arina, Chiara and Fuks, Benjamin and Mantani, Luca",
    title = "{A universal framework for t-channel dark matter models}",
    eprint = "2001.05024",
    archivePrefix = "arXiv",
    primaryClass = "hep-ph",
    reportNumber = "CP3-20-01, MCNET-20-01",
    doi = "10.1140/epjc/s10052-020-7933-7",
    journal = "Eur. Phys. J. C",
    volume = "80",
    number = "5",
    pages = "409",
    year = "2020"
}

@article{Ghosh:2024boo,
    author = "Ghosh, Anupam and Konar, Partha",
    title = "{Unveiling desert region in inert doublet model assisted by Peccei-Quinn symmetry}",
    eprint = "2407.01415",
    archivePrefix = "arXiv",
    primaryClass = "hep-ph",
    doi = "10.1007/JHEP09(2024)104",
    journal = "JHEP",
    volume = "09",
    pages = "104",
    year = "2024"
}
\end{document}